\pdfoutput=1

\documentclass[a4paper,11pt]{article}
\usepackage{jheppub}

\usepackage{bm}

\usepackage{color}
\usepackage[usenames,dvipsnames,svgnames,table]{xcolor}

\usepackage{amsmath}
\usepackage{amssymb}
\usepackage{graphicx}
\usepackage{slashed}
\usepackage{soul}
\usepackage{lscape}
\usepackage{graphicx}
\usepackage{xcolor}
\usepackage{multirow}
\usepackage{placeins,hyperref}
\usepackage[separate-uncertainty=true]{siunitx}
\usepackage{caption}
\usepackage{subcaption}
\usepackage{mathrsfs}
\usepackage[version=3]{mhchem}
\usepackage{amsmath, amssymb, amscd, amsthm, amsfonts}

\usepackage{lineno}
\begin{document}

\title{Mineral Detection of Neutrinos and Dark Matter. \\ A Whitepaper}

%----------------------------------------------
\author[1]{Sebastian~Baum,}
\affiliation[1]{Stanford Institute for Theoretical Physics, Department of Physics, Stanford University, Stanford, CA 94305, USA}

\author[2]{Patrick~Stengel;}
\affiliation[2]{Istituto Nazionale di Fisica Nucleare, Sezione di Ferrara, via Giuseppe Saragat 1, I-44122 Ferrara, Italy}

\author[3]{Natsue~Abe,}
\affiliation[3]{Mantle Drilling Promotion Office, Japan Agency for Marine-Earth Science and Technology (JAMSTEC), Yokohama, Kanagawa 236-0001, Japan}

\author[4]{Javier~F.~Acevedo,}
\affiliation[4]{SLAC National Accelerator Laboratory, Stanford University, 2575 Sand Hill Road, Menlo Park, CA 94025, USA}

\author[5,a]{Gabriela~R.~Araujo,}
\affiliation[5]{Department of Physics, University of Zurich, Winterthurerstrasse 190, CH-8057 Zurich, Switzerland}
\note[a]{on behalf of the PALEOCCENE collaboration}

\author[6]{Yoshihiro~Asahara,}
\affiliation[6]{Department of Earth and Environmental Sciences, Graduate School of Environmental Studies, Nagoya University, Furo-cho, Chikusa-ku, Nagoya, 464-8601, Japan}

\author[7]{Frank~Avignone,}
\affiliation[7]{Department of Physics and Astronomy, University of South Carolina, Columbia, SC 29208, USA}

\author[8]{Levente~Balogh,}
\affiliation[8]{Mechanical and Materials Engineering, Queen's University, 130 Stuart Street, Kingston, ON, K7L 2V9, Canada}

\author[5]{Laura~Baudis,}

\author[9]{Yilda~Boukhtouchen,}
\affiliation[9]{Department of Physics, Engineering Physics, and Astronomy, Queen's University, 64 Bader Lane, Kingston, Ontario, K7L 2S8, Canada}

\author[9,10]{Joseph~Bramante,}
\affiliation[10]{Perimeter Institute for Theoretical Physics, Waterloo, ON N2J 2W9, Canada}

\author[4]{Pieter~Alexander~Breur,} 

\author[11]{Lorenzo~Caccianiga,}
\affiliation[11]{Istituto Nazionale di Fisica Nucleare, Sezione di Milano, Via Celoria 16, Milan, Italy}

\author[12]{Francesco~Capozzi,}
\affiliation[12]{Dipartimento di Scienze Fisiche e Chimiche, Universit{\`a} degli Studi dell'Aquila, 67100 L’Aquila, Italy}

\author[13]{Juan~I.~Collar,}
\affiliation[13]{Department of Physics, University of Chicago, Chicago IL 60637, USA}

\author[14,15]{Reza~Ebadi,}
\affiliation[14]{Department of Physics, University of Maryland, College Park, Maryland 20742, USA}
\affiliation[15]{Quantum Technology Center, University of Maryland, College Park, Maryland 20742, USA}

\author[16]{Thomas~Edwards,}
\affiliation[16]{The William H. Miller III Department of Physics and Astronomy, The Johns Hopkins University, Baltimore, MD 21218, USA}

\author[17]{Klaus~Eitel,}
\affiliation[17]{Institute for Astroparticle Physics, Karlsruhe Institute of Technology, 76021 Karlsruhe, Germany}

\author[17]{Alexey~Elykov,}

\author[18]{Rodney~C.~Ewing,}
\affiliation[18]{Earth and Planetary Sciences, Stanford University, Stanford, CA 94305-2115, USA}

\author[19,20]{Katherine~Freese,}
\affiliation[19]{Texas Center for Cosmology and Astroparticle Physics, Weinberg Institute for Theoretical Physics, Department of Physics, The University of Texas at Austin, Austin, TX 78712, USA}
\affiliation[20]{The Oskar Klein Centre, Department of Physics, Stockholm University, AlbaNova, SE-10691 Stockholm, Sweden}

\author[9]{Audrey~Fung,}

\author[21]{Claudio~Galelli,}
\affiliation[21]{Department of Physics, Universit{\`a} degli Studi di Milano, via Celoria 16, 20133 Milano, Italy}

\author[22]{Ulrich~A.~Glasmacher,}
\affiliation[22]{Institute of Earth Sciences, Heidelberg University, Im Neuenheimer Feld 234, 69120 Heidelberg, Germany}

\author[4]{Arianna~Gleason,}

\author[23]{Noriko~Hasebe,}
\affiliation[23]{Institute of Nature and Environmental Technology, Kanazawa University, Kanazawa 920-1192, Japan}

\author[24]{Shigenobu~Hirose,}
\affiliation[24]{Center for Mathematical Science and Advanced Technology (MAT), Japan Agency for Marine-Earth Science and Technology (JAMSTEC), Yokohama, Kanagawa 236-0001, Japan}

\author[25,26]{Shunsaku~Horiuchi,}
\affiliation[25]{Center for Neutrino Physics, Department of Physics, Virginia Tech, Blacksburg, VA 24061, USA}
\affiliation[26]{Kavli IPMU (WPI), UTIAS, The University of Tokyo, Kashiwa, Chiba 277-8583, Japan}

\author[27]{Yasushi~Hoshino,}
\affiliation[27]{Department of Physics, Kanagawa University, Rokkakubashi, Kanagawa-ku, Kanagawa-ken 221-8686, Japan}

\author[25,a]{Patrick~Huber,}

\author[28]{Yuki~Ido,}
\affiliation[28]{Department of Physics, Toho University, 2-2-1 Miyama Funabashi Chiba 2748510, Japan}

\author[29]{Yohei~Igami,}
\affiliation[29]{Division of Earth and Planetary Sciences, Kyoto University, Kitashirakawa-Oiwakecho, Sakyo-ku, Kyoto, 606-8502, Japan}

\author[30]{Norito~Ishikawa,}
\affiliation[30]{Japan Atomic Energy Agency (JAEA), Tokai-mura, Ibaraki 319-1195, Japan}

\author[31]{Yoshitaka~Itow,}
\affiliation[31]{Institute for Space-Earth Environmental Research, Nagoya University, Furo-cho, Chikusa-ku, Nagoya, 464-8601, Japan}

\author[32]{Takashi~Kamiyama,}
\affiliation[32]{Faculty of Engineering, Hokkaido University, Kita 13 Nishi 8, Kita-ku, Sapporo 060-8628, Japan}

\author[31]{Takenori~Kato,}

\author[33]{Bradley~J.~Kavanagh,}
\affiliation[33]{Instituto de F\'isica de Cantabria (IFCA), UC-CSIC, Avenida de Los Castros s/n, 39005 Santander, Spain}

\author[24]{Yoji~Kawamura,}

\author[34]{Shingo~Kazama,}
\affiliation[34]{Kobayashi-Maskawa Institute for the Origin of Particles and the Universe, Nagoya University, Furo-cho, Chikusa-ku, Nagoya, 464-8601, Japan}

\author[4]{Christopher~J.~Kenney,}

\author[5]{Ben~Kilminster,}

\author[6]{Yui~Kouketsu,}

\author[35]{Yukiko~Kozaka,}
\affiliation[35]{Center for Advanced Marine Core Research, Kochi University, B200 Monobe, Nankoku, Kochi, 783-8502, Japan}

\author[4,36]{Noah~A.~Kurinsky,}
\affiliation[36]{Kavli Institute for Particle Astrophysics and Cosmology, Stanford University, Stanford, CA 94035, USA}

\author[9]{Matthew~Leybourne,}

\author[9]{Thalles~Lucas,}

\author[37,38,39]{William~F.~McDonough,}
\affiliation[37]{Research Center for Neutrino Science, Tohoku University, Sendai, Miyagi, 980-8578, Japan}
\affiliation[38]{Department of Earth Sciences, Tohoku University, Sendai, Miyagi, 980-8578, Japan}
\affiliation[39]{Department of Geology, University of Maryland, College Park, MD 20742, USA}

\author[15,40]{Mason~C.~Marshall,}
\affiliation[40]{Department of Electrical and Computer Engineering, University of Maryland, College Park, Maryland 20742, USA}

\author[41]{Jose~Maria~Mateos,}
\affiliation[41]{Center for Microscopy and Image Analysis, University of Zurich, Switzerland}

\author[16]{Anubhav~Mathur,}

\author[6]{Katsuyoshi~Michibayashi,}

\author[9]{Sharlotte~Mkhonto,}

\author[42,43,44]{Kohta~Murase,}
\affiliation[42]{Department of Physics, Department of Astronomy \& Astrophysics, Center for Multimessenger Astrophysics, Institute for Gravitation and the Cosmos, The Pennsylvania State University, University Park, PA 16802, USA} 
\affiliation[43]{School of Natural Sciences, Institute for Advanced Study, Princeton, NJ 08540, USA}
\affiliation[44]{Center for Gravitational Physics and Quantum Information, Yukawa Institute for Theoretical Physics, Kyoto University, Kyoto 606-8502, Japan}

\author[28]{Tatsuhiro~Naka,}

\author[24]{Kenji~Oguni,}

\author[16]{Surjeet~Rajendran,}

\author[45]{Hitoshi~Sakane,}
\affiliation[45]{SHI-ATEX Co., Ltd., Saijo, Ehime 799-1393, Japan}

\author[11]{Paola~Sala,}

\author[46]{Kate~Scholberg,}
\affiliation[46]{Department of Physics, Duke University, Durham, NC, 27708, USA}

\author[9]{Ingrida~Semenec,}

\author[28]{Takuya~Shiraishi,}

\author[47]{Joshua~Spitz,}
\affiliation[47]{Department of Physics, University of Michigan, 450 Church. St, Ann Arbor, MI 48109, USA}

\author[48]{Kai~Sun,}
\affiliation[48]{Department of Materials Science and Engineering, University of Michigan, Ann Arbor, MI 48109, USA}

\author[49]{Katsuhiko~Suzuki,}
\affiliation[49]{Submarine Resources Research Center, Japan Agency for Marine-Earth Science and Technology (JAMSTEC), Yokosuka, Kanagawa 237-0061, Japan}

\author[16]{Erwin~H.~Tanin,}

\author[9]{Aaron~Vincent,}

\author[50]{Nikita~Vladimirov,}
\affiliation[50]{URPP Adaptive Brain Circuits in Development and Learning (AdaBD), University of Zurich, Winterthurerstrasse 190, Zurich, Switzerland}

\author[14,15,40]{Ronald~L.~Walsworth,}

\author[37]{and Hiroko~Watanabe}
%----------------------------------------------

\emailAdd{sbaum@stanford.edu}
\emailAdd{pstengel@fe.infn.it}

%\preprint{Report Number}

%*********************************************************
\abstract{Minerals are solid state nuclear track detectors -- nuclear recoils in a mineral leave latent damage to the crystal structure. Depending on the mineral and its temperature, the damage features are retained in the material from minutes (in low-melting point materials such as salts at a few hundred $^\circ$C) to timescales much larger than the 4.5\,Gyr-age of the Solar System (in refractory materials at room temperature). The damage features from the $\mathcal{O}(50)\,$MeV fission fragments left by spontaneous fission of $^{238}$U and other heavy unstable isotopes have long been used for {\it fission track dating} of geological samples. Laboratory studies have demonstrated the readout of defects caused by nuclear recoils with energies as small as $\mathcal{O}(1)\,$keV. This whitepaper discusses a wide range of possible applications of minerals as detectors for $E_R \gtrsim \mathcal{O}(1)\,$keV nuclear recoils: Using natural minerals, one could use the damage features accumulated over $\mathcal{O}(10)\,{\rm Myr} \textit{--} \mathcal{O}(1)\,$Gyr to measure astrophysical neutrino fluxes (from the Sun, supernovae, or cosmic rays interacting with the atmosphere) as well as search for Dark Matter. Using signals accumulated over months to few-years timescales in laboratory-manufactured minerals, one could measure reactor neutrinos or use them as Dark Matter detectors, potentially with directional sensitivity. Research groups in Europe, Asia, and America have started developing microscopy techniques to read out the $\mathcal{O}(1) \textit{--} \mathcal{O}(100)\,$nm damage features in crystals left by $\mathcal{O}(0.1) \textit{--} \mathcal{O}(100)\,$keV nuclear recoils. We report on the status and plans of these programs. The research program towards the realization of such detectors is highly interdisciplinary, combining geoscience, material science, applied and fundamental physics with techniques from quantum information and Artificial Intelligence.}
%*********************************************************

\maketitle
\flushbottom

%*********************************************************
\section{Introduction} \label{sec:Intro}
%*********************************************************
%{\color{blue} Coordinator: Sebastian Baum}

Minerals have been used as solid state nuclear track detectors for more than 50\,years~\cite{Fleischer:1964,Fleischer383,Fleischer:1965yv,GUO2012233}. If an atomic nucleus travels through a mineral, its interactions with the electrons and nuclei of the crystal lattice stop the ion, leaving behind latent damage to the crystal structure. Depending on the nature of the mineral and the ion's charge, mass, and kinetic energy, this damage can take multiple forms, including mechanical stress in the crystal lattice, changes to the electron density, local amorphization of the crystal, and vacancy defects. Crystal defects may be erased by re-crystallization in a process known as ``self-annealing'' on time-scales that depend on the crystal's temperature and composition. The time-scale for self-annealing ranges from minutes for low-melting-point materials such as salts at temperatures of a few hundred $^\circ$C to timescales orders of magnitude larger than the age of the Solar System in refractory materials (a geological class of materials resistant to heat), e.g., in diopside, $t_{\rm ann} \sim 10^{59}\,$yr at room temperatures~\cite{Fleischer:1965yv}. The defects in crystals can be read out using a number of different microscopy techniques, e.g., Transmission Electron Microscopy (TEM); Scanning Electron Microscopy (SEM); Scanning Probe Microscopy (SPM) techniques, such as Atomic Force Microscopy (AFM); X-ray microscopy; and optical microscopy.

The property of minerals to record and retain the traces of nuclear recoils is useful for a wide range of applications. Perhaps the best-known application is {\it fission track dating} of geological samples: any mineral contains trace amounts of heavy unstable nuclei such as $^{238}$U and $^{232}$Th. Such heavy nuclei undergo spontaneous fission, i.e., the heavy nucleus can split into (typically two) fission fragments which recoil with energies $\mathcal{O}(50)\,$MeV, giving rise to $\mathcal{O}(10)\,\mu$m-long damage tracks. By independently measuring the concentration of heavy radioactive elements in a sample and the density of fission tracks one can establish the {\it fission track age} of a sample (see, e.g., Refs.~\cite{Wagner:1992,Malusa:2018} for reviews). This dating technique is typically applied to minerals such as apatite, zircon, monazite, and titanite. The oldest established fission track ages are $\sim 0.8\,$Gyr in precambrian apatite~\cite{Murrell:2003,Murrell:2004,Hendriks:2007} and $\sim 2\,$Gyr in zircon~\cite{Montario:2009}, demonstrating the existence of geological environments that have been sufficiently cold and stable for fission tracks to be preserved over billion-year timescales on Earth. A less widely used radiogenic dating technique is {\it $\alpha$-recoil dating}: measuring the density of the $\mathcal{O}(30)\,$nm-long damage tracks left by the $\mathcal{O}(100)\,$keV recoils of the daughter nuclei in $\alpha$-decays in the $^{238}$U, $^{235}$U, $^{232}$Th, and $^{147}$Sm decay chains, one can determine the {\it $\alpha$-recoil track age} of minerals (see, e.g., Refs.~\cite{Goegen:2000,Glasmacher:2003}). Calibration studies in the laboratory using both low-energy ion implantation and fast neutron irradiation have demonstrated that damage features from nuclear recoils with energies as low as few keV in micas~\cite{Snowden-Ifft:1995zgn,Snowden-Ifft:1995rip} (see also section~\ref{sec:Studies-JAMSTEC}) can be read out.

In this whitepaper, we discuss a number of possible applications of minerals as nuclear track detectors: using either natural or laboratory-manufactured crystals, minerals could be used as passive, tamper-proof detectors for fast neutrons or nuclear-reactor neutrinos scattering off the nuclei in the crystal~\cite{Cogswell:2021qlq,Alfonso:2022meh}. Particular laboratory-grown crystals, e.g., diamond implanted with quantum defects, could be used as directional detectors for Dark Matter (DM), the missing 85\,\% of our Universe's matter budget~\cite{Rajendran:2017ynw,Marshall:2020azl,Ebadi:2022axg}. The ability of minerals to record and retain defects caused by nuclear recoils over geological timescales opens up the possibility of using natural minerals with ages of $\mathcal{O}(10)\,{\rm Myr} \textit{--} \mathcal{O}(1)\,$Gyr as {\it paleo-detectors}: reading out the nuclear damage tracks in $1\,$kg of material that has recorded defects from nuclear recoils for $1\,$Gyr, one would achieve the same exposure as a conventional laboratory-based detector with $10^6\,$tonnes of target mass would achieve in 1\,yr. Hence, paleo-detectors are promising as detectors for rare-event searches, e.g.\ to search for nuclear recoils caused by Dark Matter~\cite{Price:1986ky,Snowden-Ifft:1995zgn,Collar:1994mj,Engel:1995gw,Snowden-Ifft:1997vmx,Baum:2018tfw,Drukier:2018pdy,Edwards:2018hcf,Sidhu:2019qoa,Ebadi:2021cte,Acevedo:2021tbl,Baum:2021jak} or neutrinos from a number of astrophysical sources such as our Sun~\cite{Tapia-Arellano:2021cml}, supernovae in our Galaxy~\cite{Baum:2019fqm,Baum:2022wfc}, or produced by the interactions of cosmic rays with Earth's atmosphere~\cite{Jordan:2020gxx}. 

Besides the raw exposure enabling the search for rare events, the long time-scales over which paleo-detectors could record nuclear recoils open up another exciting possibility: by measuring the number of signal events in a series of minerals that have been recording and retaining tracks for different times, e.g., $100\,$Myr, $200\,$Myr, \ldots, $1\,$Gyr, one could measure the time-dependence of the signal rate from, e.g., Dark Matter or astrophysical neutrinos over hundred-Myr timescales~\cite{Baum:2019fqm,Jordan:2020gxx,Tapia-Arellano:2021cml,Baum:2021chx,Bramante:2021dyx}. In the context of Dark Matter searches, such time scales mean that one would no longer be sensitive to the density of Dark Matter in the Solar System today, but rather could infer the density of Dark Matter on the Solar System's path around the Milky Way, which has an orbital period of $\sim 250\,$Myr. In the context of astrophysical neutrino searches, one could, e.g., measure the time-dependence of the galactic supernova rate over 100\,Myr timescales, a proxy for the Milky Way's star formation history. From measuring solar neutrinos, one could infer the temperature-evolution of the Solar core on timescales comparable to the Sun's age, $t_\odot \sim 4.5\,$Gyr.

Of course, minerals have long been used as nuclear recoil detectors in Dark Matter and neutrino searches, see, for examples, Refs.~\cite{Ahlen:1987mn,Majorana:2013cem,EDELWEISS:2017lvq,Bernabei:2018jrt,SABRE:2018lfp,CRESST:2019jnq,COHERENT:2020iec,GERDA:2020xhi,Amare:2021yyu,LEGEND:2021bnm,COHERENT:2021xmm,SuperCDMS:2022kse,COSINE-100:2021zqh}. These detectors use active instrumentation to measure phonons or photons excited by nuclear recoils in the material or bolometrically measure the energy deposited in the material. The focus of this whitepaper is to instead use the latent damage to the crystal structure caused by nuclear recoils as a probe.

The readout of the damage in crystals left by nuclear recoils has been demonstrated with a number of microscopy techniques, including TEM, SEM, AFM, X-ray microscopy and optical microscopy~\cite{Fleischer:1964,Fleischer383,Fleischer:1965yv,Snowden-Ifft:1995zgn,Snowden-Ifft:1995rip,GUO2012233,BARTZ2013273,RODRIGUEZ2014150,Kouwenberg:2018}. In order to unleash the full potential of mineral detectors as detectors for neutrinos and Dark Matter, the throughput of existing microscopy techniques has to be scaled up to allow for the efficient readout of larger sized samples. To exemplify the challenge, note that the interactions of reactor, solar, or supernova neutrinos as well as from canonical Weakly Interacting Massive Particle (WIMP)-like Dark Matter particles in the $m_\chi \sim 0.1 \textit{--} 10^4\,$GeV mass range would give rise to $\mathcal{O}(0.1-100)\,$keV nuclear recoils. Such nuclear recoils cause damage features in minerals that are $\mathcal{O}(1)\textit{--}\mathcal{O}(100)\,$nm long. Scanning $\mathcal{O}(1)\,$kg of material, corresponding to a volume with linear dimensions of order 10\,cm, with the required spatial resolution is clearly an enormous task, that will require combining a host of microscopy techniques. As we will discuss further below, one promising approach is to use optical (superresolution) fluorescent microscopy to identify color centers (vacancy defects) in the crystal, and then investigate such candidate sites for nuclear damage tracks with a microscopy technique that allows for nm-scale resolution, e.g., TEM. During the last years, research groups in Europe, Asia, and America have started studying the feasibility of different aspects of this program, and we will report on the status and plans of these studies below.

There are two reasons why one can today envisage a successful program towards reading out damage features at the nm-scale in macroscopic volumes of minerals: first, a number of microscopy techniques driven by applications in the bio-sciences, chemistry, materials science and the miniaturization of integrated electronic circuits have made enormous progress during the last decades. To name a few examples, modern optical microscopy techniques such as confocal laser scanning, light-sheet, or structured illumination microscopy are now standard tools in many university laboratories as well as in commercial applications. The hard X-ray microscopy capabilities at synchrotron and Free-Electron Laser (FEL) light sources are rapidly increasing and tomography at the $\mathcal{O}(10)\,$nm resolution scale is within reach in the near future. In the sub-nm resolution regime, SPM techniques can scan samples with ever-increasing speed, and He-ion beam microscopy is a newly developed technique that is now commercially available. Second, searching for nm-sized features in an $\mathcal{O}(1)\,$kg sample is an enormous data analysis challenge. Modern Machine Learning techniques are now available, and are ideally suited to automatize this process of identifying patterns in image data. 

As discussed above, fission track dating and $\alpha$-recoil track dating are important tools in geoscience. Besides inferring the fission/$\alpha$-recoil  track age of samples, one can also use the confined length distribution of fission tracks to study the temperature history of samples. Currently, these applications are limited by the standard read-out techniques of the nuclear recoil tracks used in the geoscience community, mainly relying on preparing clean sample surfaces by e.g. cleaving, and then enlarging the damage features via chemical etching in order for them to be visible in optical microscopes. While this technique is well established, it fundamentally limits the range of samples to which these techniques can be applied, since the density of fission/$\alpha$-recoil tracks must be high enough such that they can be chemically etched at the prepared surface. If successful, the program of developing microscopy techniques for mineral detectors for neutrinos and Dark Matter discussed here would instead allow for a volumetric read-out of damage tracks. As we will discuss, this will enable a number of applications in geoscience, including the fission/$\alpha$-recoil track dating of low-U/Th samples and much more detailed studies on the temperature history of samples.

The remainder of this whitepaper is structured as follows: 
section~\ref{sec:MineralDetectors} gives an overview of minerals as nuclear track detector. We discuss in detail what is known about minerals as nuclear track detectors from their fission-track and $\alpha$-track dating applications in geoscience. 
In section~\ref{sec:MineralDetectors_DM} we discuss previous applications of natural minerals as paleo-detectors for Dark Matter by Snowden-Ifft and collaborators, and also give a brief discussion of some of the most important sources of backgrounds in paleo-detectors. 
Section~\ref{sec:Physics} discusses a wide range of possible applications of minerals as nuclear track detectors ranging from fundamental particle physics over astrophysics- and geoscience-applications to applied-science uses such as nuclear safeguarding. In particular, we describe the use of minerals to search for Dark Matter, astrophysical neutrinos, and reactor neutrinos. We also discuss how the readout techniques for the nuclear damage features in minerals developed for such applications could revolutionize fission-track and $\alpha$-track dating in geoscience. 
Section~\ref{sec:ReadOut} describes a number of promising microscopy techniques for the readout of the latent damage to the crystal structure in minerals, ranging from optical microscopy to microscopy techniques with sub-nm spatial resolution such as Transmission Electron Microscopy or Scanning Probe Microscopy techniques.
Section~\ref{sec:Studies} describes the status and plans of a number of experimental studies towards probing the feasibility of mineral detection of neutrinos and Dark Matter. Groups at research institutes in Europe, Asia, and America are pursuing such studies. In this whitepaper, we collect, for the first time, these studies and show some of their first results.
In section~\ref{sec:LaundryList} we discuss a number of near-future steps towards using minerals as detectors for astrophysical and reactor neutrinos as well as as Dark Matter detectors. 
Finally, we summarize in section~\ref{sec:Summary}.

%*********************************************************
\section{Mineral detectors} \label{sec:MineralDetectors}
%*********************************************************
%{\color{blue} Coordinator: Patrick Stengel, Ulrich Glasmacher}

Minerals record signatures of nuclear recoils from various sources, potentially over geological timescales. We primarily focus on nuclear recoil tracks in mineral detectors. As a recoiling nucleus passes through the crystal lattice of a mineral detector, the energy deposition due to ionization losses of the nucleus along the trajectory of the recoil can leave persistent damage in non-conducting materials. The damage tracks can subsequently be read out using a variety of microscopy techniques with vastly different spatial resolution. For example, if a chemical etchant is used to enhance the damage along the recoils of relatively heavy and energetic ions associated with spontaneous fission and $\alpha$-decay, then the tracks can be read out using optical microscopes. Similar techniques have also been applied in searches for much softer recoils associated with Dark Matter-nucleus scattering, but nanoscale imagining using scanning probe microscopy has been necessary to uncover the more subtle damage features. 

The lattice damage from nuclear recoils can also manifest as other features, for example, stable defects in the lattice substructure from nuclear recoils have been identified by local increases in the dark current in silicon semiconductors~\cite{Lee:2022sxx}. In certain minerals, the damage also manifests as color centers. Color centers are produced in single site vacancy defects caused by lower energy nuclear recoils. When an anionic vacancy in the crystal lattice of a mineral is occupied by unpaired electrons, the absorption of light in a transparent mineral can be altered to emit color from around the vacancy. As described in section~\ref{sec:ReadOut}, an appealing aspect of searching for color centers is the possibility to use optical microscopy techniques which are able to scan large volumes of target minerals. However, it is not clear how well color centers can be used to resolve more extended damage features such as tracks from higher energy nuclear recoils.

Mineral detectors can potentially be sensitive to a variety of signals which are detailed in section~\ref{sec:Physics}, producing damage features ranging from color centers arising from the $\lesssim 100\,{\rm eV}$ recoils induced by solar or reactor neutrinos to composite Dark Matter (DM) causing macroscopic damage to the crystal along its trajectory. Atmospheric neutrinos with energies $\gtrsim 10 \, {\rm GeV}$ can also produce recoil tracks of many different nuclei in Deep Inelastic Scattering (DIS) interactions with the target mineral. In addition to all of these potential signals, nuclear recoils can be induced by a variety of radiogenic and cosmogenic backgrounds. For recoils induced by any of these sources, the (etched) damage track length can be used as a proxy for the energy of the nuclear recoil and the associated track length spectra present in the target mineral can potentially be used to discriminate signal from background.

\subsection{Petrological overview}
Several minerals have been proposed as detectors of elementary particles. In the 1980s, there were searches for signatures of magnetic monopoles in muscovite mica~\cite{Price:1986ky, Guo:1988,Ghosh:1990ki}. As discussed in section~\ref{sec:MineralDetectors_DM}, these studies were followed by searches for Dark Matter interactions with nuclei in ancient mica~\cite{Snowden-Ifft:1995rip,Snowden-Ifft:1995zgn,PhysRevLett.70.2348}. More recently, a wider selection of minerals have been proposed as potential detectors of Dark Matter and neutrino interactions. For example, Ref.~\cite{Drukier:2018pdy} theoretically investigated the possibility of mineral detection of Dark Matter using nchwaningite [Mn$_2$SiO$_3$(OH)$_2\cdot$H$_2$O], halite (NaCl), epsomite [MgSO$_4\cdot$7(H$_2$O)], nickelbischofite [NiCl$_2\cdot$6(H$_2$O)] and olivine [(Mg, Fe)$_2$SiO$_4$] as targets. In addition, Ref.~\cite{Edwards:2018hcf} proposed sinjarite [CaCl$_2\cdot$2(H$_2$O)]. These minerals were chosen since they can be found in marine evaporite deposits and ultra-basic rocks.  These geological environments can have sufficiently low concentrations of radioactive isotopes to suppress the radiogenic neutron background described in more detail below.

To establish mineral detection techniques which are viable from a experimental perspective, a more comprehensive set of criteria for selecting minerals must be clearly defined. First, we need a sufficient amount of mineral grain to be available for a mineral detection experiment to be sensitive to rare events, such as nuclear recoils induced by Dark Matter and neutrino interactions. Furthermore, minerals must be stable and suitable during experimentation, including sample preparation. Another crucial requirement is the establishment of suitable mineral dating techniques. With these criteria in mind, we review the occurrence and properties of the minerals mentioned above.

The occurrence of nchwaningite, epsomite, nickelbischofite and sinjarite is limited, and it is not easy to obtain sufficient samples of various ages. For example, nchwaningite was reported only from South Africa. Muscovite and olivine are major rock-forming minerals that commonly occur worldwide. Nchwaningite, epsomite, nickelbischofite, sinjarite, and gypsum are hydrates. Hydrate minerals and rock salt are or may be water soluble and deliquescent. Therefore, (1) it is impossible to use water during sample preparation, and (2) it is not easy to handle them in the atmosphere. When deliquescence occurs, the surface condition is altered, and tracks from nuclear recoils caused by the interactions of Dark Matter and other particles may be erased. Therefore, they are not suitable, at least for surface observation. On the contrary, muscovite and olivine are much more ``stable'' in the atmosphere,\footnote{It should be noted that olivine is a mineral easily altered at the surface of the Earth. Therefore, even if long-term interactions were recorded in olivine, it can not have been exposed to the surface for too long.} making muscovite and fresh olivine good candidates for mineral detectors.

Muscovite, a mica group mineral, has perfect cleavage along one planar direction defined by the crystal structure [basal (001) cleavage], thus, it can easily be ``peeled'' into thin (elastic) sheets. This property may enable various readout methods. First, obtaining a good surface that has not been artificially damaged is possible. It is a significant factor when surface observation is used. Muscovite sheets are highly transmissive, facilitating observation by transmitted light. A sufficient mass of muscovite can be ensured by using many sheets. Although olivine is a major constituent mineral in the Earth’s upper mantle, only a few studies reported the concentrations of trace elements, including U and Th (see, e.g., Refs.~\cite{Heier:1964, Cargnan:1996, DeHoog:2010}. Preliminary measurement showed U concentrations in mantle-derived olivine samples ranging from sub-ppb (parts per billion) to several tens of ppb~\cite{Kato:unpublished}. Reference~\cite{Eggins:1998} performed laser-ablation Inductively Coupled Plasma Mass Spectrometry (ICP-MS) measurements of uranium and thorium concentrations in olivine samples, reporting upper limits of $< 200\,$ppt (parts per trillion) and $< 30\,$ppt for the cleanest samples, see also Ref.~\cite{McIntyre:2021} for a recent study confirming these results. Concentrations of U and Th in muscovite are still unclear; Ref.~\cite{Pyle:2001} reported upper limits on uranium and thorium concentrations of $< 0.1\,$ppm (parts per million) in muscovite samples, and Ref.~\cite{Snowden-Ifft:1995zgn} used muscovite mica samples with uranium concentrations as low as $\mathcal{O}(0.01)\,$ppb for their Dark Matter search~\cite{Snowden-Ifft:1996dug}.

Finally, it should be noted that real samples are usually heterogeneous. It is not uncommon for minerals to be chemically heterogeneous (e.g., solid solution), showing zonal structure. They may also contain inclusions of other mineral phases and fluids. Cracks and dislocations can also be present. Distortions also exist generally. For example, the ideal crystal structure is distorted in real muscovite due to a tetrahedral-octahedral-tetrahedral layer misfit. It is necessary to distinguish between tracks from Dark-Matter- and neutrino-induced nuclear recoils and the artefacts from heterogeneity in the development of readout methods.

\subsection{Geochronological overview}
Here we review the various dating techniques available for the target minerals proposed for mineral detectors. Radiometric dating compares the abundances of naturally occurring radioactive isotopes in a sample to the respective abundances of their decay products. Since the half-life of each parent isotope is known, rocks can be dated given the relative abundance of the associated decay products. Muscovite contains K as a major element. Therefore, K-Ar dating is widely applied. In addition, muscovite is also used for Rb-Sr dating due to the similar ionic radii of K and Rb. Fission track dating, as discussed in section~\ref{sec:MineralDetectors-FissioTrackAnalysis}, is also possible in muscovite. In contrast, olivine contains few radioisotopes, and there are no widely used dating methods. Fission track dating of olivine is possible but has few geological applications due to the relatively low concentrations of heavy radioactive contaminants.\footnote{As discussed in section~\ref{sec:Physics-Geoscience}, note that the imaging techniques discussed in this whitepaper for applications to rare event searches could also be applied to fission track dating of minerals with characteristically low concentrations of radioactive contaminants.} Direct radioactive dating of halite is also difficult because of the lack of conventionally datable material. Reference~\cite{Belmaker:2013} applied $^{10}$Be dating to ancient halite formation. However, the short half-life of $^{10}$Be (1.39\,Myr) makes it unsuitable for mineral detector dating. Several studies (see, e.g., Refs.~\cite{Fitzpatrick:1994, Sanna:2010, Sanna:2011, Obert:2022}) applied $^{230}$Th/U dating to gypsum, but this method is also unsuitable for mineral detector dating. The possibility of dating nchwaningite, epsomite, nickelbischofite and sinjarite is unknown.

Closure temperature and mineralization process must be considered to find suitable mineral detectors. Reference~\cite{Dodson:1973} introduced the concept of closure temperature of a cooling system. At closure temperature, a system can be considered closed so that, below this temperature, there is no longer any significant diffusion. Each system has its own closure. Closure temperatures of the Rb-Sr and K-Ar systems are about $500\,^{\circ}$C and $350\,^{\circ}$C, respectively~\cite{Jaeger:1967, Purdy:1976, Dodson:1985, Harrison:2009}. Alternatively, the closure temperature of fission track dating in a given mineral is the temperature above which fission tracks are completely annealed. Compared to conventional isotopic dating methods, the closure temperature of fission track dating is generally lower. For example, the closure temperature of the fission trace age of muscovite and olivine is about $150\,^{\circ}$C~\cite{Lal:1975, James:1988}. Since the tracks associated with nuclear recoils relevant for mineral detection of Dark Matter and neutrinos are similar to fission tracks, the closure temperature of tracks from such nuclear recoils may be similar as well. However, the details of the closure temperature of Dark-Matter- and neutrino-induced crystal damage are unclear on geological timescales. 

The relationship between geological setting and closure temperature is a crucial factor in determining whether the period over which nuclear recoil tracks have formed can be estimated in mineral detection. To be concrete, we consider the example cases of a mineral detector sample found in igneous rocks (e.g., muscovite and olivine) and of olivine in peridotite. In the case of volcanic rocks, the time between eruption and crystallisation of minerals is instantaneous. In the case of plutonic rocks, the time from the emplacement of magma to crystallisation is probably on the order of a million years. Therefore, the estimated time of track formation would not be significantly affected by the closure temperature of the applied dating method. In the case of olivine in peridotite, the temperature of the sample at the time of track formation relative to the closure temperature of the dating method is essential. The temperature of the upper mantle is $\mathcal{O}(1000)\,^{\circ}$C. The olivine then interacts with Dark Matter and neutrinos in the mantle for a long time. The sample then cools before moving up to the Earth’s surface. Therefore, if the closure temperature of the dating method is significantly higher than that of the nuclear recoil tracks then the exposure time of the sample to Dark Matter and neutrinos could be overestimated.\footnote{For a mineral detector with track length resolution and exposure  sufficient to be sensitive to e.g. WIMP-like Dark Matter interactions not already ruled out by conventional direct searches, the age of the mineral can potentially inferred from the radiogenic background recoil spectrum without external constraints on the age of the mineral~\cite{Baum:2021jak}.} 

\subsection{Color Centers}
\label{sec:MineralDetectors-ColorCenters}

As discussed above, the latent damage to the crystal structure from a recoiling nucleus can take the form of {\it color centers}. These defects are of particular interest because, although the physical size of the vacancy causing the color center is the lattice spacing (typically a few {\AA}ngstr{\"o}m), color centers can be detected and localized to order 10\,nm with optical superresolution microscopy which can process large sample volumes, see sections~\ref{sec:ReadOut-Optical} and~\ref{sec:Studies-VT}. 

Color centers are point defects (or clusters of point defects) associated with the absence of an atom from the crystal lattice of a material~\cite{Hosch:2002,Tilley:2014}. This vacancy leads to the trapping of an electron or hole\footnote{Perhaps the best understood type of color center is the so-called F-centre (from the German \textit{Farbe}, or `color') in which the vacancy of a negatively charged ion traps an electron.}. The trapped particle may be excited by the absorption of certain colors of visible light (or infrared or ultraviolet radiation)~{\cite{Fowler:1975, Hayes:2004, Nassau:2001, Tilley:2008, Tilley:2011}, leading to a characteristic color in solids which would otherwise be colorless. Dopant atoms which give rise to color in otherwise colorless crystals are also often referred to as color centers, though here we use to term to refer only to those associated with lattice vacancies.

It is known for long time that the interaction of accelerated particles or radiation with solid matter can create color centers~\cite{Bertel:1982, Schwartz:2006, Schwartz:2008, Schwartz:2015, Manzano-Santamaria:2012}. Luminescence dating utilizes centers that have been created by the interaction of alpha and beta particles and/or gamma irradiation in minerals such as quartz or feldspar. In particular, fluorescence microscopes such as mesoscale Selective Plane Illumination Microscopy (mesoSPIM)~\cite{mesoSPIM} allow for the visualization of color centers in 3D deep within crystals, due to the long working distances of the objectives and the light-sheet optical sectioning. Such  microscopes might be capable of detecting color centers created by accelerated particles in matter, such as the daughter nuclei from the fission of radioisotopes in mineral detectors. Near future research will focus on testing the visualization of latent fission-tracks (MeV-tracks) in apatite and keV-tracks in salt minerals and olivine with mesoSPIM. Apatite is known for color centers created by the fission products~\cite{Bertel:1982}; halite and olivine are known for forming color centers while irradiated with accelerated ions.

\begin{figure}
    \centering
    \includegraphics[width=1\linewidth]{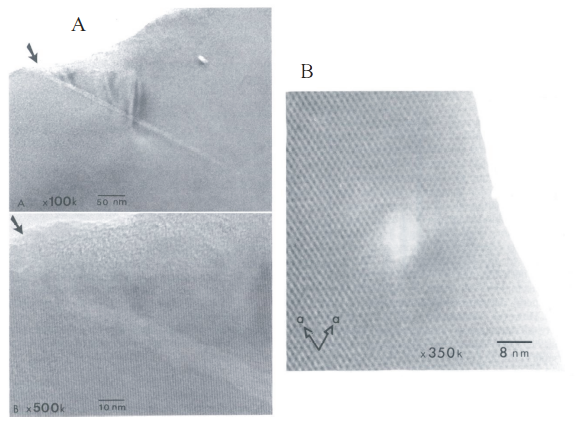}
    \caption{TEM-images of latent fission-tracks in apatite. {\it Left (A)}: Images taken parallel to the flight trajectory (light grey). {\it Right (B)}: Image taken perpendicular to the flight trajectory. Core of a fission-track is visible in the central part of the image. Figure taken from Ref.~\cite{Paul:1992}.}
    \label{fig:FT_TEM}
\end{figure}

\begin{figure}
    \centering
    \includegraphics[width=1\linewidth]{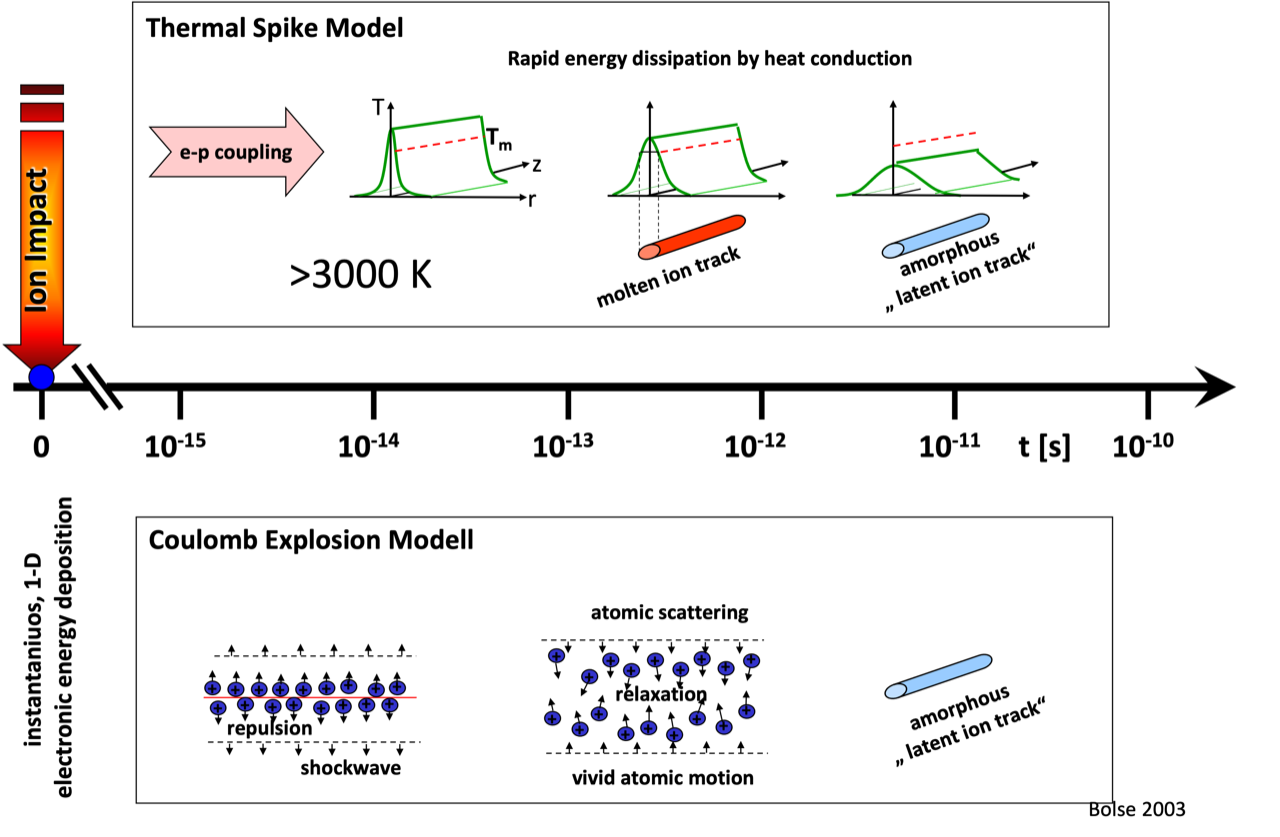}
    \caption{Two different models for the formation of latent fission tracks in condensed matter. See Ref.~\cite{Toulemonde:2000} for a discussion of the ``Thermal Spike Model'' (illustrated in the top panel) and Ref.~\cite{Fleischer:1965a} for the ``Coulomb Explosion Model'' (bottom).}
    \label{fig:TrackModels}
\end{figure}

\begin{figure}
    \centering
    \includegraphics[width=0.5\linewidth]{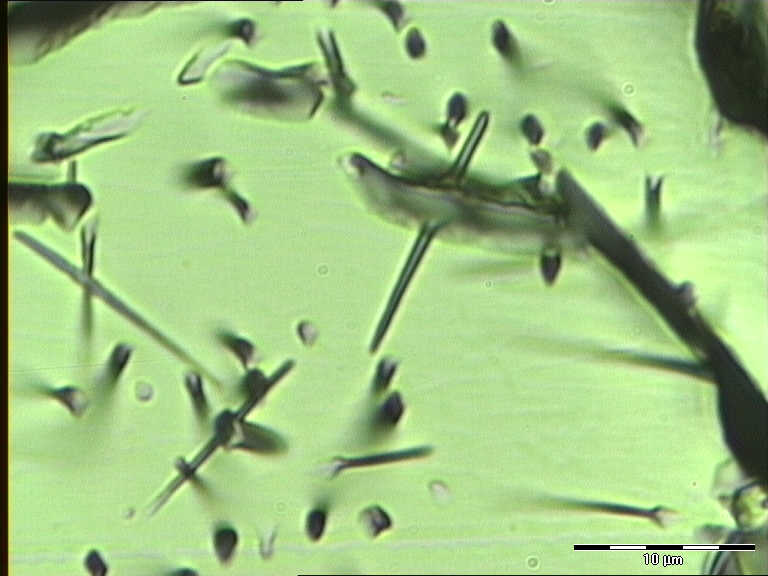}
    \caption{Etched spontaneous fission tracks in apatite visualized by optical microscopy. Apatite samples obtained from southern Uralides, Russia~\cite{UG_unpublished}.}
    \label{fig:EtchedApatite}
\end{figure}
    
\subsection{Fission Track Analysis} \label{sec:MineralDetectors-FissioTrackAnalysis}

Latent fission tracks are cylindrical volumes ($\mathcal{O}(10)\,$nm diameter and $\mathcal{O}(20)\,\mu$m length) of damage in a crystal produced during the spontaneous fission decay of $^{238}$U, $^{235}$U, and $^{232}$Th, see Fig.~\ref{fig:FT_TEM}. Due the the half lives and relative abundances of $^{238}$U, $^{235}$U, and $^{232}$Th, $^{238}$U is the most relevant radioisotope fissioning within the last 1\,billion years. Each fission event causes two fission fragments that travel in exactly opposite directions, producing a single trail of crystal damage (defects) with the length determined by the energy loss (typically of order keV/nm) of the fragments in condensed matter along the trajectory. The initial kinetic energy of the fission fragments amounts in average to 170\,MeV with a distribution between 160\,MeV and 190\,MeV. 

The damage caused by the fission fragments on the nanoscale depends on the material. For example, the spontaneous fission track in apatite (Ca$_5$(PO$_4$)$_3$(F, Cl, OH, REE) is a trail of amorphous material within the normal crystal lattice (Fig.~\ref{fig:FT_TEM}). Two models of latent spontaneous fission track formation are discussed in literature: the thermal spike model and a Coulomb explosion model, see Fig.~\ref{fig:TrackModels}. These latent spontaneous fission tracks can be visualized with an optical microscope after enlarging and stabilizing the tracks by etching (Fig.~\ref{fig:EtchedApatite}). The etchable length is shorter than the latent length. Every mineral has a threshold of energy loss above which the crystal defect can be enlarged by chemical etching. At the ends of the latent damage trail the energy loss is below the threshold. 

Furthermore, the etching conditions and the etchant depend strongly on the material under investigation. A large list of etchants exists in literature for numerous minerals, for more details, see Ref.~\cite{Wagner:1992}. We briefly discuss several examples of fission track analysis which are relevant for the mineral detection of neutrinos and Dark Matter. 

\subsubsection{Fission tracks in apatite}

The fission-track dating technique has two archives that store information pertaining to the thermal history of mineral grains. The first uses the temperature-time ($T$--$t$) path related changes of the etch pit areal density at an artificially polished internal surface, and the second applies the $T$--$t$ path related change of length distribution of horizontal confined tracks. Confined tracks are tracks within the crystal that do not intersect the polished surface of the grain. These tracks are etched completely by etchant that has travelled along other tracks or fractures. Ref.~\cite{WagnerReimer:1972} analyzed apatite fission-track etch pit areal density data by applying their own annealing model and found that the fission-track age of apatite corresponds to a temperature at which about 50\,\% of the tracks are stable.

The fission-track data obtained from track annealing experiments are usually presented in the form of an Arrhenius plot~\cite{Fleischer:1965b} in which the straight line (iso-density contours or iso-length contours) describes the relationship between annealing time and annealing temperature. For various degrees of annealing, lines of different slopes (i.e. different activation energies) are valid. These plots are used to extrapolate annealing data from laboratory to geological time scales. Different empirical and semi-empirical equations were defined by researchers to model the annealing process. For temperatures where track annealing exceeds track production at a given time interval, no tracks are preserved in the crystal. With decreasing temperature, the velocity of annealing decreases until finally a value is reached where the annealing of tracks is very minor even over geological time scales. Below this temperature all formed tracks keep their original length in the crystal. 

Reference~\cite{WagnerReimer:1972} estimated 50\% fission-track retention temperatures of $215\,^{\circ}$C, $185\,^{\circ}$C, $155\,^{\circ}$C, $125\,^{\circ}$C and $100\,^{\circ}$C for monotonic constant cooling rates of $10\,^\circ$C/yr, $1\,^\circ$C/$10\,$yr, $1\,^{\circ}$C/$10^3\,$yr,  $1\,^{\circ}$C/$10^5\,$yr, $1\,^{\circ}$C/$10^7\,$yr, respectively, in apatite. They also discussed the potential problems which can arise when applying an Arrhenius equation to describe the annealing experiments of fission-tracks in apatite. While the physical properties of track annealing are not known, the assumption of first-order reaction kinetics might be wrong. With increasing degrees of annealing, the reaction kinetics might change. Reference~\cite{Green:1988} further stated that the diffusion process is complicated in an anisotropic, polyatomic crystal with a spectrum of defect species. In addition, a monotonic constant cooling rate is not applicable in geological environments because the change of temperature with time is highly variable.

Nowadays, the time and temperature related change of the etch pit areal density is no longer used as the main data archive of the time-temperature history. In the 1980s, annealing models based on the measurement of horizontal confined track length distributions replaced the earlier annealing models. The first fission-track length measurements in apatite were published by Wagner and Storzer~\cite{WagnerStorzer:1972}. They determined the projected track length distributions and found that the distributions of fossil tracks were different from those of induced tracks. 

In the years following, measurements of confined track-length reduction in samples artificially heated in a laboratory furnace were combined with confined track-length distributions in samples with a reasonably known geological $T$--$t$ history~\cite{Green:1986, Crowley+:1991, Carlson:1999, Barbarand:2003a}. Based on this annealing data, algorithms were constructed that describe the time- and temperature-dependence of the track annealing process~\cite{Laslett:1987, Crowley+:1991, Galbraith:1996, Ketcham:1999}. The extension of isothermal laboratory annealing data to variable temperature annealing and extrapolation to geological time periods enabled the use of these algorithms in numerical models to predict fission-track length distributions and ages for a specific $T$--$t$ scenario~\cite{Duddy:1988, Green:1989}. The predicted fission-track length distributions and ages were statistically compared with the real fission-track length distributions and ages measured from a field sample~\cite{Lutz:1991, Gallagher:1995, Willett:1997, Ketcham:1999}. The following critical statement of Ref.~\cite{Barbarand:2003b} might elucidate the danger of careless use of computer-based numerical models: ``In a geological application, maximum palaeotemperatures, periods of heating and cooling, estimates of the amount and timing of missing sections, denudation amounts and rates, `uplift' [\ldots] all are postulated from the goodness-of-fit of such predictions with measured sample [fission track] data, constrained by other geological information." 

References~\cite{Gleadow:1986, Gleadow:2003} define three important properties pertaining to the use of spontaneous fission-tracks as a data archive of the temperature history of apatite below about $120\,^{\circ}{\rm C}/10\,$yr:
\begin{itemize}
    \item ``All tracks in apatite have a very similar length when first produced, which is controlled by the energies of the fission decay and the nature of the track recording material''~\cite{Gleadow:1986}. Note that this statement is an assumption. It has never been physically proven that the track lengths of newly formed fission tracks are independent of the temperature conditions at the formation time of the tracks.
    \item ``Fission-tracks become progressively shorter during thermal annealing in a way that is controlled principally by the maximum temperature that each track has experienced''.
    \item ``New tracks are continually added to the sample through time so that each one has experienced a different fraction of the total thermal history''~\cite{Gleadow:2003}. This is the important statement. Each individual track records its own temperature-time history. In a pure cooling environment tracks that form during the cooling will start annealing at a different temperature and will therefore be longer than tracks formed earlier on the cooling path.
\end{itemize}

Reference~\cite{Gleadow:1986} investigated different geological environments where the thermal history was reasonably well constrained. For longer cooling times, the mean confined-fission-track length was shortened and the length distribution was shown to be broader with a number of relatively short tracks. Furthermore, Ref.~\cite{Gleadow:1986} showed that the confined track length distribution in apatite of volcanic rocks, with very fast cooling to ambient (surface) temperatures and an undisturbed thermal history afterwards, have a slightly shorter mean confined track length than the confined length distribution of induced fission tracks in apatite. This is an indication that track shortening occurs at ambient temperatures~\cite{Gleadow:2003}. This effect has not been considered in the thermal annealing models generated from analyses of freshly induced track distributions in different apatites. Furthermore, research by Donelick {\it et al.}~\cite{Donelick:1990} indicated that an initial phase of track annealing (the first $0.5\,\mu$m of the fission track) occurs at room temperature ($\sim 23\,^{\circ}$C) on a remarkably short timescale ($< 1\,$month).

Even though fission track analysis has been used to constrain thermal histories in a variety of settings, the physical properties that govern the annealing of fission tracks in minerals are still unknown. Experiments conducted by Ref.~\cite{Green:1986} showed that the tracks first shorten axially from both ends and to a lesser amount from the sides. Further annealing results in a stage where the track is broken into segments~\cite{Paul:1992, Hejl:1995}. The lattice between the segments cannot be etched. The damaged relic parts of the former track can be etched but the length cannot be considered as an archive for the thermal history. Two effects on the annealing of fission-tracks in apatite are important to consider: First, crystallographic effects, and second, chemical composition effects: 

The crystallographic effects on the annealing of fission-tracks in apatite were first described by Ref.~\cite{Green:1977}. Tracks orthogonal to the $c$-axis anneal more rapidly than tracks parallel to the $c$-axis~\cite{Green:1988}. This anisotropy increases as annealing progresses~\cite{Green:1981, Laslett:1984, Donelick:1990, Donelick:1999, Galbraith:1990, Donelick:1991}. References~\cite{Donelick:1990, Donelick:1991, Donelick:1999} further extended the database on crystallographic effects of annealing in apatite and integrated the results into the recent annealing model of Refs.~\cite{Ketcham:2003, Ketcham:2017}. Reference~\cite{Barbarand:2003b} confirmed the strong influence of crystallographic orientation by presenting a large annealing dataset of apatites with different chemical composition.

The first geological observation elucidating that the chemical composition of apatite might influence the fission-track annealing rate was described by Ref.~\cite{Gleadow:1981} for drill-core samples from the Otway basin in Australia. References~\cite{Green:1985, Green:1986} demonstrated that the annealing of fission-tracks in apatite is dependent on the chlorine/fluorine ratio, where fluorine-rich apatites show more annealing than chlorine-rich samples. The effects of composition have been described for sedimentary and magmatic environments~\cite{Burtner:1994, OSullivan:1995}. Fluorine-rich apatites such as Durango apatite show complete annealing in geological environments at temperatures of $90\,^{\circ}$C--$110\,^{\circ}$C/10\,Myr. In contrast, chlorine-rich apatites completely anneal at temperatures of $110^{\circ}$C--$150^{\circ}$/10\,Myr~\cite{Burtner:1994}. A recently published extensive study~\cite{Barbarand:2003a} gave further indications that the chlorine content dominantly controls fission-track annealing in apatite. Within the temperature range between total annealing (see above) and $60\,^{\circ}$C/10\,Myr, the old and newly formed spontaneous fission tracks are partly annealed. Below $60\,^{\circ}$C/10\,Myr, no significant annealing has been reported so far.

Numerical annealing algorithms used to simulate the time-temperature related evolution of the fission-track length distribution have been developed over the last 20\,years. The first approach was published by Ref.~\cite{Bertagnolli:1983} and has been continuously further developed. This approach used a convection-type equation~\cite{Igli:1998}. Reference~\cite{Crowley:1985} used a semi-analytical solution. The so-called `Laslett {\it et al.} model'~\cite{Laslett:1987} was based on a fanning Arrhenius relationship used to model the extensive data set of Ref.~\cite{Green:1986}. The disadvantage of this model is that it was developed based on the Durango apatite composition and does not account for the compositional effects. This model was further developed with additional parameters by Ref.~\cite{Laslett:1996}. Further annealing experiments with apatites of different chemical compositions lead to the model of Ref.~\cite{Crowley:1993}. This model was also based on a fanning-linear Arrhenius-type relationship. Reference~\cite{Carlson:1990} proposed a numerical algorithm to model fission-track annealing data on a physico-chemical basis and tested his algorithm with already published annealing data. Discussion of the approach by Refs.~\cite{Crowley:1993, Green:1993, Carlson:1993a, Carlson:1993b} lead to an improved version the multikinetic model~\cite{Ketcham:2017} that was based on a substantial annealing dataset~\cite{Carlson:1990, Donelick:1999}. The multikinetic model included mixed-composition apatites and accounted for crystallographic effects~\cite{Ketcham:2000, Ketcham:2005, Ketcham:2007a, Ketcham:2007b, Ketcham:2009}. 

These models do not describe the real physical process of annealing. The physical meaning of the successive annealing is still under debate. Restoration of the crystal lattice at the radiogenic defect site might mainly be governed by the law of diffusion. This process could be called an ``in-crystal diffusion process''. The most important open questions to solve this problem are: 
\begin{itemize}
    \item What is the chemical composition of the radiogenic defect? 
    \item How does crystal lattice restoration happen? 
    \item Is the annealing process a double process described by element (F, Cl) or/and electron diffusion and re-crystallization of the amorphous internal part of the track? 
    \item Does re-crystallization only occur at the boundary between crystal lattice and amorphous area? What is the diffusion distance? 
    \item What energy is necessary for the diffusion process and what energy is necessary to restore the radiogenic crystal lattice at the defect site? 
\end{itemize}
Detailed Transmission Electron Microscopy (TEM) studies combined with laboratory experiments and advances in understanding of the track formation process might help to answer some of these questions in the future. A first important physical approach was presented by Ref.~\cite{Belton:2002}. That work applied the laws of solid-state diffusion and steadily changed the radius of the amorphous internal part of the track. Applying his model to laboratory annealing data, Ref.~\cite{Belton:2002} explained part of the annealing. Further research is necessary before this model can be applied to real geological data. 

More recently, Refs.~\cite{Afra:2011, Kluth:2012} applied synchrotron small-angle X-ray scattering to determine the latent track morphology and the track annealing kinetics in Durango apatite. They describe that structural relaxation followed by re-crystallization occurred during the annealing experiments. Further investigations used TEM investigations to characterize the processes during annealing~\cite{Li:2011, Li:2012}. A similar study using TEM to measure fission and ion tracks (partially) annealed in the laboratory in epidote was recently performed by Nakasuga {\it et al.}~\cite{Nakasuga:2022}.

\begin{figure}
    \centering
    \includegraphics[height=5.5cm]{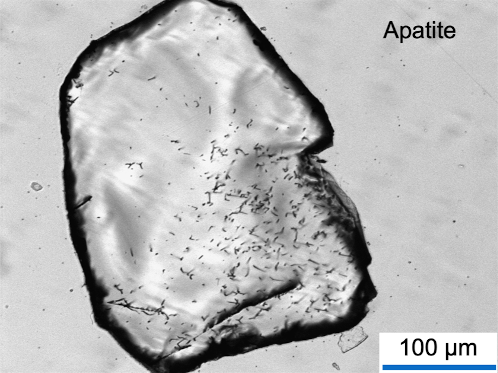}
    \includegraphics[height=5.5cm]{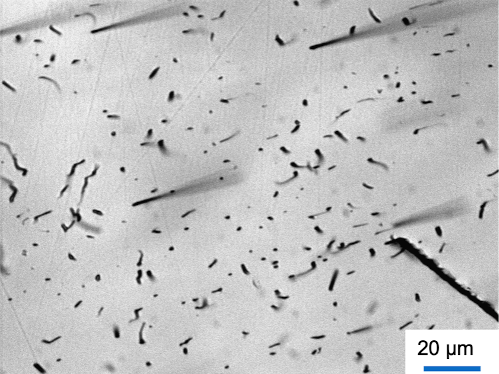}
    \caption{Etched dislocation, spontaneous latent fission tracks, and mineral inclusions in apatite. Image taken by optical microscopy~\cite{UG_unpublished}.}
    \label{fig:etchedApatite}
\end{figure}

Besides latent spontaneous fission tracks, minerals also store etchable crystal defects such as dislocations. We do understand that the same etching conditions that reveal fission tracks etch those dislocations as well, see Fig.~\ref{fig:etchedApatite}.

\subsubsection{Fission tracks in zircon}

Since zircon was proposed as a material for disposal of actinides, highly-enriched nuclear waste and weapons-grade plutonium, a wide range of studies on its structure, ion-induced amorphization and re-crystallization have been carried out~\cite{Holland:1955, Ewing:1987, Ewing:1993, Ewing:1995, Ewing:2003, Weber:1990, Weber:1991, Weber:1993, Murakami:1986, Murakami:1991, Wang:1992a, Wang:1992b, Burakov:1993, Anderson:1993, Ewing:1999, Ewing:2001, Weber:1994, Weber:1996, Weber:1997, Meldrum:1996, Meldrum:1998a, Rios:1999, Capitani:2000, Rios:2000a, Rios:2000b, Ewing:2000, Nasdala:2001, Ewing:2011}. Due to the frequent $\alpha$-decay of uranium and thorium, the metamictization/amorphization of natural zircon is mainly due to the crystal changes caused by the interactions of the associated helium nuclei and heavy recoil nuclides. The formation of fission tracks, which occurs less frequently, also has a subordinate influence on the degree of metamictization. Natural zircon can usually carry up to 5000\,$\mu$g/g UO$_2$+ThO$_2$. However, 7\,\% by weight U+Th have also been detected~\cite{Speer:1982}. Newly formed crystalline zircon with a uranium content of 6.1\,\% to 12.9\,\% by weight was found in the ``Chernobyl lava''~\cite{Anderson:1993}. 

The effects of the ion-induced changes in zircon are reflected in the systematic changes in its physical properties as follows:
\begin{itemize}
    \item expansion of cell parameters and broadening of X-ray diffraction peaks~\cite{Holland:1955, Murakami:1991, Weber:1993},
    \item decrease in intensity, and a significant broadening of the infrared and Raman bands~\cite{Vance:1975, Woodhead:1991a, Nasdala:1995, Nasdala:1996, Nasdala:1998, Nasdala:2004, Nasdala:2010, Nasdala:2011, Marsellos:2010}, 
    \item decrease in the refractive index and bireflexion~\cite{Holland:1955, Vance:1972}, 
    \item absorption of OH groups~\cite{Aines:1986, Woodhead:1991b},
    \item increase in fracture toughness~\cite{Chakoumakos:1987},
    \item decrease in density~\cite{Holland:1955, Murakami:1991},
    \item change in high-resolution transmission electron microscopy (HRTEM) diffraction spectra~\cite{Yada:1987, Murakami:1991}, 
    \item change in $^{29}$Si-NMR-behavior, decrease in hardness~\cite{Chakoumakos:1991}, 
    \item change in the diffuse X-ray scattering of single crystals~\cite{Salje:1999},
    \item and the occurrence of Huang-type diffuse X-ray diffraction~\cite{Rios:1999}.
\end{itemize}
Two models of amorphization/metamictization are distinguished. The first model is called the ``direct impact model'' or ``direct amorphization model'' and assumes that every $\alpha$-decay leads to an amorphization cascade~\cite{Morehead:1970, Gibbsons:1972}. Reference~\cite{Carter:1979} extended this model to the ``elaborate-overlap model''. Reference~\cite{Weber:2000} formulated the ``double-overlap model''. In this model, a critical concentration of point defects is required before the amorphization process begins. Studies and calculations by Refs.~\cite{Farnan:1999, Rios:2000b, Zhang:2001, Palenik:2003} showed that the ion-induced changes in zircon can be explained using the ``direct impact model''. 

Despite intensive investigations, the exact structural composition of metamict zircon is still a matter of controversy. Reference~\cite{Holland:1955} explained their results with an intermediate polycrystalline phase. Reference~\cite{Murakami:1991}, on the other hand, concluded that there were two crystalline phases. Reference~\cite{Stott:1946} describes the occurrence of monoclinic ZrO$_2$ based on X-ray diffraction spectra. Cubic and possibly tetragonal ZrO$_2$ have been reported by Refs.~\cite{Anderson:1962, Vance:1972}. Reference~\cite{Wasilewski:1973} proposed a two-stage change process. First, in undamaged zircon, the recoil core will create areas where the lattice is severely strained and expanded, and where twisted SiO$_4$ tetrahedra appear. In the second stage, the recoil core forms ZrO$_2$, SiO$_4$ and a-periodic ZrSiO$_4$. Reference~\cite{Vance:1975} describes that Si-O bonds that occur in strongly ion-changed zircon do not differ significantly from bonds in amorphous SiO$_2$. References~\cite{Meldrum:1998a, Meldrum:1998b, Meldrum:1999a, Meldrum:1999b} describe the change in synthetic zircon that has been irradiated with heavy ions at high temperatures. They propose the formation of a ``melt-like'' state along the ion-track and discuss that slow cooling allows formation of crystalline ZrO$_2$. In contrast, Refs.~\cite{Zhang:2000a, Zhang:2002, Zhang:2003} describe the formation of SiO$_2$ and ZrO$_2$ from the decomposition of a strongly metamict ZrSiO$_4$ at a temperature of approximately 1100\,K. At temperatures between 1400\,K and 1500\,K, zirconium (ZrSiO$_4$) is formed again from the individual components SiO$_2$ and ZrO$_2$. Based on an infrared spectroscopic analysis of metamict zircons, Ref.~\cite{Zhang:2001} concluded that the local or short-range order of Zr is still present in the amorphous phase. The additional IR absorption bands (800--1300\,cm$^{-1}$ and $\sim 400\textit{--}650\,$cm$^{-1}$) are attributed by Ref.~\cite{Zhang:2001} to the reformation of Si-O-Si bonds and a local reorganization of the normal Si-O-Si attributed to bonds in zircon. The detailed investigations were followed by modeling of the ion-induced material changes on the atomic scale using the techniques of molecular dynamics~\cite{Crocombette:1998, Crocombette:1999, Crocombette:2001, Trachenko:2001, Trachenko:2003}.

Micro-Raman spectroscopic studies on natural metamict zircons from Sri Lanka in Refs.~\cite{Zhang:2000a, Palenik:2003} gave no indication of the formation of crystalline ZrO$_2$ and SiO$_2$. Reference~\cite{Zhang:2000b} described Raman spectra indicating crystalline zircon areas with partially twisted lattice and amorphous areas. The results of the first extensive Raman investigations on natural zircons were reported by Ref.~\cite{Nasdala:1995}. Reference~\cite{Palenik:2003} extended the measurements by Ref.~\cite{Nasdala:1995} and describe a relationship between the full-width half-maximum of the B$_{1{\rm g}}(v3)$-band and the number of defects caused by natural radioactive decay. Raman spectra of synthetic zircons were reported by Ref.~\cite{Syme:1977}.

\begin{figure}
    \centering
    \includegraphics[height=3.7cm]{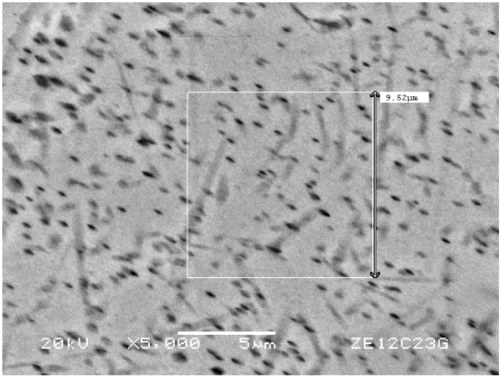}
    \includegraphics[height=3.7cm]{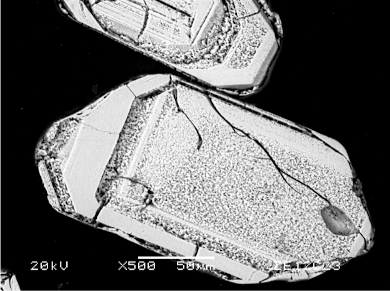}
    \includegraphics[height=3.7cm]{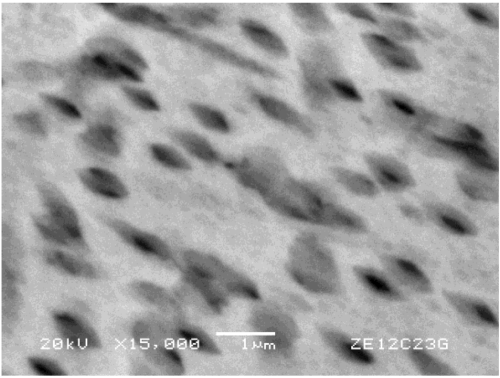}
    \caption{Etched spontaneous fission tracks in zircon (image taken by SEM-technique)~\cite{UG_unpublished}.}
    \label{fig:zirconSEM}
\end{figure}

In addition to apatite, Zircon and titanite are the two other important minerals used in fission-track thermochronology. In natural environments, zircon [Zr(SiO$_4$)] and titanite [CaTi(SiO$_4$)(O,OH,F)] show latent spontaneous fission tracks, $\alpha$-recoil tracks, and point defects caused by the $\alpha$-particles ($^4$He). 
As the density of zircon is higher than that of apatite, the latent spontaneous fission tracks in zircon are smaller in size ($\sim 2\,{\rm nm} \times 13\,\mu$m). Besides the etching and visualization of spontaneous fission tracks in zircon by optical microscopy, visualization using etched spontaneous fission tracks in zircon by SEM techniques is also possible (Fig.~\ref{fig:zirconSEM}). The volume density of $\alpha$-recoil tracks is responsible for the degree of amorphization or so called metamictization in zircon crystals. Annealing of fission-tracks in zircon is a function of temperature, time, and $\alpha$-radiation damage (here understood as the combination of processes caused by $\alpha$-decay and the $\alpha$-recoil nucleus)~\cite{Kasuya:1988, Weber:1999, Trachenko:2002,Li:2021}. The amount of $\alpha$-radiation damage in zircon increases with time and the uranium and thorium concentrations.

Based on the investigation of fission tracks in zircon with an average degree of metamictization in geological environments, Ref.~\cite{Hurford:1986} argued for a complete annealing temperature of zircon at about $240\,^{\circ}$C/10\,Myr. Reference~\cite{Rahn:2001} found a different reset temperature of the zircon fission-track system, arguing based on mineralogical and petrological investigations in the Swiss Alps and the Olympic Mountains, USA, that complete annealing of fission-tracks in zero damage zircons requires a temperature of $320 \pm 20\,^{\circ}$C/10\,Myr. Similarly, Refs.~\cite{Green:1996, Tagami:1998} stated for natural zircon samples from various environments that total annealing would be achieved in a reasonable time above about $320\,^{\circ}$C. In both geological cases, the duration of the thermal event was in the range of 10\,Myr. Several annealing models are also described by Refs.~\cite{Zaun:1985, Kasuya:1988, Tagami:1990, Tagami:1995, Carpena:1992, Yamada:1995, Tagami:1996, Galbraith:1997, Tagami:1998, Brandon:1998}.

Reference~\cite{Tagami:1996} studied the zircon fission-track system around a granitic pluton. They discussed a Zircon Partial Annealing Zone (ZPAZ) between $\sim 230\,^{\circ}$C and $320\,^{\circ}$C for a heating duration of about $10^6$\,years. Fission-tracks in zircon grains from Miocene to Pleistocene sandstones and Miocene to Pliocene rhyolites of two drill holes in a sedimentary basin in Japan indicated a temperature above $200\,^{\circ}$C for the lower limit of the ZPAZ at a stable temperature over about 1\,Myr without knowing the degree of metamictization~\cite{Hasebe:2003}. Furthermore, results of laboratory annealing experiments point towards a similar temperature range of the ZPAZ~\cite{Yamada:1995, Tagami:1998}. This temperature range of the ZPAZ covers the anchi- to epizone of natural environments in the continental crust. Fission-tracks in zircon grains with high $\alpha$-radiation damage start to anneal at temperatures between $150\,^{\circ}$C and $200\,^{\circ}$C/10\,Myr~\cite{Garver:2001, Riley:2002}. Complete annealing of fission-tracks in zircon would need temperatures above $320\,^{\circ}$C/10\,Myr~\cite{Tagami:1996, Tagami:1998, Rahn:2001, Brix:2002}. References~\cite{Green:1996, Tagami:1998, Rahn:2001, Rahn:2004, Brix:2002} stated for natural zircon samples from various environments that total annealing would be achieved in a reasonable time above $\sim 320\,^{\circ}$C/10\,Myr. Reference~\cite{Garver:2002} stated that the color of zircon might be an indication for the degree of amorphization. White zircon exhibits a relatively low degree of amorphization, which increases in yellow zircon and reaches a maximum in red zircon. However, the chemical composition of zircon can also influence the color. Therefore, more comprehensive studies are necessary applying spectroscopic techniques to further quantify the degree of amorphization in zircon.

Summarizing, fission tracks in zircon are stable up to a temperature of about $150\,^{\circ}$C~\cite{Yamada:1995, Tagami:1996, Green:1996, Tagami:1998, Garver:2001, Riley:2002, Hasebe:2003}. An important influencing parameter for this limit temperature is the volume density of the $\alpha$-recoil defects in the crystal~\cite{Kasuya:1988, Rahn:2001}. In the temperature interval of $\sim 200\,^{\circ}$C/10\,Myr to $\sim 330\,^{\circ}$C/10\,Myr, the spontaneous fission tracks heal under geological conditions~\cite{Wagner:1992, Coyle:1997}. Above the temperature of about $330\,^{\circ}$C, any spontaneous fission track formed is healed independently of the $\alpha$-recoil track density~\cite{Tagami:1996, Tagami:1998, Rahn:2001, Brix:2002}. With geothermal gradients of $\sim 10\textit{--}30\,^{\circ}$C/km, these temperatures correspond to depths in the Earth of $\sim 30\textit{--}10\,$km and thus lithostatic pressure ratios of $\approx 0.36\textit{--}1.1\,$GPa. Similar stability conditions are assumed for fission tracks in titanite~\cite{Coyle:1998, Jonckheere:2000}. 

\subsubsection{Fission tracks in titanite}

Studies on defects in natural titanites caused by radioactive decay were carried out in Refs.~\cite{Hawthorne:1991, Zhang:2002}. Both papers concluded that there are amorphous areas in the natural titanites, but that there is no phase separation in CaTiO$_3$, CaSiO$_3$ and SiO$_2$. Various analytical methods such as X-ray diffraction analysis, NMR spectroscopy~\cite{Farnan:1999, Farnan:2001, Larsen:2002}, optical microscopy, infrared, Raman and M{\"o}ssbauer spectroscopy were used to determine the degree of amorphization and re-crystallization of zircon and titanite~\cite{Hawthorne:1991, Zhang:2002}. Experimental annealing data suggests temperatures of $340\,^{\circ}$C to $430\,^{\circ}$C/10\,Myr for a 50\,\% reduction in track density (see Ref.~\cite{Wagner:1992}). In comparison to closure values for other dating systems (K-Ar, Rb-Sr), these values are too high. Biotite dated with the K-Ar system revealed equal or higher ages than titanite from the same samples dated with the fission track technique~\cite{Kohn:1993}. Several studies~\cite{Gleadow:1978, Harrison:1979, Gleadow:1979, Fitzgerald:1988} estimate the titanite closure temperature to be in the range of $250\,^{\circ}$C/10\,Myr. Based on a detailed titanite study from samples of different depth and therefore temperature from the German ``Main Continental deep drill hole'', Ref.~\cite{Coyle:1998} assumed the partial annealing zone of titanite to be in the temperature range of $265\textit{--}310\,^{\circ}$C/10\,Myr. 

\subsubsection{Fission tracks in olivine}

Olivine has also been tested for the use in thermochronology. Unfortunately (for fission-track-dating applications), olivine typically has very low uranium and thorium concentrations in the range of $\mathcal{O}(0.01)\textit{--}\mathcal{O}(1)\,$\,ng/g for U and similarly for Th (see, e.g., Ref.~\cite{McIntyre:2021}). Olivine has two end members, forsterite [Mg$_2$(SiO$_4$)] and fayalite [Fe$^{+2}_2$(SiO$_4$)]. Applying the published etching solution and conditions~\cite{Krishnaswami:1971}, latent spontaneous fission tracks can be enlarged and visualized by optical microscopy. The etching solution is as follows: 1\,g of oxalic acid, 1\,ml of orthophosphoric acid (85\,\%), and 40\,g of the disodium salt of EDTA (ethylenediaminetetraacetic) acid in 100\,ml of distilled water; the pH of the solution is adjusted to $8.0 \pm 0.3$ by adding NaOH pellets. Temperature: $105 \pm 1\,^{\circ}$C, 4\,h to reveal very flat etch pits and more than 4\,h to reveal fission tracks or tracks of super-heavy elements. If the etching solution is applied for 4 hours and the surface is surveyed using the Normaski-Differential-Interference-Contrast technique of an optical microscope, very small etch pits are visible, see Fig.~\ref{fig:Olivine_fissionTracks}. These etch pits are similar to those known from etched alpha-recoil-tracks in biotite or muscovite.  

\begin{figure}
    \centering
    \includegraphics[width=0.49\linewidth]{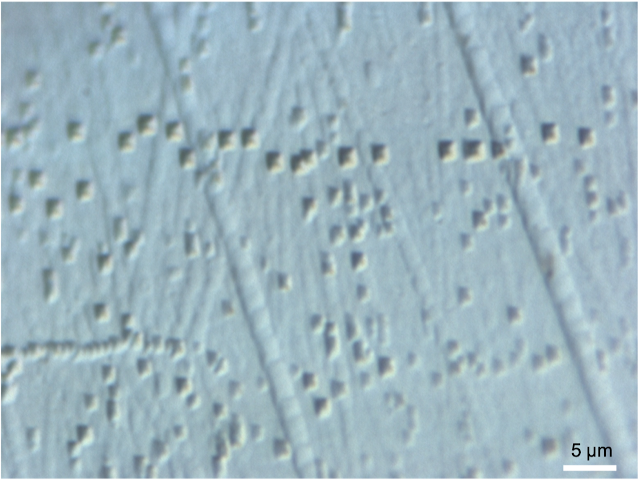}
    \includegraphics[width=0.49\linewidth]{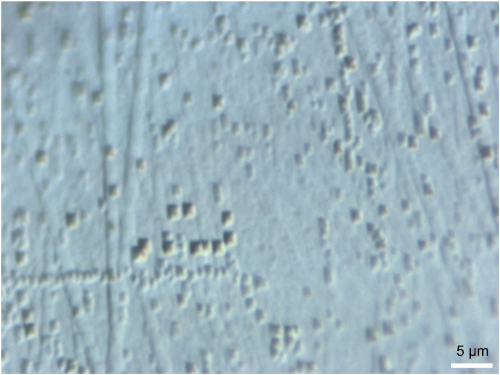}
    \caption{Very flat etch pits on the surface of olivine visualized by Normaksi-Differential-Interference-Contrast technique of an optical microscope~\cite{UG_unpublished}.}
    \label{fig:Olivine_fissionTracks}
\end{figure}

References~\cite{Fleischer:1965a, Fleischer:1975} describe fission track fading in olivine after 1\,hour at $500\,^{\circ}$C. Etching experiments of Davie and Durrani~\cite{Davie:1978} indicated an anisotropic etching behaviour of fission tracks caused by $^{252}$Cf irradiation of olivine in relation to the crystallographic orientation. The follow-up investigations were done by Ref.~\cite{Dersch:1991}. They confirmed the results of Ref.~\cite{Davie:1978} and enlarged the data base. The formation of fission tracks in olivine and other minerals was well described by Ref.~\cite{Pellas:1984}. Reference~\cite{Jakupi:1990} provided evidence for tracks caused by cosmic rays in olivine from the Soko Banja meteorite. They claim that the tracks are formed by nuclei of the iron group. The study of superheavy elements was initiated in the 1970's~\cite{Fowler:1977}, followed by numerous works, see Refs.~\cite{Fleischer:1967, Fowler:1977, Shirk:1978} and others. Most important are the publication of the group of Perelygin between 1977 and 2003. This research detected etched tracks in olivine from the meteorites Marjalahti and Eagle Station of unusual length ($120\textit{--}350\,\mu$m)~\cite{Perelygin:1977, Perelygin:1986, Perelygin:1991}. The very long etched tracks are interpreted as caused by super-heavy elements. To separate fission tracks caused by the fissioning of uranium and thorium, the olivine samples were annealed for a certain time. The annealing causes complete annealing of fission tracks and shortening of long super-heavy element tracks in olivine. A long list of publications exists dealing with etching, annealing, length of tracks in olivine from meteorites, tracks caused by super-heavy elements, and ion irradiation to simulate tracks in olivine. A more comprehensive list of literature on the subject is maintained by an author of this whitepaper, U.~A.~Glasmacher (Heidelberg University).

\subsection{Alpha-recoil tracks} \label{sec:MineralDetectors-ART}

Alpha-Recoil Tracks (ARTs) are the second group of radiogenic crystal defects caused by daughter products of the radioactive decay process in nature. They form during the $\alpha$-decay of $^{238}$U, $^{235}$U, $^{232}$Th and daughter products as well as $^{147}$Sm. The $\alpha$-decay of $^{238}$U is $10^6$ times more frequent than spontaneous fission. ARTs in mica were first described by Refs.~\cite{HuangWalker:1967, Huang+:1967}. They discovered a background of small and shallow etch pits as they tried to etch spontaneous fission-tracks in mica. Reference~\cite{HuangWalker:1967} proposed that the quantification of ART densities, combined with the analysis of uranium and thorium could provide a numeric dating technique. Methodological problems and difficulties related to the shallow and faint appearance of ARTs hampered the development of a practicable dating technique. Nevertheless, Refs.~\cite{Garrison:1978, Wolfman:1978} succeeded in dating millimeter-sized muscovite inclusions in pre-Columbian pottery. References~\cite{Hashemi-Nezhad:1981, Hashemi-Nezhad:1983} addressed shortcomings of the ART system in mica and improved the understanding of the physical model. Reference~\cite{Goegen:2000} first described an etching model for the increase of etch pit areal density ($\rho_a$) at a given 001-cleavage plane of dark mica (phlogopite) with increasing etching time. They applied the dating technique based on the etching model to Quaternary volcanic rocks of the Eifel region, Germany. 

\begin{figure}
    \centering
    \includegraphics[height=6.6cm]{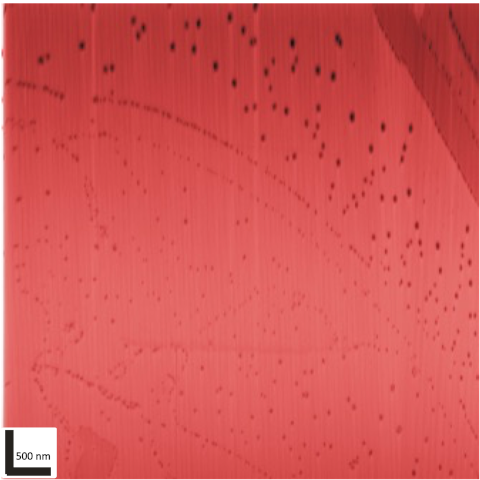}
    \includegraphics[height=6.6cm]{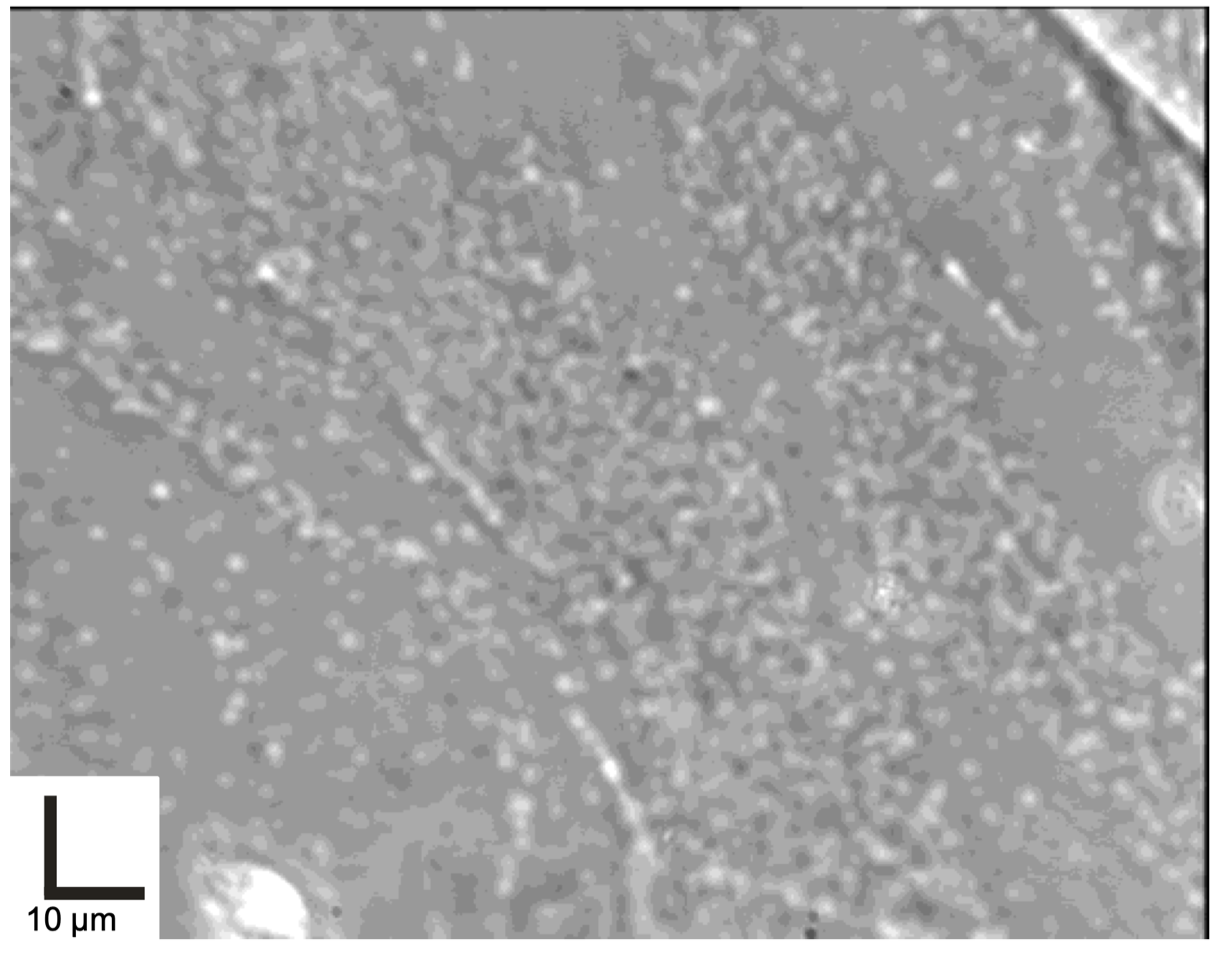}
    \caption{{\it Left}: AFM-image (topographic mode) of latent ART with hillocks above on a 001-plane of muscovite~\cite{UG_unpublished}. {\it Right}: Optical microscopy image of etched ART in the same muscovite on a 001-plane of muscovite as used for the AFM-image.}
    \label{fig:ART_muscovite}
\end{figure}

\begin{figure}
    \centering
    \includegraphics[width=0.5\linewidth]{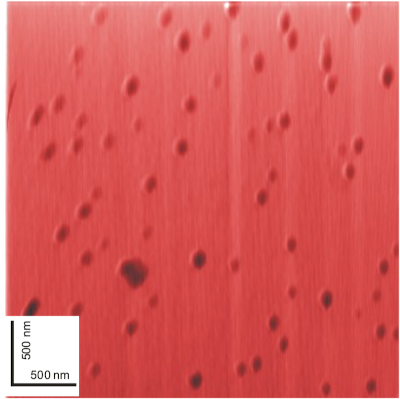}
    \caption{AFM image (topographic mode) of hillocks on top of latent ART on a 001-plane of phlogopite~\cite{UG_unpublished}.}
    \label{fig:ART_phlogopite}
\end{figure}

 The $\alpha$-disintegration releases energies of several MeV, part of which is transferred to the daughter-nucleus as recoil-energy ($\sim 10^2\,$keV). The $\alpha$-recoil nucleus slows down when interacting with the lattice atoms producing $\sim 10^3$ lattice defects which together constitute one single ART of about $30 \pm 5\,$nm in size~\cite{Jonckheere:2001, Stuebner:2006}. Within the $\alpha$-decay chain, the first $\alpha$-disintegration forms the ART~\cite{Lang:2003} and subsequent decays do not produce additional ARTs easily distinguishable from already existing ART left by the initial parent nucleus~\cite{Hashemi-Nezhad:1981}. Further decays in the decay chain generate a nest-shaped dislocation area in minerals, which have linear dimensions larger than 100\,nm~\cite{Jonckheere:2001, Stuebner:2006}. The latent size of the defects caused by the $\alpha$-decays in the time range between $0$ and $1\,$Myr was numerically simulated by applying Monte-Carlo techniques, leading to an increased estimate of feature size to about $125 \pm 50\,$nm. As expected, the amount of small-size crystal defects saturates at about $1\,$Myr, whereas the numbers of larger defects increase linearly with time~\cite{Jonckheere:2001}. The size of latent ARTs in dark mica is in the range of $30$ to $100\,$nm. The most important step in ART studies is the visualisation of nm-size defects in natural crystals caused by the $\alpha$-decay of $^{238}$U, $^{235}$U, $^{232}$Th and daughter products, as well as $^{147}$Sm. Investigations of spontaneous crystal defects caused by the $\alpha$-recoil nucleus in dark mica using scanning force microscopy has been described by Ref.~\cite{Lang:2002a}, see also Figs.~\ref{fig:ART_muscovite} and~\ref{fig:ART_phlogopite}. 

\begin{figure}
    \centering
    \includegraphics[width=1\linewidth]{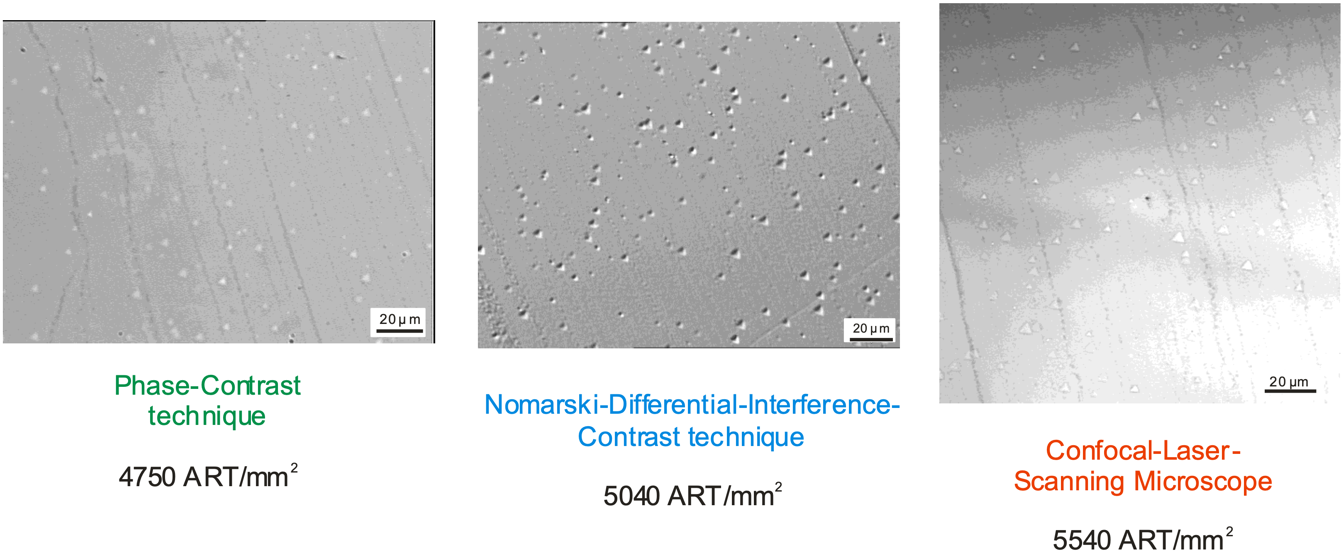}
    \caption{Different visualization techniques are applied by using etching and optical microscopy. Figure taken from Ref.~\cite{Glasmacher:2002}.}
    \label{fig:ART_etching_comparison}
\end{figure}

\begin{figure}
    \centering
    \includegraphics[width=1\linewidth]{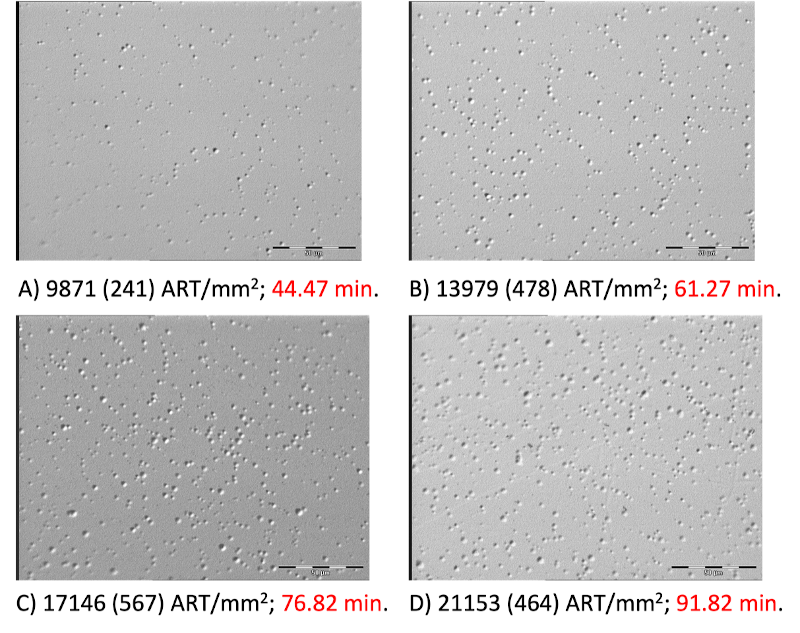}
    \caption{Optical microscopy images ($206 \times 156\,\mu{\rm m}^2$) displaying ART etch pit distributions at different etching times on the 001-plane of phlogopite from the Kerguelen Islands (OB-93-214). The numbers below each image display the areal densities, the errors representing $1\,\sigma$ of the mean. Etching was performed with 4\,\% HF at $21\,^{\circ}$C and etching times as indicated. Figure taken from Ref.~\cite{Glasmacher:2003}.}
    \label{fig:VolumeDensity-ART_1}
\end{figure}

\begin{figure}
    \centering
    \includegraphics[width=1\linewidth]{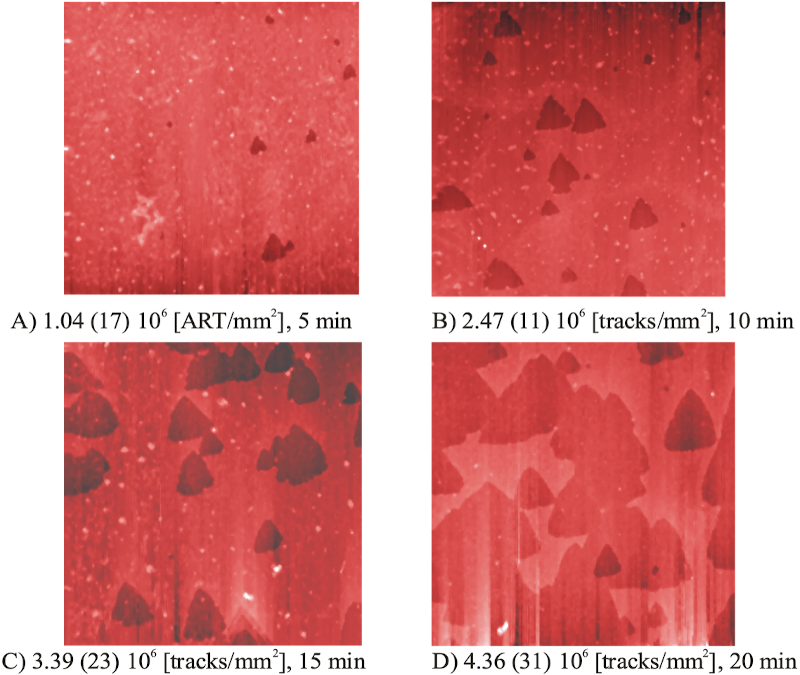}
    \caption{SFM images ($3 \times 3\,\mu{\rm m}^2$) displaying ART etch pit distributions at increasing etching times on the 001-plane of phlogopite from the Kovdor magmatic complex. The numbers below each image display the areal density. The errors of the areal densities are $1\,\sigma$ of the mean. Etching was performed with 4\,\% HF at $21\,^{\circ}$C and etching times as indicated. Figure taken from Ref.~\cite{Glasmacher:2003}}
    \label{fig:volume_AFM}
\end{figure}

To visualize latent ARTs, the easiest technique put into practice so far uses etchants to enlarge the crystal defects caused by the spontaneous recoils before visualizing the etch pits by optical, Scanning Electron (SEM), and Scanning Force Microscopy (SFM); see Figs.~\ref{fig:ART_etching_comparison}--\ref{fig:volume_AFM}. Application of the different techniques depends on the formation and cooling ages of dark micas, such as phlogopite and biotite. In samples less that $6\,$Myr old, the areal density ($< 2 \times 10^4\,{\rm mm}^{-2}$) can be quantified by optical microscopy. Above this age, the high ART track densities require SEM or SFM to determine densities of $5 \times 10^6\,{\rm mm}^{-2}$. Depending on the visualization technique, different etching conditions are applied to the dark mica sheet (grain size $\geq 500\,\mu$m). For example, ARTs can be read out using optical microscopy after etching with 40\,\% or 4\,\% HF at $23\,^{\circ}$C for a certain time, producing small shallow triangular etch pits on the cleavage plane. By their shape, ART-etch pits can be distinguished from etched fission-tracks and dislocations~\cite{Jonckheere:2001, Lang:2002a, Lang:2002b, Lang:2004, Stuebner:2008, Stuebner:2015}. 

The etch pits are visible and countable in transmitted and reflected light. In transmitted light, phase-contrast microscopy is used~\cite{Huang+:1967, Goegen:2000}. Until recently, this technique restricted ART observations to transparent samples. For instance, dark mica samples had to be cleaved until they became translucent. However, the Normansky-Differential-Interference-Contrast-technique now allows for the visualization of shallow ART etch pits in reflected light (Figs.~\ref{fig:ART_etching_comparison} and~\ref{fig:VolumeDensity-ART_1}). Thus, the only limit on visualization which remains is the etchability of the ARTs. Application of the Normansky-Differential-Interference-Contrast-technique with an Olympus BX~50 microscope with objective of 100x and a lens of 2x, a Panasonic F15 HS video camera, and the analySIS\textsuperscript{\textregistered} software of Soft Imaging System lead to a resolution of $0.7\,\mu$m. In addition, confocal laser scanning microscopy in reflected mode was tested. It has the advantage that the spatial resolution is half of the applied laser wavelength~\cite{Glasmacher:2001, Glasmacher:2002, Glasmacher:2003}. SFM requires that dark mica is etched with 4\,\% HF at $23^{\circ}$C between 1 to 20\,min~\cite{Glasmacher:2002}. These etching conditions also  create shallow triangular, but much smaller etch pits on the cleavage plane, see Fig.~\ref{fig:volume_AFM}. 

The appropriate dating technique depends on how the ARTs are visualized, the volume density of ARTs, and the uranium and thorium concentrations. The age equation combines the volume density ($\rho_V$) of ARTs with the U and Th contents~\cite{Goegen:2000}. Etching latent $30\textit{--}100\,$nm ARTs in phlogopite by HF (40\,\%; 4\,\%) at various etching times and measuring the areal density of triangular etch pits by optical microscopy or SFM reveals linear ART-growth (areal density versus etching time), see Figs.~\ref{fig:VolumeDensity-ART_1} and~\ref{fig:volume_AFM}. The volume density of ARTs is derived from the slope of the ART-growth and the effective etch rate ($v_{\rm eff}$). Using ART analytical techniques, single grain ages ranging from $10^3$ to $10^8\,$years can be determined. A linear relationship of the surface density of ART etch pits with etching time ($t_e$) was observed by optical microscopy for phlogopite of Quaternary and Neogene volcanics from the Eifel region (Germany), the East African Rift system and the Kerguelen Islands (Indian Ocean). This linear relationship was also observed using SFM for phlogopite from the Middle Devonian Kovdor magmatic complex (Russia) and a Triassic dike (Central Spain)~\cite{Goegen:2000, Glasmacher:2001, Glasmacher:2003}.

Based on phlogopite from Quaternary volcanic rocks of the Eifel region, Ref.~\cite{Goegen:2000} proposed an etching model for the increase of etch pit areal density at a given cleavage plane of mica with increasing etching time. The volume density ($\rho_V$) of ARTs, which is necessary to evaluate an age, is a function of the slope of the linear ART-growth (areal density versus etching time) and the efficient etch rate ($v_{\rm eff}$). This etch rate is not identical to the rate $v_v$ parallel to the $c$-axis, as had been assumed by Ref.~\cite{Goegen:2000}. This discrepancy becomes obvious when calculating independent ART ages from the U and Th contents, the slope of the ART-growth, and $v_v$. The resulting ages turn out to be too high by a factor of 5 to 10. To account for this discrepancy, the effective etch rate has been calculated by solving the age equation for $v_{\rm eff}$ after inserting the independently known ages of samples from Laacher See (12.9\,kyr) and Bausenberg ($150 \pm 40\,$kyr), both located in the Eifel area. Within the respective errors, both values of $v_{\rm eff}$ agree with the previously determined etch rates $v_h$ perpendicular to the $c$-axis~\cite{Goegen:2000}. Therefore, $v_h$-values can be applied for ART age calculations. The $v_h$ parameter is particularly easy to measure since it is directly derived from the size of the largest ART etch pits.

The uranium and thorium concentration in dark mica can be quantified using Laser Ablation Inductively Coupled Plasma Mass Spectrometry (LA-ICP-MS). This technique allows for the analysis of dark mica volumes with areas defined by the laser beam size and the depth determined by that of the laser pit. Typically, the beam size of the laser has linear dimensions of $80\,\mu$m and the pit depth is $120\,\mu$m. The U and Th contents of the sample are continuously recorded. The time-concentration profile reflects the change of uranium and thorium concentration with depth. Furthermore, concentrations of Zr, Ce, Sm, and Hf are determined since these elements indicate mineral inclusions in dark mica, see Fig.~\ref{fig:etchedApatite}. Two to five analytical points for each grain are used to characterize and quantify the uranium and thorium concentrations. Parts of the sample characterized by anomalously high concentrations of Zr, Ce, Sm, and Hf indicate the presence of inclusions and need to be avoided when calculating the uranium and thorium contents of the mica.

As all crystal defects due to radiation damage in minerals are metastable, the retention characteristics of ARTs are also likely controlled by temperature and time. When phlogopite is kept at $500\,^{\circ}$C for 15\,min, no etchable track is left~\cite{Goegen:1999}. Extrapolating the experimental data to geologic time scales (1\,Myr) implies that the ART density is reduced by 50\,\% at a temperature of $50\,^{\circ}$C. The annealing behaviour also seems to be affected by the chemical composition of dark micas, as demonstrated by experiments on Ti-rich phlogopite from the Kerguelen. Extrapolating data from these experiments suggest ART track retention in the range of $60\,^{\circ}$C to $70\,^{\circ}$C. In the case of volcanic rocks which cool rapidly, the ART retention temperatures are significantly higher and are passed within hours or days after eruption so that the ART age represents the time of rock formation. On the other hand, during the exhumation of mica-bearing rocks, very slow cooling rates prevail. The relatively low ART retention temperatures of $50\textit{--}70\,^{\circ}$C are reached long after rock formation so that the ART ages correspond to the low-temperature cooling ages. Although considerably less data is available for tracks associated with nuclear recoils from weak interactions, the vast literature pertaining to both fission track and ART analyses can provide important lessons moving forward for the mineral detection of neutrinos and Dark Matter.

\subsection{Recoils from weak interactions} \label{sec:MineralDetectors_DM}

In sections~\ref{sec:Physics} and~\ref{sec:Studies}, we discuss recent work on searches for various signals in mineral detectors from weak interactions. In this section, we summarize earlier studies focused on detecting nuclear recoils from the weak interactions of Dark Matter, followed by a brief discussion of nuclear recoils induced by fast neutrons. 

Snowden-Ifft and Chan~\cite{Snowden-Ifft:1995rip} and Snowden-Ifft {\it et al.}~\cite{Snowden-Ifft:1995zgn} published results of first experiments related to the use of muscovite mica as a paleo-detector for Dark Matter. They performed experiments with accelerated ions and neutrons that clearly describe the variation in etch pit depth in relation to the particle energy. Within this publication they also provided information on the etch pit depth of defects caused by nuclear recoils that would be induced by the interaction of Dark Matter with atoms of the crystal lattice of muscovite mica. The visualization after etching the mica by 49\% HF for 1\,h at room temperature was performed by Atomic Force Microscopy (AFM)~\cite{Snowden-Ifft:1995zgn}. The mode for the AFM is not provided in these papers. Also, they changed the etching conditions between the two papers. Two years earlier, Snowden-Ifft {\it et al.}~\cite{PhysRevLett.70.2348} described etch pits of alpha-recoil-tracks in muscovite micas. The etching was performed with 49\% HF for 4\,h at $20\,^{\circ}$C. Muscovite mica was annealed at $450\,^{\circ}$C for 1\,h to erase the natural alpha-recoil-tracks. Thereafter, the muscovite mica was irradiated with 200\,keV Ag-ions, etched at the same conditions than before and visualized by AFM-technique. Despite the limited sensitivity to Dark Matter scattering off nuclei, the neutron scattering calibration performed in these earlier studies demonstrated that lower energy nuclear recoils associated with Dark Matter scattering can indeed form persistent damage features in rock forming minerals.

Snowden-Ifft and Westphal~\cite{Snowden-Ifft:1997vmx} discussed a unique signature of Dark Matter in muscovite mica. They connected the movement of the Earth around the center of the Galaxy with the location of Dark Matter interaction induced tracks in muscovite mica. They stated that Dark-Matter-induced tracks would have a preferred orientation in mica. Such a preferred orientation would not be seen if neutrons create tracks in muscovite mica. In a follow-up paper, Baltz {\it et al.}~\cite{Baltz:1997dw} describe their Monte Carlo-simulations related to the Dark Matter halos in mica. An interesting paper by Engel {\it et al.}~\cite{Engel:1995gw} discussed the response of mica to Dark Matter in detail. They analysed the spin-dependent Dark Matter--nucleus scattering within the mica. The most important result is that ``the efficiency with which each element in mica can be detected when it recoils is different''. An overall cross section for representative WIMPs cannot easily be described. Related to Ref.~\cite{Snowden-Ifft:1995zgn}, Collar~\cite{Collar:1995aw} provided some discussion, see also the reply in Ref.~\cite{Snowden-Ifft:1996dug}. In the rest of this section, we describe recoils induced by radiogenic and cosmogenic neutrons, which are common to many of the searches which discussed below.

Since neutrons are only weakly interacting, neutrons scattering with nuclei inside of a mineral detector can yield nuclear recoils similar to scattering induced by neutrinos or Dark Matter. Neutrons produced by spontaneous fission and ($\alpha,n$)-interactions typically begin with $\sim 1 \, {\rm MeV}$ kinetic energies. For neutrons energies down to $\sim {\rm keV}$, which can yield nuclear recoils similar to various signals in mineral detectors, elastic scattering is the dominant interaction. However, the interaction length for the elastic scattering of neutrons off nuclei at such energies is $\sim 1 \, {\rm cm}$. Thus, a single radiogenic neutron will typically yield between $\sim 10 - 1000$ nuclear recoils in the energy range of interest, depending on the target material. In addition, these recoils will be distributed over volumes much larger than the target sample sizes considered for many mineral detector applications. Due to the macroscopic interaction length and lack of electromagnetic interactions, neutron scattering with nuclei cannot be traced back to the original radioactive decay. Unlike conventional rare event searches with large instrumented volumes of target material which are monitored in real time for multiple scattering events, mineral detectors simply record nuclear recoils from radiogenic neutrons as an irreducible background.

On the other hand, nuclear recoils induced by cosmogenic neutrons could be used as a probe of cosmic ray (CR) interactions with the Earth's atmosphere over geological timescales. A primary CR particle, typically a proton, interacting in the atmosphere produces a cascade of secondary particles, including charged pions. The subsequent decays of the charged pions yield a flux of muons and neutrinos which can penetrate the Earth's surface. The muons and neutrinos can then interact with nuclei in the vicinity of a mineral detector and yield fast neutrons which can induce nuclear recoils similar to the radiogenic neutrons described above. As discussed in sections~\ref{sec:Physics-AstroNu} and~\ref{sec:Physics-CosmicRays}, the atmospheric neutrinos and muons interacting within the mineral detector volume are a less ambiguous probe of CR interactions when compared to the signal associated with cosmogenic neutrons. Thus, we consider cosmogenic neutrons as a background for the searches considered here. The flux of the muons associated with most of the cosmogenic neutron production is exponentially dependent on the depth of the mineral detector due to the absorption of charged particles by the overburden of the Earth. Thus, a relatively detailed history of how the depth of a mineral detector has changed with time is necessary to be able to accurately model the cosmogenic neutron background.

While it is not clear if modeling of the geological history for a ``paleo-detector'' application of minerals could be sufficiently accurate to account for the cosmogenic neutron background, the radiogenic neutron interactions are relatively straightforward to model in a given target material and calibrate in laboratory setting. Thus, we generally require that mineral detectors are known to have been buried sufficiently deep in the Earth on timescales relevant for the various signals of interest such that the cosmogenic neutron background can be safely ignored. For example, in a mineral detector with $\sim {\rm cm}$ linear dimensions, the number of cosmogenic neutron recoils expected over $\sim 1 \, {\rm Gyr}$ is $\lesssim 100$ at depths of $\gtrsim 5 \, {\rm km}$. Here it is worth emphasizing one of the most important differences between a rare event search with a ``paleo-detector'' and more conventional experiments. Due to the relatively large exposure possible with a mineral detector, the number of signal and background events can be up to $\mathcal{O}(10^4)$ at the sensitivity threshold compared to $\mathcal{O}(1)$ for the same signal at conventional rare event searches. Thus, the number of cosmogenic neutron events observed in a mineral detector retrieved from a borehole at a depth of $\sim 5 \, {\rm km}$ is typically only a $\sim 1 \%$ contribution to the total number of events. 
 
For mineral detectors in which the cosmogenic neutron backgrounds are sufficiently suppressed, the sensitivity to a given signal is then largely determined by the number of nuclear recoils induced by radiogenic neutrons in the recoil energy range of interest. The normalization of the radiogenic neutron background is controlled by the concentration of radioactive isotopes in the material surrounding the mineral detector. In order to be sensitive to nuclear recoil signals associated with WIMP-like Dark Matter, as well as solar and supernova neutrinos, mineral detectors must be extracted from rocks in very radiopure geological environments, several examples of which are discussed in this section and below. For mineral detector searches for reactor neutrinos and geoneutrinos, in addition to lower mass WIMPs, radiopurity is typically less important due to the overwhelming number of lower energy nuclear recoils induced by the solar neutrino flux. Alternatively, mineral detector searches for atmospheric neutrinos, atmospheric muons, composite Dark Matter and other heavy exotic astroparticles are largely unaffected by the radiogenic neutron background. In short, the recoil energies relevant for such searches are $\gg {\rm MeV}$, the scale associated with the most energetic recoils induced by radiogenic neutrons. The sensitivity of mineral detectors in searches for a variety of potential signals and backgrounds, including cosmogenic and radiogenic neutrons, are discussed in more detail below.

%*********************************************************
\section{Applications of Mineral Detectors} \label{sec:Physics}
%*********************************************************
There are a wide range of applications for natural as well as laboratory-fabricated minerals as solid state nuclear track detectors, encompassing various aspects of fundamental particle physics, astrophysics and geoscience, as well as applied-science uses such as nuclear safe-guarding. Depending on the particular mineral and the readout technology used, minerals could serve as detectors with nuclear recoil energy thresholds as low as $\mathcal{O}(0.1)\,$keV. The key differences to consider when compared to more conventional nuclear recoil detectors are that mineral detectors are (i) passive and (ii) can retain the crystal defects caused by recoiling nuclei for geological timescales. Depending on the particular application of mineral detectors, these differences can manifest as either challenges to be overcome or advantages to be exploited. For all of the applications described in this section, mineral detectors have potential sensitivity to signals which are either very difficult or impossible to measure with conventional detectors.

Section~\ref{sec:Physics-WIMPs} discusses the prospects of using minerals as detectors for the most well-studied Dark Matter candidate, Weakly Interacting Massive Particles (WIMPs). Natural minerals used as paleo-detectors could compete with or even surpass the discovery reach of conventional current and next-generation direct detection experiment for a wide range of Dark Matter masses. Furthermore, paleo-detectors would be highly complementary to conventional detectors: as we will see, thanks to the enormous possible exposure, paleo-detectors might be much more powerful than conventional detectors for reconstructing the properties of a WIMP such as its mass. Due to the long effective integration times of $100\,{\rm Myr}\textit{--}1\,$Gyr that might be possible with paleo-detectors, rather than being sensitive to the Dark Matter density today on Earth, paleo-detectors would measure the average Dark Matter density over their long integration times which could, e.g., open up the exciting possibility of directly detecting Dark Matter substructure in the Milky Way's halo. Section~\ref{sec:Physics-CompositeDM} describes possible uses of mineral detectors for heavier (composite) Dark Matter candidates and other exotic high energy astroparticles. The main difficulty  in searching for such particles is that their expected flux through a detector on Earth would be tiny. For example, if Dark Matter is composed of Planck-mass states ($m_{\rm DM} \sim 10^{19}\,$GeV), the expected flux is less than one per year through a meter-sized detector. Converting this flux into units appropriate for a paleo-detector, one instead finds $\Phi_{\rm DM} \sim 10^4\,{\rm cm}^{-2}\,{\rm Gyr}^{-1} \times \left(10^{19}\,{\rm GeV}/m_{\rm DM}\right)$.

In section~\ref{sec:Physics-AstroNu} we turn to guaranteed signals: mineral detectors could be used to measure the neutrinos from a number of astrophysical sources such as our Sun, supernovae, or Cosmic Rays interacting with Earth's atmosphere. Perhaps of particular interest, such measurements could not only infer the neutrino fluxes associated with these sources, but also probe the time-evolution of the fluxes over $10\textit{--}10^3\,$Myr timescales. Section~\ref{sec:Physics-CosmicRays} discusses ideas to use mineral detectors as latent detectors for showers from high-energy cosmic rays. In section~\ref{sec:Physics-ReactorNu} we discuss possible uses of mineral detectors as passive, room-temperature reactor-neutrino detectors via coherent elastic neutrino-nucleus scattering. As such, laboratory-manufactured minerals could constitute a readily field-deployable system for nuclear non-proliferation safeguards. 

In section~\ref{sec:Physics-GeoNu} we turn to geoscience applications of mineral detectors: Geo-neutrinos have energies similar to those of solar, supernova, and reactor neutrinos, and give rise to nuclear recoils via coherent elastic neutrino-nucleus scattering. While searching for geo-neutrinos with mineral detectors is a particularly challenging application due to the solar-neutrino background, it would open up the exciting possibility of using a series of natural mineral detectors from different locations and of different ages to map the geo-neutrino flux in time and space. Finally, section~\ref{sec:Physics-Geoscience} takes us back to the fission-track and $\alpha$-recoil track dating applications of natural minerals in geoscience. If successful, the program of developing microscopy techniques capable of detecting the defects from neutrino- or Dark-Matter-induced nuclear recoils in minerals would also revolutionize the geoscience applications of mineral detectors. These techniques could be used to infer the ages of geological samples with much lower uranium/thorium concentration than what is currently possible and allow for much more detailed studies on the temperature-history of samples.

Before entering the more detailed discussion of the various possible applications of mineral detectors, let us stress some general points of how one would differentiate the signal from the background in mineral detectors. While the details of how such a search would work are particular to each possible application, for most physics cases discussed here, including searches for WIMP-like Dark Matter, astrophysical neutrinos, reactor neutrinos, or geoneutrinos, the situation would be very different to what the reader might be familiar with from more conventional experiments. For example, the typical strategy for the direct detection of Dark Matter is to construct a signal region with zero (or a small number of) background events, and search for an excess of a small number of (signal) events in that region. In most applications of mineral detectors, instead, at the projected sensitivity limit, one would observe a large number of background and signal events. In order to claim the (non-)observation of a signal, one must reconstruct additional information about each event, e.g., the length of damage tracks as a proxy for the nuclear recoil energies, to obtain a differential distribution of the events. One would then compare the predicted differential distribution for the various backgrounds and the signal to the data, and perform a statistical analysis (for example, a maximum likelihood ratio test) to test if the observed distribution is better matched by the background-only or the signal+background models. Additional evidence for possible signals can be obtained by using other mineral detector samples with different chemical composition, exposure times, etc. and testing if the various background sources and the claimed signal scale as expected in such different samples.

Let us also briefly comment on the prospects of using mineral detectors as {\it directional} nuclear recoil detectors. In principle, the latent lattice damage left by recoiling nuclei does contain information about the direction of the nuclear recoil. This directional information  can be used to differentiate the signal from the background in some applications of mineral detectors, for example, when looking for signals from reactor neutrinos (see section~\ref{sec:Physics-ReactorNu}) or in ideas to use laboratory-manufactured crystals for the directional detection of Dark Matter (see section~\ref{sec:Studies-MD-Directional}). However, when minerals are used as paleo-detectors, the signal (and background) is accumulated over Myr -- Gyr timescales, over which the crystal would re-orient relative to the sources due to Earth's rotation (and the axis of rotation nutates and precesses), Earth's orbit around the Sun (and the obliqueness of planetary orbits can change on Myr -- Gyr timescales), the Solar System's orbit around the Milky Way, and tectonic drifts of the geological deposits hosting the paleo-detector. These movements do wash out most of the differences in directionality between signals and background in paleo-detector applications of mineral detectors.

%*********************************************************
\subsection{WIMP-like Dark Matter} \label{sec:Physics-WIMPs}
%*********************************************************
%{\color{blue} Coordinator: Bradley Kavanagh}

There is overwhelming cosmological and astrophysical evidence for the existence of Dark Matter (DM)~\cite{Bertone:2004pz}. Though the exact nature of DM is unknown, the dominant paradigm in recent decades has been that DM takes the form of an as-yet-undiscovered Weakly Interacting Massive Particle (WIMP), whose interaction cross section with Standard Model particles is small. A wide range of low-threshold, low-background detectors have been deployed (or are being developed) to search for WIMP DM scattering with nuclei~\cite{Schumann:2019eaa,Billard:2021uyg}. These experiments aim to reconstruct the energy of recoiling nuclei by measuring ionisation, scintillation or heat signals in the detector. By looking for an excess of events over the expected backgrounds, these \textit{direct detection} experiments aim to set limits on the WIMP-nucleon elastic scattering cross section $\sigma_n^\mathrm{SI}$, or perhaps one day make a discovery. 

Mineral detectors could be used to search for WIMP DM in a similar way, with a WIMP-nucleus scattering event leading to a nucleus recoiling through the crystal. Information about the energy of the recoil would be recovered by measuring the length of the \textit{damage track} produced by the recoiling nucleus (known as the primary knockout atom, or PKA)~\cite{Baum:2018tfw,Drukier:2018pdy,Edwards:2018hcf,Baum:2021jak}. Depending on the PKA, a WIMP-induced nuclear recoil would have a typical energy on the order of $\mathcal{O}(0.1-100\mathrm{s})\,\mathrm{keV}$. Assuming that the PKA produces a track along the full length of its stopping distance, this would correspond to tracks around $1-1000\,\mathrm{nm}$ in length. Figure~\ref{fig:Range} shows the relationship between the recoil energy and the range (i.e.~the nuclear stopping distance) for different recoiling nuclei in sinjarite.

\begin{figure}[t]
    \begin{center}
        \includegraphics[width=0.7\linewidth]{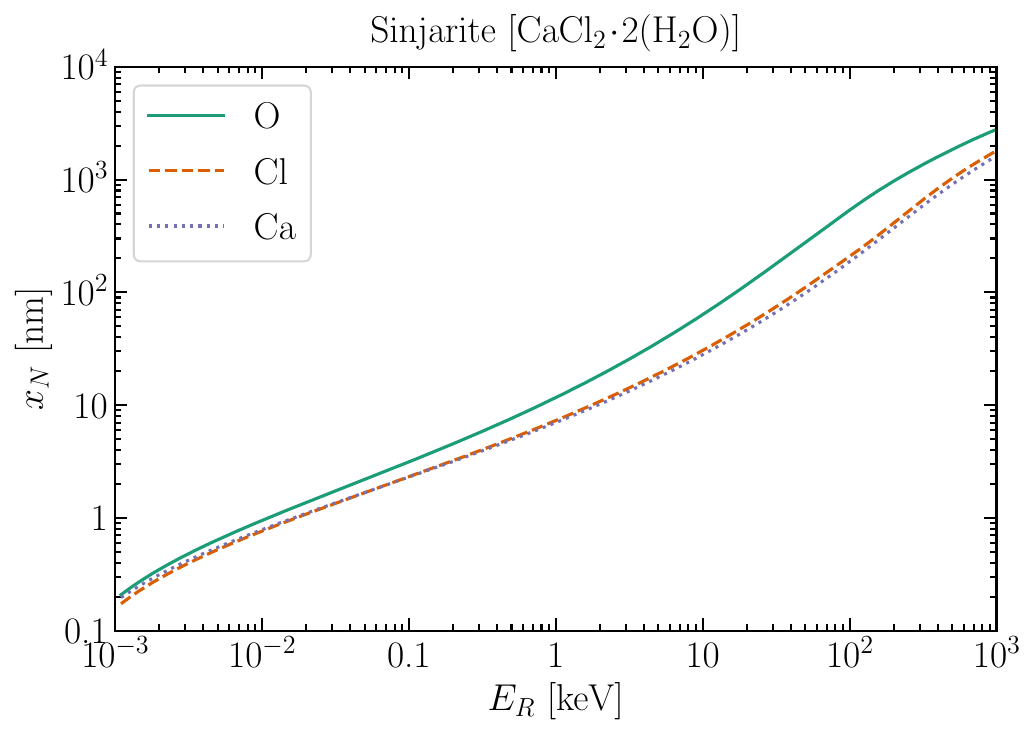}
        \caption{Range $x_N$ of recoiling nuclei in sinjarite. The range depends on the primary knockout atom, but nuclei recoiling with typical energies of $0.1-100\,\mathrm{keV}$ are stopped over a distance $1-1000\,\mathrm{nm}$. These estimates of the range are obtained using \texttt{SRIM}~\cite{Ziegler:1985}.}
    \label{fig:Range}
    \end{center}
\end{figure}

The signature of WIMP DM would be an excess of tracks with these typical lengths, with the exact distribution depending on the WIMP mass. One of the key advantages of mineral detectors is that a large experimental exposure can be achieved by examining rock samples with a sufficiently large age (that is, the time over which damage tracks have been being recorded). Even if only a small sample mass can be analysed, this is compensated by the $\mathcal{O}(\mathrm{Gyr})$ exposure times which would be possible with existing rock samples, which we call `\textit{paleo-detectors}'. However, unlike conventional direct searches, `paleo-detectors' are inherently \textit{passive}, with no active background mitigation. It is therefore crucial to understand other non-WIMP sources which could lead to tracks of a similar length. The track-length distribution for non-WIMP sources is illustrated in Fig.~\ref{fig:Spectra} and we briefly summarize the most important contributions below.

\begin{figure}[t]
    \begin{center}
        \includegraphics[width=0.6\linewidth]{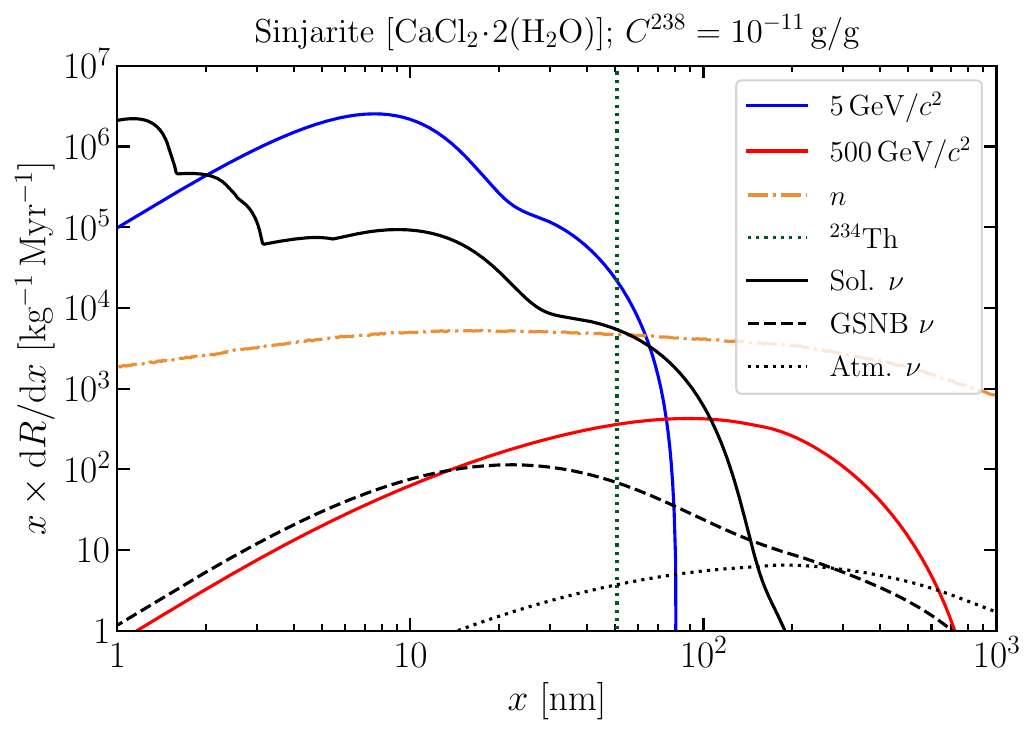}
        \caption{Predicted track-length spectra for nuclear recoils in sinjarite. Black curves show the contribution from coherent elastic neutrino-nucleus scattering, produced by solar neutrinos (black solid), the background of galactic supernova neutrinos (black dashed) and atmospheric neutrinos (black dotted). Note that contribution from deep inelastic scattering of atmospheric neutrinos (as described in section~\ref{sec:Physics-AstroNu}) is not included here. Radiogenic backgrounds include thorium-234 tracks produced in single-alpha decays (whose energy is marked as a vertical green dotted line) and radiogenic neutrons from spontaneous fission and ($\alpha$, $n$) reactions (orange dot-dashed). The solid blue and red lines show the spectra of tracks produced by WIMPs with masses $m_\chi = 5 \,\mathrm{GeV}$ and $m_\chi = 500 \,\mathrm{GeV}$ respectively. These track-length spectra are obtained by converting the expected distribution of recoil energies, using the ranges shown in Fig.~\ref{fig:Range}.}
    \label{fig:Spectra}
    \end{center}
\end{figure}

Solar, atmospheric and diffuse supernova neutrinos will easily penetrate any rock overburden and -- if they interact -- will produce single nuclear recoils, just like WIMPs. Neutrinos therefore represent a key background for WIMP searches, with different neutrino sources being the more relevant background for different WIMP masses~\cite{OHare:2021utq}. Of course, mineral detectors may also be used to search for neutrino-induced tracks themselves, as discussed in sections~\ref{sec:Physics-AstroNu},~\ref{sec:Physics-ReactorNu} and~\ref{sec:Physics-GeoNu}.

Radiogenic backgrounds will also be substantial, depending strongly on the uranium-238 concentration ($C^{238}$) in the mineral sample. Uranium-238 decays to thorium-234 (${ }^{238} \mathrm{U} \rightarrow{ }^{234} \mathrm{Th}+\alpha$) with a half-life of around $4.5\,\mathrm{Gyr}$. The thorium-234 child nucleus recoils with an energy of $72\,\mathrm{keV}$ and would in principle be indistinguishable from a WIMP-induced recoil of that energy~\cite{Baum:2018tfw,Drukier:2018pdy}. Subsequent steps in the decay chain occur much more rapidly and so most ${ }^{238} \mathrm{U}$ decays would produce a cluster of tracks which can be rejected as background. However, a small fraction of nuclei in the chain have not undergone a second $\alpha$-decay and leave behind a single track. In absolute terms, the number of such `single-$\alpha$' tracks can be large, though these tracks should have a fixed length (fixed by the 72 keV recoil energy) and so do not pollute wide ranges of the track-length spectrum. A more difficult background to reject comes from radiogenic neutrons, produced in the spontaneous fission of  ${ }^{238} \mathrm{U}$ and in ($\alpha$, $n$) reactions. These fast neutrons can scatter with nuclei 10s-100s of times, producing a relatively flat spectrum of recoil tracks over the range of track-lengths relevant for WIMP searches. Obtaining minerals which have a small uranium-238 concentration is crucial to minimize backgrounds from both single-$\alpha$ decays and radiogenic neutrons.

With such a large number of tracks coming from neutrinos and radioactive decays, event-by-event discrimination between DM-induced events and background will not be possible. Instead, by studying and modelling the \textit{distribution} of track-lengths in a mineral detector, it may be possible to distinguish an excess of DM-induced tracks. There are therefore two crucial requirements for the success of a WIMP DM search with mineral detectors:
\begin{itemize}
\item Precise track-length measurement: in order to find evidence for an excess of DM-induced tracks, it will be necessary to measure the full 3-dimensional lengths of tracks, so that the track-length spectrum can be compared to theoretical expectations. As a concrete example,  single-$\alpha$ tracks do not represent a major hindrance to a WIMP search if the full length of the track can be precisely measured; these tracks appear with a characteristic length and can therefore be rejected. 
\item Control of systematics: DM-induced tracks are likely to be a subdominant population and therefore a precise understanding of the expected background spectra is crucial. This includes a good understanding of the normalization of backgrounds (for example, by estimating $C^{238}$) but also the expected distribution of tracks produced by radiogenic neutrons. 

\end{itemize}

\begin{figure}[t]
    \begin{center}
        \includegraphics[width=0.7\linewidth]{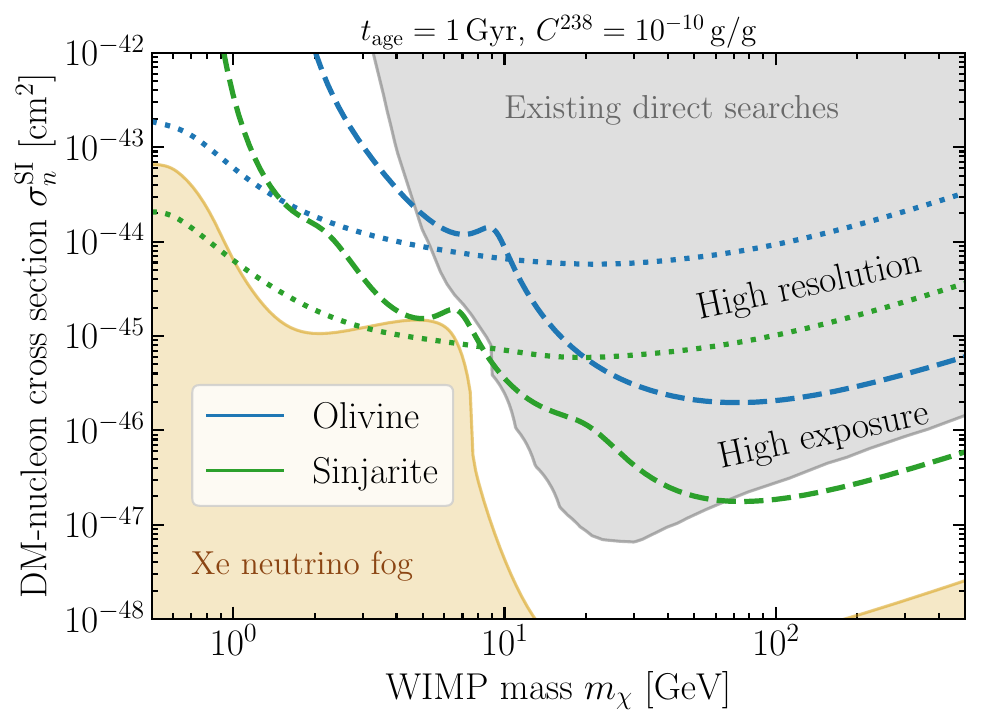}
        \caption{Projected WIMP Dark Matter discovery reach for olivine (blue) and sinjarite (green) mineral detectors. We assume a mineral age of $1\,\mathrm{Gyr}$ and a uranium concentration of $C^{238} = 10^{-10}\,\mathrm{g}/\mathrm{g}$. We consider two possible scenarios for the track length resolution $\sigma_x$ and the sample mass $M_\mathrm{sample}$: \textbf{High resolution} ($\sigma_x = 1\,\mathrm{nm}$, $M_\mathrm{sample} = 10 \,\mathrm{mg}$, dotted lines) and \textbf{High exposure} ($\sigma_x = 15\,\mathrm{nm}$, $M_\mathrm{sample} = 100 \,\mathrm{g}$, dashed lines). The projections were produced using \texttt{PaleoSens}~\cite{Baum:2021jak}. For comparison, we show in grey the envelope of existing constraints from direct WIMP searches (taken from Ref.~\cite{OHare:2021utq}, with the addition of recent limits from LUX-ZEPLIN~\cite{LZ:2022ufs}). We also show in yellow the position of the neutrino fog \textit{for Xenon-based experiments}, as calculated in Ref.~\cite{OHare:2021utq}.}
    \label{fig:WIMP_reach}
    \end{center}
\end{figure}

Figure~\ref{fig:WIMP_reach} shows the projected discovery reach for mineral detectors, assuming that 3-dimensional track-length reconstruction is possible, where we consider olivine and sinjarite as concrete examples. These projections show that mineral detectors may be competitive with existing dedicated direct searches for WIMP Dark Matter. In particular, if track-lengths can be modeled and measured at the level of $\sigma_x = 1\,\mathrm{nm}$, sensitivity to GeV-mass WIMPs can be improved by several orders of magnitude compared to current constraints from low-threshold detectors. With a higher exposure (which may be achievable at the expense of a lower track-length resolution), some mineral detectors may be competitive with existing high-mass searches. In fact, these searches are highly complementary. For WIMP masses above around 200 GeV, a signal in a standard WIMP search would translate only into a lower bound on $m_\chi$, rather than a precise reconstruction~\cite{Bozorgnia:2018jep}. This is because the spectrum of recoil energies produced by heavy WIMPs is \textit{almost} insensitive to the precise WIMP mass~\cite{Green:2008rd}. In the event of a discovery in a mineral detector, however, the WIMP mass could be well-constrained for masses as large as $1\,\mathrm{TeV}$~\cite{Edwards:2018hcf}, owing in part to the large number of signal tracks, which would allow a very precise estimate of the track-length spectrum. Mineral detectors offer further unique possibilities compared to conventional detectors:

First, if the flux of DM varies on timescales longer than $\mathcal{O}(\mathrm{yr})$, this cannot be detected by conventional searches. This would be the case particularly if the DM halo contains a substantial amount of substructure, for example, in the form of a Dark Disk~\cite{Read:2008fh,Fan:2013yva}, or if a large fraction of the DM is bound up in sub-halos~\cite{Berezinsky:2007qu,Erickcek:2011us,Buckley:2017ttd}. In extreme cases, currently-running searches may expect to see a strongly suppressed signal if the Earth is currently passing through a region where the DM density is much lower than the average. Instead, mineral detectors have been recording tracks over $\mathcal{O}(\mathrm{Gyr})$ timescales, alleviating this issue~\cite{Bramante:2021dyx}. Moreover, by examining mineral samples with different ages, it should be possible to infer information about the time-dependence of the DM flux~\cite{Baum:2021chx}. Thus, the spectrum of DM-induced nuclear recoil tracks accumulated in a mineral detector as the solar system rotates around the galaxy every $\sim 250 \, \mathrm{Myr}$ could potentially be used as a probe of substructure even if most of the DM is part of the smooth galactic halo. 

Second, if the flux of DM is very small, then the probability of a DM particle encountering the detector during operation is substantially reduced. Fixing the local DM mass density, the number flux of DM particles scales as $\Phi_\chi \sim \rho_\chi/m_\chi$. Assuming a DM mass of $10^{18}\,\mathrm{GeV}$, a meter-scale detector would expect to encounter only one DM particle during a year of operation. Thus, conventional detectors rapidly lose sensitivity to ultraheavy DM at the Planck mass-scale and above ($m_\chi \gtrsim 10^{19}\,\mathrm{GeV}$)~\cite{Kavanagh:2017cru,Bramante:2018qbc}. Instead, the long exposure time of mineral detectors provides sensitivity to heavy but rare DM candidates. Going to such low fluxes would mean that the DM must be strongly-interacting in order to allow for a detection. In this case, the signature is distinct from an excess of single-recoil-induced tracks, as described in the next subsection.

%*********************************************************
\subsection{Composite Dark Matter and other high energy astroparticles} \label{sec:Physics-CompositeDM}
%*********************************************************
%{\color{blue} Coordinator: Joe Bramante}

Minerals provide unique sensitivity for composite Dark Matter and other cosmogenic particles that deposit large amounts of energy into the mineral, resulting in damage that is more easily read out than single nuclear recoils, and that is not expected from a Standard Model background process ($e.g.$ background from decays of nuclear isotopes). 

In particular, some simple composite models of Dark Matter with a large cross-section can leave tracks in minerals that are strikingly distinct from radiogenic backgrounds: these tracks can have widths ranging from nanometers to microns and extend along a straight trajectory for very long distances through mineral samples. This long straight track provides a unique signature of strongly interacting cosmogenic particles. In principle, these tracks could extend for multiple kilometers, because the initial kinetic energy of the particle is extremely large, compared to the energy it deposits in each nuclear interaction. For many composite models the damage track will be straight, because the Dark Matter has a mass tens of orders of magnitude greater than nuclei, resulting in a very minute angular deflection, even after many interactions with nuclei \cite{Bramante:2018qbc,Acevedo:2020avd,Davoudiasl:2019xeb}. Note that due to the size of such tracks, microscopy readout is much less demanding than searching for the $1\textit{--}1000\,$nm long tracks expected from WIMP-like Dark Matter.

Recent work exploring composite Dark Matter models \cite{Wise:2014jva,Nussinov:1985xr,Bagnasco:1993st,Alves:2009nf,Kribs:2009fy,Lee:2013bua,Krnjaic:2014xza,Detmold:2014qqa,Jacobs:2014yca,Bramante:2018tos,Ibe:2018juk,Coskuner:2018are,Bai:2018dxf,Bai:2019ogh,Bramante:2019yss,Wise:2014ola,Hardy:2014mqa,Hardy:2015boa,Gresham:2017zqi,Gresham:2017cvl,Gresham:2018anj} has shown that relatively simple models for the cosmological formation of composite Dark Matter predict composite Dark Matter masses that range up to and beyond $10^{34}\,$GeV for composites bound together by a scalar field, and to masses up to $10^{43}\,$GeV for composites bound together by a new confining force in the dark sector \cite{Bai:2018dxf}. These composites are typically formed through a straightforward cosmological assembly, where large composite states are built out of fermions or bosons, much like nuclei are built from nucleons during big bang nucleosynthesis. It has recently been appreciated that large composite states can have a large variety of interactions with nuclei, that include accelerating nuclei in the interior of the composite to produce nuclear recoils and Migdal electrons \cite{Acevedo:2021kly}, nuclear capture \cite{Bai:2019ogh}, and even X-ray Bremsstrahlung and nuclear fusion \cite{Acevedo:2020avd}. 

Minerals have two distinct advantages in searching for heavy Dark Matter. First, the destructive tracks left by very massive states will usually be straight for kinematic reasons described above. It is usually the case that there is no ``Standard Model" background process which would produce such extremely long tracks, so depending on the mineral readout methods, these searches are essentially background free. Second, because minerals have extraordinarily long exposure times, they greatly surpass the mass-reach of terrestrial Dark Matter detection experiments. The mass-reach, or largest mass to which a detector is sensitive, is determined by the flux of cosmogenic particles through the detector during its observation time. As mentioned above, the flux of Dark Matter through a detector scales inversely with the Dark Matter mass, $\Phi_{\rm DM} \propto 1/m_{\rm DM}$. The relative mass-reach of mineral versus non-mineral terrestrial detection techniques will scale with the exposure time. As a consequence, a meter-scale Dark Matter detector can, in principle, find heavy Dark Matter with mass up to only a microgram in a year [$\Phi_{\rm DM} \sim 1\,{\rm m}^{-2}\,{\rm yr}^{-1} \times \left( 1\,\mu{\rm g}/m_{\rm DM} \right)$]~\cite{Bramante:2018qbc}, while a meter-scale mineral detector could find heavy Dark Matter states up to a kilogram in mass, assuming a gigayear-long exposure time [$\Phi_{\rm DM} \sim 1\,{\rm m}^{-2}\,{\rm Gyr}^{-1} \times \left( 1\,{\rm kg}/m_{\rm DM} \right)$]. 

Because they can be high mass, and are equally capable of depositing energy into minerals that exceeds and propagates differently than radiogenic backgrounds, monopoles \cite{Weinberg:1983bf}, Q-balls \cite{Coleman:1985ki,Kusenko:1997si}, self-destructive Dark Matter \cite{Grossman:2017qzw}, and other highly energetic cosmogenic particles \cite{Bai:2019ogh,Acevedo:2021kly} can also be sought using more coarse readouts of minerals than are required for identifying single nuclear recoils. Indeed, minerals have already been employed in searches for monopoles in muscovite mica \cite{Price:1986ky}. This search has recently been recast as a search for heavy composite Dark Matter \cite{Acevedo:2021tbl}. For details on highly energetic cosmogenic Dark Matter searches in Quartz, see section~\ref{sec:Studies-MD}. For details on preliminary investigations of composite Dark Matter discovery in olivine and galena, see section~\ref{sec:Studies-Queens}. 

%*********************************************************
\subsection{Astrophysical Neutrinos} \label{sec:Physics-AstroNu}
%*********************************************************
%{\color{blue} Coordinator: Shunsaku Horiuchi, Paola Sala}

The application of mineral detectors for astrophysical neutrinos is rich. Neutrinos already hold unique roles in astrophysics: due to their weak interactions, they are important energy sinks, and observationally they allow one to see into regions which are otherwise hidden from other cosmic probes like photons and cosmic rays. It is in these contexts that mineral detectors bring a further unique property: namely, the ability to probe neutrino fluxes over geological time scales. Supporting this is the secondary, but still very important, ability of mineral detectors to track neutrinos across a wide range of energies and flavor sensitivities. Here, we introduce three major sources of astrophysical neutrinos---solar, supernova, and atmospheric neutrinos, see Fig.~\ref{fig:nu_flux}---and we review their connections with mineral detectors. 

\begin{figure}
    \centering
    \includegraphics[width=0.8\linewidth]{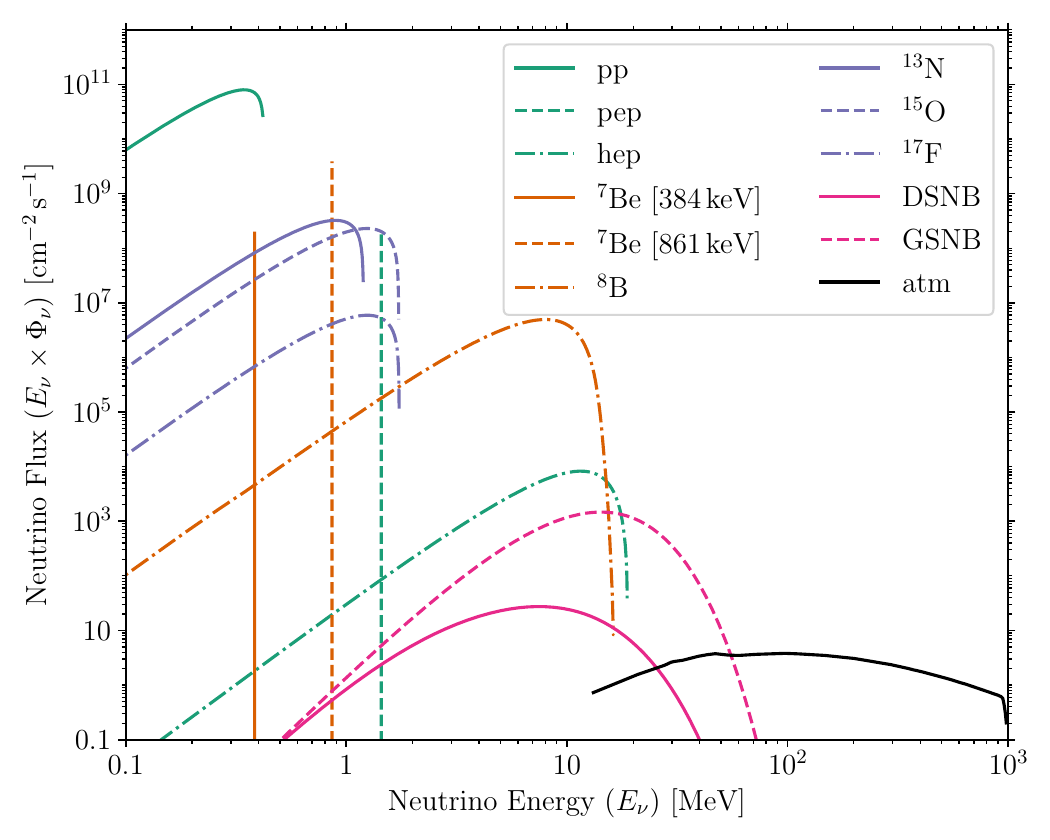}
    \caption{Fluxes of astrophysical neutrinos at Earth as labeled in the legend. The green, orange, and purple lines show different components of the solar neutrino flux (taken from Ref.~\cite{Vahsen:2020pzb}). The solid and dashed magenta lines show the (time-averaged) fluxes of the diffuse supernova neutrino background (DSNB) and from galactic supernova neutrinos (GSNB), respectively, taken from Ref.~\cite{Baum:2019fqm}. The black line shows the atmospheric neutrino flux (atm, taken from Ref.~\cite{Vahsen:2020pzb}).}
    \label{fig:nu_flux}
\end{figure}

In the context of our Sun, solar neutrinos have a history of more than half a century and have been sought after to understand the physics of the solar interior \cite{Haxton:2012wfz,Wurm:2017cmm}. On the theory side, a Standard Solar Model has been developed; and on the experimental side, multiple neutrino experiments have now detected various components of solar neutrinos, most recently the CNO neutrinos by Borexino \cite{BOREXINO:2020aww}. All measurements so far have been of the solar neutrino emission at the present time. However, stars evolve, and our Sun is no exception. Being on the main sequence our Sun evolves slowly, nevertheless solar neutrino components such as the boron-8 flux are strongly dependent on the core temperature and vary over geological time scales of paleo-detectors (millions to billion years). Solar neutrinos span from below 1 MeV to the higher energy component of boron-8 flux exceeding $\sim 10$ MeV; therefore the nuclei recoils are keV energies, and can be picked up by damage tracks in geological minerals~\cite{Tapia-Arellano:2021cml}, see also Fig.~\ref{fig:Spectra}. There are various motivation for further studies of solar neutrinos: despite theoretical and experimental progress, there still remains some outstanding questions. For example, the neutrino measurements are $\sim 2 \sigma$ off theoretical expectations of best-fitting large-mixing-angle MSW solution; and Solar models fit to metallicity measurement of the Sun's atmosphere are in tension to models fit to helio-seismology data.

Neutrinos also hold varied and special roles in the context of core-collapse supernovae \cite{Mirizzi:2015eza}. These occur throughout the Universe as massive stars enter the last stages of stellar fusion and gravitationally collapse into compact objects (neutron stars, black holes). In addition to allowing to probe the interior of a supernova and various aspects of physics beyond the Standard Model~\cite{Horiuchi:2018ofe}, neutrinos are also crucial as they energetically dominate the explosion: some 99\% of the released gravitational binding energy is emitted as neutrinos. All neutrino flavors are emitted, with nearly Fermi-Dirac spectra and typical temperatures of 3--8\,MeV, in a short burst lasting tens of seconds. Since the historic detection of the $\bar{\nu}_e$ flux from the nearby supernova SN1987A~\cite{Bionta:1987qt,Totsuka:1988iyh}, there has yet to be another supernova neutrino burst detected. The occurrence rate of a nearby supernova is not very high: most estimates fall in the range of a few per century in the Milky Way~\cite{Adams:2013ana}. Paleo-detectors are a clear win in this regard. Since they record supernova neutrinos over durations much longer than the inverse of the supernova rate, they have ``observed'' millions of supernova neutrino bursts, and there is no waiting for a supernova. Averaged over timescales long compared to the inverse galactic supernova rate, the flux of neutrinos from supernovae in our Galaxy at the location of Earth is about two orders of magnitude larger than the Diffuse Supernova Neutrino Background (DSNB, the flux of neutrinos arriving at Earth from supernovae exploding in the far-away Universe), see Fig.~\ref{fig:nu_flux}. Furthermore, due to their longer exposures, paleo-detectors could probe the time-evolution of the supernova rate over the history of the Milky Way, which remains poorly constrained otherwise \cite{Baum:2019fqm}. Finally, paleo-detectors are sensitive to all flavors of neutrinos, including the heavy-lepton neutrinos ($\nu_\mu$, $\nu_\tau$ and their antiparticles), which are typically elusive to neutrino experiments with sensitivity to the supernova neutrino burst (tens of MeV energies). Paleo-detectors remain perhaps the sole known method of measuring the average heavy-lepton neutrinos from a large number of core-collapse supernovae \cite{Baum:2022wfc}. 

Atmospheric neutrinos provide indirect but reliable information about cosmic rays (CRs). Indeed, decades of neutrino oscillation studies have allowed to refine and validate  the models to predict neutrino production by CR showers in the atmosphere~\cite{Kajita:2004ga}. CR fluxes and their elemental compositions have been extensively measured at Earth, both in and outside of Earth's atmosphere~\cite{ParticleDataGroup:2022pth}. However, many questions remain open about their origin and their propagation in the interstellar medium~\cite{Blasi:2013rva}.   

Paleo-detectors are excellent candidates to  provide measurements of the time evolution of CR fluxes over geological timescales, through detection of neutrino interactions~\cite{Jordan:2020gxx}. To date, we have measurements of the evolution of the CR flux over the last $\sim 1.5\,$Gyr extracted from data of isotope abundances in meteorites; however, the interpretation of these data is controversial, suggesting increases of the flux in the mentioned period between 50\% and a factor three~\cite{Alexeev2016, Alexeev2017, Ammon2009, Wieler2013, Hedman2019}, either with smooth variations~\cite{Alexeev2017} or step wise variations~\cite{Ammon2009}, and possibly periodic modulations~\cite{Alexeev2016}. Paleo-detectors could offer the unique opportunity of directly measuring the CR flux over timescales as long as a few gigagyears by measuring the damage tracks caused by nuclear recoils from the interaction of atmospheric neutrinos with nuclei in the minerals~\cite{Jordan:2020gxx}. In turn, this independent measurement of CR variations will significantly improve the determination of the age of meteorites or even icy objects~\cite{Hedman2019}.

Atmospheric neutrinos have larger energies than solar or supernova ones: their spectra peak in the GeV range and extend far beyond. This allows for larger energy transfer to the target nuclei. In this neutrino energy range, other processes beyond coherent elastic neutrino nucleus scattering become relevant, in particular, quasi-elastic interactions and Deep Inelastic Scattering (DIS). Such interactions produce not only primary nuclear recoils, but also lighter and more energetic interaction products which can either directly cause damage tracks or give rise to further nuclear recoils via their interactions with nuclei in the crystal lattice. The resulting damage track spectrum extends to track lengths much longer than those caused by the radiogenic neutron background (cf. Fig.~\ref{fig:Spectra}), potentially allowing for an atmospheric neutrino signal isolated from any known background~\cite{Jordan:2020gxx}. The rate of signal tracks from atmospheric neutrino interactions in the background-free track-length region has been estimate to be of the order of 10$^4$\,tracks/100\,g/Gyr~\cite{Jordan:2020gxx}. We note that choosing samples that have been buried deeper than $\sim 5\,$km ensures sufficient shielding from cosmic muons, see also section~\ref{sec:MineralDetectors_DM}.

When inferring the CR flux from the atmospheric neutrino flux, systematic errors could arise from backgrounds and variations in the geomagnetic field that affects the propagation of cosmic rays. The selection of long energetic tracks produced by high energy neutrinos allows to decouple the neutrino flux measurement from possible variations of the Earth geomagnetic field. Neutrinos with energies above a few GeV are produced by CRs with magnetic rigidity large enough to be insensitive to the terrestrial field in its present conditions~\cite{Lipari:2000wu} and thus also to its variations, both in orientation and up to $\sim 100$\% in magnitude.

Finally, cosmogenic neutrinos, not produced in the atmosphere but originating directly from galactic or extra-galactic events, are the only messengers that can reach the Earth without being deflected or absorbed by the microwave background and interstellar medium. Searches for these Ultra High Energy neutrinos are ongoing at experiments such as IceCube~\cite{IceCube:2018cha}. The possibility to detect Ultra High Energy neutrinos in paleo-detectors has not yet been investigated.

%*********************************************************
\subsection{Cosmic Rays} \label{sec:Physics-CosmicRays}
%*********************************************************
%{\color{blue} Coordinator: Lorenzo Caccianiga}

Cosmic Rays are charged particles that reach the Earth from outer space. They can reach energies of more than $10^{20}\,$eV, the most energetic particles ever recorded. If they have enough energy, when interacting with the atmosphere, they create cascades of secondary particles called extensive air showers that can reach the ground and interact with minerals causing nuclear recoils and thus leaving tracks. Although some of the secondary particles, such as muons, can penetrate up to several km into the Earth, the majority of the shower is absorbed in the first few meters. Thus, rock samples best suited for studies of extensive air showers have geological histories different from what is expected to be optimal for detecting neutrinos or Dark Matter (cf. section~\ref{sec:MineralDetectors_DM}).

Cosmic rays are easily detectable at the present time and so the interest for the use of paleo-detectors in this respect is not to measure the current flux of cosmic rays, but rather to detect the evolution of said flux in the past. Being charged particles, cosmic rays are deflected by the magnetic fields that are present within our Galaxy and in the extra-galactic space; below a certain rigidity\footnote{The rigidity is defined as the ratio E/Z with E being the energy of the cosmic rays and Z the charge. As magnetic deflections are directly proportional to the charge and inversely to the energy, particles with the same rigidity will behave in the same way in a given magnetic field.}, they are practically trapped within the Galaxy. Due to the deflection, the arrivals of cosmic rays at Earth are also delayed compared to light or neutral particles emitted from the same source. The delay depends on the rigidity of the particle, on the magnetic field strength and on the distance of the source from the Earth. For cosmic rays originating from Galactic sources with energies above that of Galactic confinement, the delay is typically between few hundreds and hundreds of thousands of years.

The use of paleo-detectors to investigate the past flux of cosmic rays requires a very detailed knowledge of the samples' exposure to the secondary cosmic rays. In particular, the best-suited samples are the ones that are created, exposed for a known, short time and then covered by an overburden of material that effectively shield the majority of secondary cosmic rays up to the present time. An example of suitable samples are the deposits of salts (in particular halite) that were produced by the temporary partial desiccation of the Mediterranean sea during the so-called \textit{Messinian salinity crisis}. During this period, around 6\,Myr ago, the strait of Gibraltar closed, causing the evaporation of a the majority of the Mediterranean sea and simultaneous production of a number of evaporitic rocks. These evaporites were directly exposed to secondary cosmic rays, or shielded by a residual thin overburden of high-salinity water. The re-opening of the strait of Gibraltar, $\sim 500\,$kyr later, caused the sudden flooding of the sea that was re-filled in possibly as short a time as few years (the so-called \textit{Zanclean flood}). This flood covered the rocks produced during the evaporation and dragged some of them to the deepest parts of the sea, most notably the region south-east of present Sicily, where they are now shielded by several km of water. This geological event is notably coincident with the current best estimates of the age of the so-called \textit{Fermi Bubbles}, large gamma-emitting lobes protruding from the Galactic Center, which might be the indication of past activity in our Galaxy. Active galaxies are amongst the best candidates for accelerating the highest-energy cosmic rays, which has been recently confirmed by the observation of high-energy neutrino emission from the nearby Seyfert galaxy M77~\cite{IceCube:2022der}. 

Similar conditions might also be obtained by datable volcanic eruptions that are then covered by subsequent eruptions creating a window for the exposure of the minerals created by the eruption itself or of xenolites brought to the surface in the volcanic event.

The nature and energy of secondary cosmic rays vary significantly, but the most interesting candidates to produce tracks in paleo-detectors are muons, neutrons and hadrons. Muons are the most abundant part of the shower apart from the electromagnetic component ($e^+, e^-,\gamma $) which is too light to produce significant tracks. High-energy muons can induce nuclear recoils up to energies of hundreds of keV, which means that even large (tens to hundreds of $\mu$m long) tracks can be produced. However, the muon spectrum decays steeply with energy and thus only few of these particles exist. Nonetheless, the track length range above $\mathcal{O}(1)\,\mu$m is very interesting for observing cosmic rays tracks, due to the low background present in this region (mostly due to radiogenic neutrons). Cosmogenic neutrons and hadrons are more efficient in producing nuclear recoils than muons, but they are fewer in number and more quickly absorbed by an even thinner overburden and thus necessitate a more careful understanding of the exposure of the sample. Finally, we note that the careful modelling of tracks induced by secondary cosmic rays can be important to better understand backgrounds for other studies, in particular, if samples suitable for other signals cannot be found at sufficient depths for the overburden to completely shield all the components of extensive air showers.

%*********************************************************
\subsection{Reactor Neutrinos} \label{sec:Physics-ReactorNu}
%*********************************************************
%{\color{blue} Coordinator: Patrick Huber}

In 2017 coherent elastic neutrino nucleus scattering (CEvNS) was observed for the first time~\cite{Akimov:2017ade} using 50\,MeV neutrinos from pion decay at rest. The interest in CEvNS from reactor neutrinos arises for several reasons: There is a significant enhancement of the cross section relative to inverse beta decay due to coherence, proportional to $N^2$, where $N$ is the number of neutrons in the nucleus. The resulting larger cross section allows, in principle, for much smaller neutrino detectors which is interesting in application contexts. Such CEvNS detectors were first explored for the detection of nuclear submarines~\cite{JASON} during the Cold War\footnote{It turns out, this is not a viable use of neutrinos.} and more recently in the context of nuclear non-proliferation safeguards~\cite{Cogswell:2016aog,Bowen:2020unj,vonRaesfeld:2021gxl,Cogswell:2021qlq,Cogswell:2021lla}. Furthermore, CEvNS events from reactor neutrinos are perfect analogs of Dark Matter events resulting in a single nuclear recoil at sub-keV energy. Thus, there is a vigorous level of effort to observe CEvNS of reactor neutrinos~\cite{Abdullah:2022zue}. 

The specific interest in using crystal defects to detect reactor CEvNS arises from the relatively low nuclear recoil energy threshold mineral detectors have and the fact that this could lead to passive, robust, room-temperature detectors. Other attempts at observing reactor CEvNS usually rely on cold detectors either at liquid nitrogen temperature or for most detectors in the  milli-Kelvin region -- the resulting systems are unlikely to meet the needs for a field-deployable system for nuclear non-proliferation safeguards~\cite{Cogswell:2021qlq}. 

Any application of mineral detectors at reactors would use manufactured crystals. The signature of reactor CEvNS in a crystal are nuclear recoils below 1\,keV and thus the resulting damage sites are small (few nm) and not very track-like. To avoid confusion with pre-existing crystal defects, crystals need to be scanned (non-destructively) prior to exposure to neutrinos and then again after. This enables event-by-event subtraction of this background. Cosmic ray neutron backgrounds have to be shielded to the extent possible (e.g., by placing the detector in the underground ``tendon galleries'' of certain commercial reactor designs). The bulk of remaining events leads to higher energy recoils and thus larger damage events and thus can be vetoed. The fraction of events of the same size as genuine CEvNS events does not change much with atomic mass $A$, whereas the CEvNS signal scales roughly as $A^2$. Therefore, using detectors made of different-$A$ materials can help to establish the CEvNS signal and discriminate neutron-induced backgrounds.

%*********************************************************
\subsection{Geoneutrinos} \label{sec:Physics-GeoNu}
%*********************************************************
%{\color{blue} Coordinator: Ingrida Semenec}

Because they are produced in the decay of heavy nuclei, neutrinos can provide some unique geological information. Specifically, geoneutrinos can provide insights into the radioactive composition of the planet, and probe the depths of the Earth that are otherwise out of the reach of current geological methods.
Geoneutrinos are electron antineutrinos coming from natural radioactive decays inside the Earth (such as uranium and thorium). Measurements of the geoneutrino flux coming from the Earth’s crust and the mantle can provide a constraint on the amount of heat-producing elements, e.g., uranium, thorium, and potassium. Such information helps determine the absolute concentration of the refractory elements (a geological class of elements resistant to heat), as well as the radiogenic heat contribution to the total surface heat flux of the Earth. So far, geoneutrinos have been observed using large liquid scintillator detectors via inverse beta decay (IBD) interactions, which are sensitive to electron antineutrinos above a threshold of $E_{\rm IBD} \approx 1.8$ MeV. Geoneutrinos  coming from \ce{U} and \ce {Th} have been measured by Borexino~\cite{Borexino:2019gps} and KamLAND~\cite{KamLAND:2022vbm}. Neutrinos from \ce{^{40}K} have yet to be observed due to their energy being below the $E_{\rm IBD}$ threshold. It might be possible to search for geoneutrinos in IBD events in minerals, if IBD events could be distinguished from other $\alpha$-, $\beta$-, and neutron-producing events in minerals. In liquid scintillator experiments, IBD events are identified in the process $\bar \nu_e + p \rightarrow e^+ + n$, where the positron is identified by annihilating with an electron, and the neutron is identified by being captured on a nucleus and creating a subsequent gamma in the nucleus's decay. It is unclear whether an IBD event could be detected through the damage it would leave in minerals.

\begin{figure}
    \centering
    \includegraphics[width=0.6\linewidth]{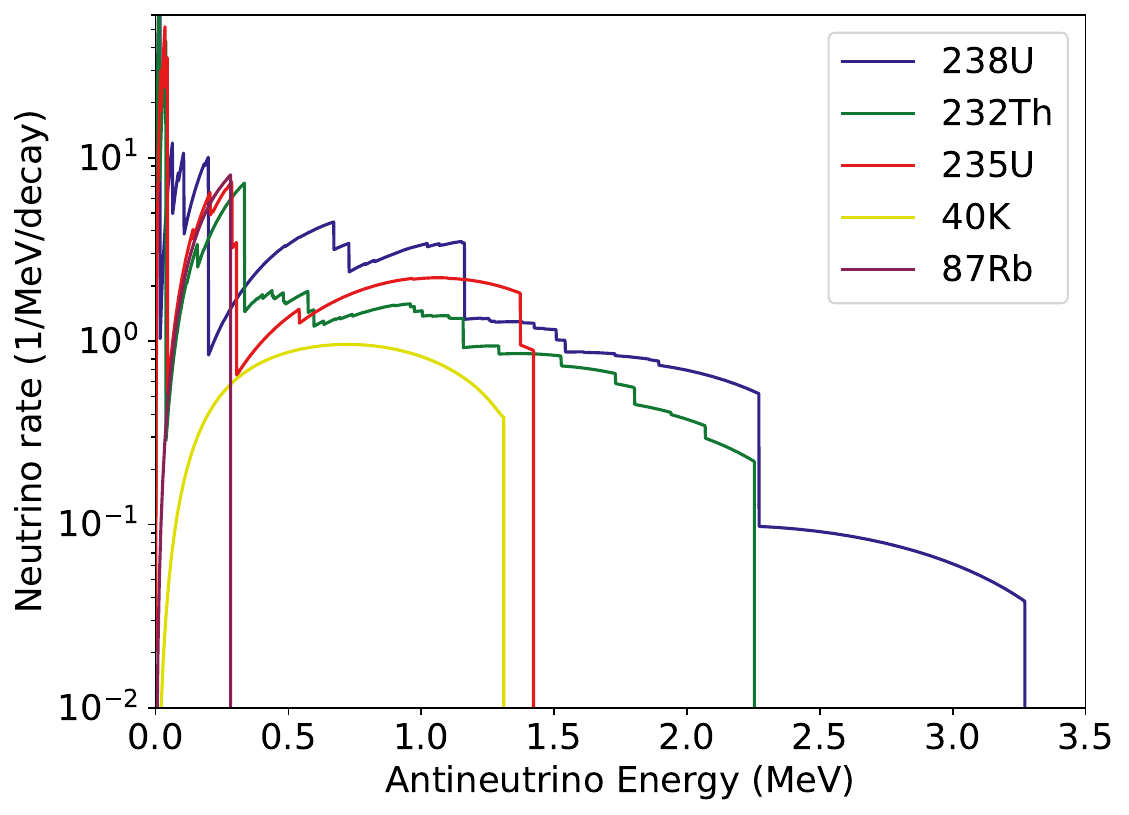}
    \caption{Geoneutrino energy spectra from \ce{U}, \ce{Th}, \ce{K}. Each spectrum is normalized to the decay of a parent isotope set by relative isotope abundances. Data taken from Ref.~\cite{enomoto}.}
    \label{fig:geosolar}
\end{figure}

On the other hand, geoneutrinos will also elastically scatter off nuclei, creating nuclear recoils. However, in this case, both geological antineutrinos and solar neutrinos would recoil against nuclei. As a consequence, the main obstacle to observing geoneutrinos using nuclear recoils in minerals is the formidably large (if not insurmountable) background provided by the flux of neutrinos from the Sun. Comparing the geoneutrino spectrum (see Fig.~\ref{fig:geosolar}) to that of solar neutrinos in (see Fig.~\ref{fig:nu_flux}), it is apparent that one promising range of neutrino energies to search at lies above $800 \,\mathrm{keV}$, which avoids the substantial background provided by \ce{^{7}Be} solar neutrinos, but still includes the contribution from \ce{^{40}K} geoneutrinos. It is important to note that the geoneutrino flux normalization shown in Fig.~\ref{fig:geosolar} will depend on the local abundance of \ce{U}, \ce{Th}, \ce{K}. Nevertheless, it is clear why using $E_\nu > 800\,$keV neutrinos to search for geoneutrinos may be a viable strategy: for \ce{U} and \ce{Th}, the geoneutrino energy spectra overlap with \ce{^{15}O}, \ce{^{17}F}, \ce{^{13}N} and \ce{^{8}B} solar neutrinos, but these fluxes are subdominant when compared to $pp$ and \ce{^{7}Be} solar neutrinos. On the other hand, the capability to cut neutrino tracks at higher energies might be useful, since $pep$ neutrinos provide another large background at $E_\nu \approx 1.5\,$MeV. If good energy resolution would be achieved, one could try to make cuts to avoid larger solar backgrounds, however the approximate signal rate from solar neutrinos would still be three orders of magnitude larger than the geoneutrino signal, if both were observed in nuclear recoils \cite{Wang:2017etb}. One could estimate that with this signal to background ratio the minimum number of observed geoneutrino events would have to be $10^4$ to achieve $5\sigma$ significance. Taking the average neutrino-nucleon interaction cross section for $1\,\mathrm{MeV}$ neutrinos to elastically scatter with carbon nuclei, $\mathcal{O}(10^{-40})\,\mathrm{cm^2}$~\cite{Abdullah:2022zue}, it would be necessary to readout 1\,kg of 1\,Gyr old mineral samples (here assumed pure carbon for simplicity), with 100\% detection efficiency and good energy resolution to observe geoneutrino nuclear recoil events with 5$\sigma$ significance. It should be noted that the nuclear recoil energy imparted by both geoneutrinos and solar neutrinos in the above analysis would be at maximum $\sim 100\,$eV, presenting a substantial challenge for mineral track detection.

However, it may be the case that the basic science involved in an astroparticle mineral detection program would greatly aid geoneutrino science, not by measuring a geoneutrino flux, but instead by providing a better understanding of the local geology at mineral excavation sites. Understanding local geology is extremely important in making any geological interpretation using a measured geoneutrino flux. Geologists are primarly interested in the flux of geoneutrinos coming from the mantle. Because geoneutrinos are produced throughout the mantle and crust, the flux from the crust is enhanced in near-surface experiments, since all geoneutrino fluxes will scale like $1/r^2$. In practice, the local crust/lithosphere's abundance of \ce{U} and \ce{Th} contributes around 50\% of the measured signal \cite{SAMMON2022117684}. Hence, even if the geoneutrino signal would be extremely difficult to extract from nuclear recoils or IBD events measured in minerals, there would be substantial benefits to having an astroparticle mineral detection program which took samples and gained a refined understanding of the local geology around current large scale liquid scintillator detectors like SNO+, since an improved understanding of the local geology at any detection site will lead to improved models of the local lithosphere. Today, modeling of the local crust/lithosphere contribution to the geoneutrino fluxes provides one of the largest uncertainties when attempting to determine what portion of the observed geoneutrino flux comes from the mantle~\cite{Borexino:2019gps,KamLAND:2022vbm}.

%*********************************************************
\subsection{Geoscience applications} \label{sec:Physics-Geoscience}
%*********************************************************
%{\color{blue} Coordinator: Ulrich Glasmacher}

Fission-track and $\alpha$-recoil track dating is based on the visualization of latent tracks after etching with optical microscopes. Counting the etch pits reveals the areal density of etch-pits at the surface of the polished mineral face. This visualization technique has its pit fall in tracks occurring in the mineral volume that do not intersect the artificially created polished mineral surface and, therefore, are not used in the dating technique. In the case of fission-track dating, several mineral grains ($\lesssim 20$) of one sample that are embedded in an epoxy mount are used for dating to overcome this problem. Furthermore, only those mineral grains are used for dating with $c$-axes parallel to the surface. Minerals with other orientations in the epoxy mounts are abundant. If the uranium content is very low, even more mineral grains ($> 40$) are used to reveal a fission-track age. Another point to consider is the mineral abundance in rocks. Some rocks do not have many mineral grains that can be used for fission-track dating, therefore, the age determined is given with a large error ($> 10\,\%$).
In addition, the power of fission-track dating is not only to reveal an age. The distribution of the confined fission-track length stores the thermal history that minerals and, therefore, the mineral-bearing rocks, have experienced in the past. Recently, etching the latent tracks reveals also confined etched fission-tracks close to the artificial surface. Confined fission-tracks are those fission-tracks which are completely in the mineral and are oriented parallel to the artificial surface. The length of etched confined fission-tracks is measured, and using the distribution of confined fission-track lengths one can numerically model the thermal evolution of the rock sample using software codes such as \texttt{HeFTy}~\cite{Ketcham:2017} or \texttt{QTQt}~\cite{Gallagher:2012}.

To overcome the obstacles described in the previous paragraph, one would have to find a visualization technique that allows to determine the number and length of latent fission-tracks in the mineral grain {\it volume} without preparation of an internal surface and etching. Such a visualization technique must quantify the volume density of fission-tracks and their length distribution in the mineral grains. In the case of minerals with higher uranium content ($> 10\,\mu$g/g) and older age, even one mineral grain could provide a significant age and length distribution using such a visualization technique. Furthermore, determination of confined length of latent fission tracks in the mineral grain volume would allow to increase the number of fission-tracks informing the length distribution that is used to numerically model the thermal history. If a mineral grain has high uranium content, it might even be possible to gain so many length data of latent fission tracks that a thermal history of a single grain might be numerically modeled. If this is possible for one sample, thermal histories of many grains of one rock sample can be combined and the significance of the thermal history is increased dramatically. An interesting technique to make progress in this direction is to read out color centers produced by fission-fragments or by $\alpha$-recoils with optical microscopes, see also sections~\ref{sec:MineralDetectors-ColorCenters} and~\ref{sec:ReadOut-Optical}. Color centers produced by nuclear recoils have been observed both in apatite and olivine (see, e.g., Refs.~\cite{Bertel:1982, Schwartz:2006, Schwartz:2008, Manzano-Santamaria:2012, Schwartz:2015}); together with their relatively high thermal annealing times, this makes apatite and olivine interesting candidates for such a color-center based technique.

%*********************************************************
\section{Read-out technologies} \label{sec:ReadOut}
%*********************************************************

The applications of mineral detectors discussed in this whitepaper all rely on the ability to detect the latent damage to the regular crystal structure produced by nuclear recoils with nuclear recoil energies $E_R \sim 0.1 \textit{--} 10^3\,$keV. For a number of applications discussed here, in particular, for WIMP-like Dark Matter (section~\ref{sec:Physics-WIMPs}) and low-energy neutrinos (sections~\ref{sec:Physics-AstroNu},~\ref{sec:Physics-ReactorNu}, and~\ref{sec:Physics-GeoNu}) the signal are damage features from isolated nuclear recoils, while for other applications such as atmospheric neutrinos (section~\ref{sec:Physics-AstroNu}) and especially composite Dark Matter and other high-energy astroparticles (sections~\ref{sec:Physics-CompositeDM} and~\ref{sec:Physics-CosmicRays}) the signal is a large number of low-energy nuclear recoils that are all spatially clustered along the trajectory of the heavy/high-energy particle giving rise to the nuclear recoils. Depending on the particular mineral, the latent damage can take many forms, including mechanical stress in the lattice, changes to the electron density, local amorphization of the crystal lattice, and color centers due to vacancy defects. A range of microscopy techniques has been used to read out such damage features, including TEM, SEM, AFM, X-ray microscopy and optical microscopy~\cite{Fleischer:1964,Fleischer383,Fleischer:1965yv,Snowden-Ifft:1995zgn,Snowden-Ifft:1995rip,GUO2012233,BARTZ2013273,RODRIGUEZ2014150,Kouwenberg:2018}. In order to unleash the full potential of mineral detectors for the wide range of applications discussed here, the throughput of microscopy techniques has to be scaled up to allow for the efficient readout of large samples. In this section, we discuss a number of promising technologies: We begin by discussing optical fluorescence microscopy in section~\ref{sec:ReadOut-Optical}. In section~\ref{sec:ReadOut-SoftX}, we turn to soft X-ray microscopy techniques possible in either a laboratory setting or at a synchrotron, and in section~\ref{sec:ReadOut-HardX} we discuss hard X-ray microscopy using accelerator light sources. In section~\ref{sec:ReadOut-NM} we discuss techniques that allow for the imaging of samples with nm-scale resolution, including Scanning Probe Microscopy (SPM) techniques such as Atomic Force Microscopy (AFM), Scanning and Transmission Electron Microscopy (SEM/TEM), and He-ion Beam Microscopy (HIM). We stress that these techniques are highly complementary; in order to achieve the goal of neutrino and Dark Matter detection with mineral detectors one will presumably need to use a combination of microscopy techniques. For example, one can imagine scanning a relatively large sample volume with optical fluorescence microscopy to identify candidate sites for nuclear recoil tracks from the observations of color centers which, in many materials, will form along the recoil track. One could then image these candidate sites with a higher-resolution technique, e.g., hard X-ray microscopy which allows for $\mathcal{O}(10)\,$nm spatial resolution or a sub-nm resolution technique such as TEM or HIM. 

Before entering the discussion of the different readout techniques, let us stress that the data analysis for any of the use-cases of mineral detectors discussed in this whitepaper will be a formidable challenge -- imaging $\mathcal{O}(1)\,$kg of material, corresponding to a volume with $\mathcal{O}(10)\,$cm linear dimensions, with a resolution of $1\,$nm would na{\"i}vely correspond to $\geq 10^{20}\,$pixels or more than a zettabyte of data. However, the damage features in the crystal are extremely sparse - almost all of the crystal lattice will be free of defects for any of the applications we are interested in here. Hence, the challenge is to identify the regions of interest for high-fidelity imaging in a large volume and then image and characterize features in these regions. A combination of different microscopy techniques paired with automated data analysis strategies will be needed to tackle this challenge. The data analysis strategy will most likely make heavy use of machine learning algorithms which are well-suited for the task at hand: recognizing patterns in image data. Developing such tools will be an important direction of work once the readout techniques of choice are firmly established. Recently, a machine learning system based on the Faster R-CNN analysis system~\cite{Ren:2015} was demonstrated to have a performance comparable to human analysis with relatively small training data sets~\cite{Shen:2021}. It proved the possibility to apply deep learning to assist the development of automated microscopy data analysis even when multiple features are present.

%*********************************************************
\subsection{Optical Fluorescence microscopy} \label{sec:ReadOut-Optical}
%{\color{blue} Coordinator: Patrick Huber, Gabriela Araujo}

Fluorescence microscopy can be used to probe nm-sized dislocations if they create fluorescent sources. In certain materials, these dislocations result in anionic vacancies which may be occupied by unpaired electrons. The resulting quantum system -- so-called color centers (CCs) -- can be excited by light in the UV to visible region and re-emit light with longer wavelengths, see also section~\ref{sec:MineralDetectors-ColorCenters}. The excitation wavelength can be filtered out when imaging, so that only the regions containing CCs yield a clear signal in fluorescence microscopy. The signal brightness can be then correlated to the density of CCs. This technique thus allows the observation of signals from nm-sized defect regions with optical microscopes and possibly also the use of pixel brightness to distinguish single-site vacancies and full-tracks.

The main advantage of optical imaging is its speed and cost-per-volume-imaged compared to  other microscopy methods, such as TEM and AFM. Among the fluorescence microscope techniques, the resolution and scan speed also largely varies: the highest scan speeds are achieved by widefield fluorescence and the selective plane-illumination microscopy (SPIM), which has been suggested by \cite{Cogswell:2021qlq} as an optimal scan method for CC passive detectors. In SPIM, a thin light sheet illuminates a $z$-section of the sample $x$-$y$ plane which is observed by a camera orthogonal to the light-sheet source. 3D images are then obtained by moving the sample in the $z$-direction.

While the observation of single CCs with widefield and confocal fluorescence microscopes have already been reported \cite{doi:10.1080/23746149.2020.1858721, D0NR05931E}, the imaging of CCs with SPIM offers a few further advantages: i) The average laser power (and consequent photodamage/bleaching) delivered to a given volume is much lower as only sections of the sample are illuminated; ii) Given the selective illumination, no background from other illuminated planes is observed; iii) Given the separation of light source and imaging objective, SPIM techniques more easily allow for large working distances (and field of view).

Although large fields of view are possible for SPIM methods, commercial setups are costly, often tailored for specific biomedical samples, and lack customization options for new applications either in hardware or software, due to their closed source. This gave rise to a recent development of SPIM: the mesoscale SPIM  microscope, which is a versatile open-source microscope that offers near-isotropic resolution imaging of cm-sized samples within minutes \cite{mesoSPIM}. Its resolution and scan speed can be tuned by the selection of detection lenses with magnifications from 0.9$\times$ to 20$\times$ (in the newest version called Benchtop mesoSPIM, manuscript in preparation), and by the software settings. The published version of mesoSPIM offered a scan rate of approximately a cubic centimeter per 5 minutes at 6 micrometer isotropic resolution \cite{mesoSPIM}. The new, currently unpublished version (Benchtop mesoSPIM)  achieves \SI{1.5}{\um} resolution in $x$ and $y$ with a scan speed up to $\sim\SI{10}{\cubic\cm}$ per hour. The data file sizes range from $\sim$ 20 GB to 10\,TB\,/\si{\cubic\cm} depending on the magnification. 

The non-destructive, fast and isotropic\footnote{While the resolution in the $z$-direction is limited by the thickness of the light-sheet of a few microns, this value can be overcome by combining scans of the same sample rotated by 90 degrees, obtaining an isotropic resolution.} scan of large 3D-volumes offered by mesoSPIM thus makes it the best method for imaging of large amounts of transparent mineral detectors which present CCs in response to nuclear recoils. 

The use of mesoSPIM in fission track dating of transparent rocks could be for instance very useful as a non-destructive 3D imaging method of large amounts of minerals without the need of etching, see section~\ref{sec:Physics-Geoscience}. In the context of neutrino and Dark Matter searches with laboratory-manufactured (passive) mineral detectors, it allows to achieve a large exposure (target mass) without the need of a long exposure (such as in paleo rocks). The latter application is described in more detail in section~\ref{sec:Studies-VT} (the PALEOCCENE concept). While in the first application the tracks are $\mathcal{O}(20)\,\mu$m long and can be resolved within the diffraction limit of optical microscopes ($\sim\lambda/2$), the tracks from Dark Matter or neutrino interactions would be $\lesssim\lambda/2$. Yet, the energy reconstruction of these events may be possible through the larger intensity of fluorescence (pixel brightness) measured from a region containing a full track in comparison to a single-site CC. Another possibility is to use the mesoSPIM as the scan method for large volumes and identification of CC regions, and then implement high-resolution techniques (such as super-resolution stimulated emission depletion (STED) technique) for the identified regions. 

%*********************************************************
\subsection{Soft X-ray microscopy} \label{sec:ReadOut-SoftX}
%*********************************************************
%{\color{blue} Coordinator: Kai Sun}

As has been mentioned previously, the data analysis for any of the use-cases of mineral detectors discussed in this whitepaper will be a formidable challenge -- imaging $\sim$1\,kg of material, corresponding to a volume with $\sim$10\,cm linear dimensions, with a resolution of $1\,$nm would correspond to $\geq 10^{20}\,$pixels or more than a zettabyte of data. The key issue here is that there is not a single instrument that can fulfill the whole duty alone. Techniques that can characterize the microstructures of large scale [$\mathcal{O}(1)$\,kg/volume with $\mathcal{O}(10)\,$cm linear dimensions] mineral detectors should be a first choice for obtaining overall structure features of the minerals, guiding further analysis using other techniques that can collect data down to the atomic scale. 

In section~\ref{sec:ReadOut-Optical}, visible-light-based techniques were presented which can provide 3D structure data down to the micron scale for a large volume. However, the techniques can only be applied for analysis of minerals that are transparent to light. X-rays can penetrate much larger volumes of minerals if higher-energy X-rays are used. In addition, X-rays do not introduce damage of mineral microstructures, especially for soft X-rays, and X-ray-based techniques can be used to study samples of of any shape. Thus, X-ray-based techniques can be a good option for the analysis of mineral microstructures. In this section, we will address laboratory-based X-ray micro Computational Tomography (Laboratory X-Ray $\mu$-CT; section~\ref{sec:ReadOut-SoftX-Lab}) and two synchrotron soft X-ray based techniques (section~\ref{sec:ReadOut-SoftX-Syn}), Small Angle X-ray Scattering (SAXS) and Scanning Transmission X-ray Microscopy (STXM). Techniques based on synchrotron hard X-ray will be discussed in section~\ref{sec:ReadOut-HardX}.   

\subsubsection{Laboratory X-ray $\mu$-CT} \label{sec:ReadOut-SoftX-Lab}

\begin{figure}
    \centering
    \includegraphics[width=0.89\linewidth]{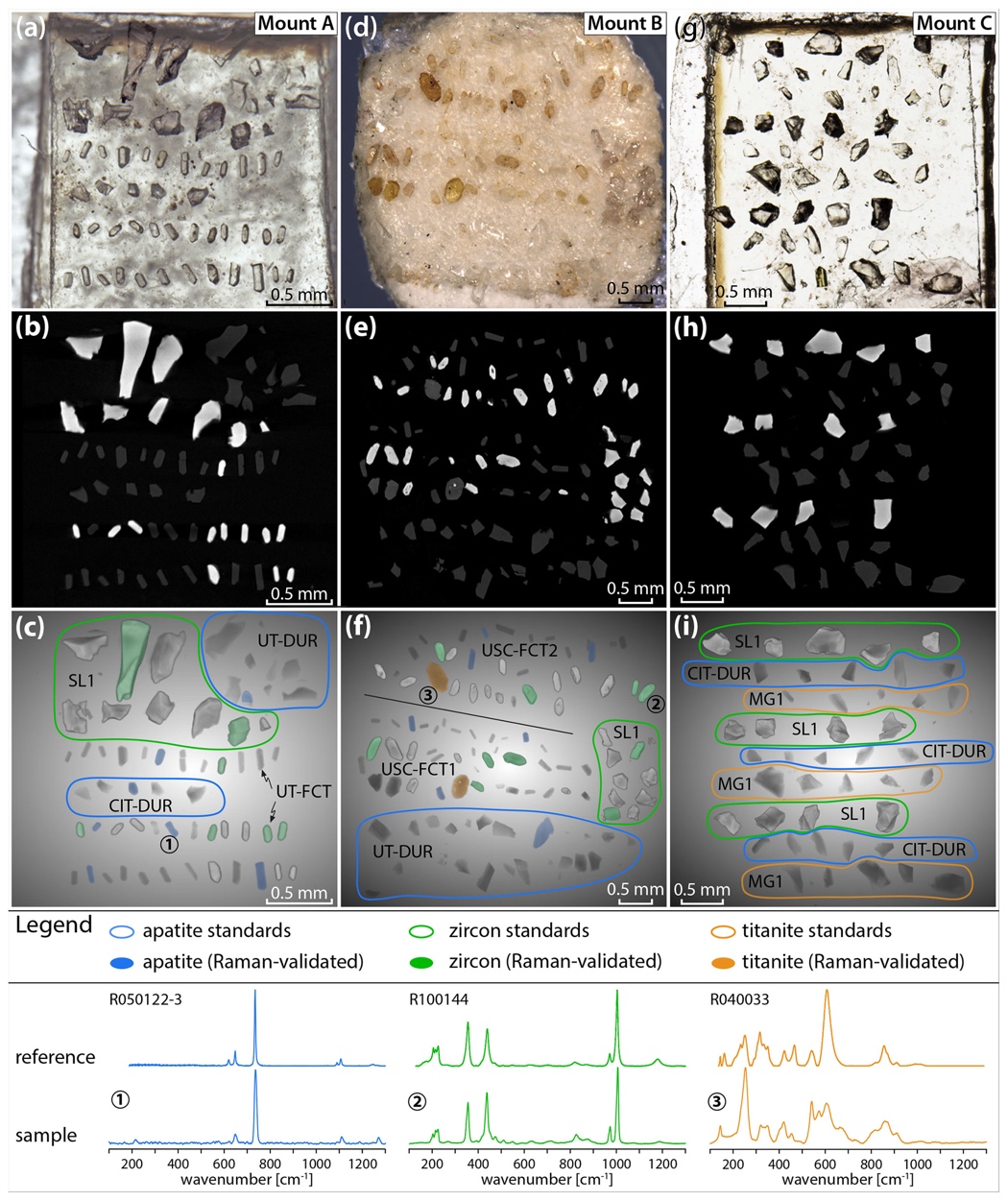}
    \caption{Transmitted light micrographs (a, d, g), $\mu$-CT slices (b, e, h), and $\mu$-CT volume renderings (c, f, i) of Mount~A, Mount~B, and Mount~C. The $\mu$-CT slices show a large contrast between zircon grains (brighter) and apatite and/or titanite grains (darker). Grayscale value and grain relief in 3D renderings are distinct for different mineral phases. The 3D renderings show Raman-validated grains highlighted and known standard shards circled in blue (apatite), green (zircon), and orange (titanite). Baseline-corrected Raman spectra of representative grains and reference spectra from the RRUFF database (including record numbers) are shown below the images. Numbers in circles indicate the grains in the volume renderings which correspond to the sample Raman spectra. Figure taken from Ref.~\cite{Cooperdock:2022}.}
    \label{fig:sX1}
\end{figure}

Laboratory-based X-ray $\mu$-CT has become commercially available in the recent years. The technique can offer similar capability to synchrotron X-ray nanotomography without requiring one to apply for very limited beamtime at a synchrotron. For example, the ZEISS X-Ray Microscopes (XRMs) family, ZEISS Versa and ZEISS Xradia, can deliver sub-micron ($\sim$0.7\,$\mu$m) and nano-scale ($\sim$50\,nm) resolution 3D data with synchrotron-like quality, respectively. These types of XRMs can be used for analyzing samples with dimension from as large as several tens of centimeter down to micrometer scales depending on the X-ray energy used. Here are some features of such XRMs:
\begin{itemize}
    \item non-destructive imaging and reconstruction of buried features (pores, particles, defects, etc.),
    \item phase contrast imaging for studying low-$Z$ or ``near-$Z$'' elements,
    \item diffraction contrast for the study of crystallographic orientation,
    \item multiscale imaging, characterization of large samples ($\sim$30\,cm linear dimension and $\sim$25\,kg weight) at high throughput.
\end{itemize}
When studying natural minerals, a first necessary step will be to find out if the sample is a pure or a mixed mineral. X-ray $\mu$-CT has been demonstrated to be able to identify different mineral phases in a mineral as shown in Fig.~\ref{fig:sX1}. Different minerals have different density and compositions. Thus, the X-ray attenuation coefficients of different minerals are different, and different minerals will have different contrasts in X-ray images~\cite{Cooperdock:2022}. 

\begin{figure}
    \centering
    \includegraphics[width=1\linewidth]{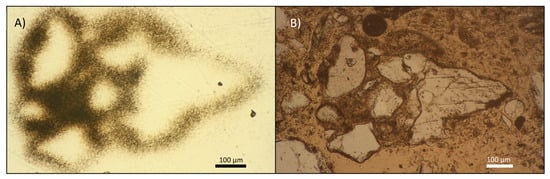}
    \caption{Fission-track and microscope images of sample GJAST-20-10 with high fission-track cementing material. (A) Fission-track image at 10$\times$ magnification. (B) Thin section in plain light at 8$\times$ magnification. Figure taken from Ref.~\cite{Johnson:2021}.}
    \label{fig:sX2}
\end{figure}

Fission tracks have been imaged by X-ray $\mu$-CT~\cite{Johnson:2021}, see Fig.~\ref{fig:sX2}. In this study, fission-track radiography was performed by placing a piece of mica over a standard thin section, irradiating the thin section and mica in a nuclear research reactor, and then examining, under a microscope (ZEISS model 472190-000/11). They found that the fission tracks were created in the mica from the radioactive decay produced by the irradiated uranium within the thin section. Areas with a higher density of fission tracks correspond to areas with more fission events and a higher concentration of uranium that was confirmed by SEM-EDS analysis. 

\subsubsection{Synchrotron Soft X-ray SAXS and STXM} \label{sec:ReadOut-SoftX-Syn}

\begin{figure}
    \centering
    \includegraphics[width=0.8\linewidth]{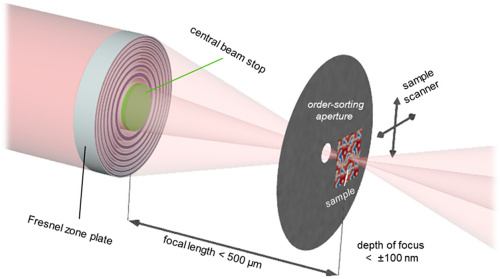}
    \caption{Schematic illustration of the setup for high-resolution Scanning X-ray Transmission Microscopy (STXM). An opaque central beam stop and an aperture are used to select the first diffraction order of the zone plate. To achieve the demonstrated spatial resolution of 7\,nm, a positioning accuracy of 3\,nm and coherent illumination are required in addition to the Fresnel zone plate design.}
    \label{fig:sX3}
\end{figure}

Considerable efforts have been put in soft X-ray microscopy techniques in recent years. Recently, the availability of intense soft X-ray beams with tunable energy and polarization has pushed the development of highly sensitive, element-specific, and noninvasive microscopy techniques to investigate condensed matter with spatial resolutions in the deep single-digit nanometer regime. R{\"o}sner {\it et al.}~\cite{Rosner:2020} reported that they had achieved an image resolution of 7\,nm in Scanning Transmission X-ray Microscopy (STXM) mode at a photon energy of 700\,eV. In their study, they combined newly developed soft X-ray Fresnel zone plate lenses with advanced precision in scanning control and careful optical design as shown in Fig.~\ref{fig:sX3}. This resolution could be used for ultimately imaging the tracks generated by nuclear recoils induced by the interaction of Dark Matter or astrophysical neutrinos.

\begin{figure}
    \centering
    \includegraphics[width=0.5\linewidth]{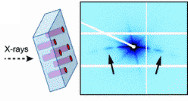}
    \caption{Schematic of the scattering geometry (left) and typical detector image (right). The arrows mark the streaky scattering patterns resulting from the ion tracks of large aspect ratio.}
    \label{fig:sX4}
\end{figure}

\begin{figure}
    \centering
    \includegraphics[width=0.8\linewidth]{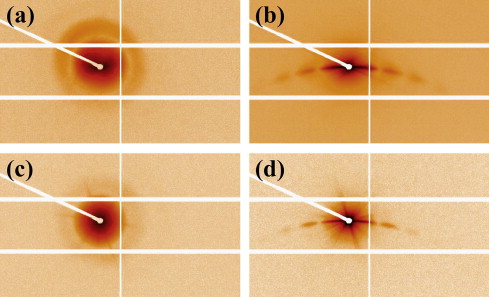}
    \caption{Scattering image of samples irradiated with 2.2\,GeV Au ions: (a) olivine with the X-ray beam parallel to the ion tracks, (b) with the tracks tilted by $\sim 10\,^\circ$, (c) apatite with the X-ray beam parallel to the ion tracks, (d) with the tracks tilted by $\sim 10\,^\circ$.}
    \label{fig:sX5}
\end{figure}

Afra {\it et al.}~\cite{Afra:2011,Afra:2012,Afra:2014} imaged and measured tracks generated by heavy ion beams in several minerals including apatite and olivine using transmission SAXS measurements performed at the SAXS/WAXS beam line of the Australian Synchrotron with X-ray energies of 12 and 20\,keV and camera lengths of approximately 1600 and 2000\,mm. In their studies, samples were mounted on a three-axis goniometer and tilted such that the axis of the tracks had different angles to the incoming X-ray beam. The spectra were collected with a Pilatus 1 M detector with exposure times of 5 and 10\,s for the X-ray energy of 12\,keV and 20 and 30\,s for 20\,keV to compensate the lower flux delivered at higher energies (Fig.~\ref{fig:sX4}). Three-dimensional structures of the ion tracks were obtained as shown in Fig.~\ref{fig:sX5}. One disadvantage of this technique is that scattering from un-irradiated samples was measured as a reference for background removal. 

%*********************************************************
\subsection{Hard X-ray microscopy} \label{sec:ReadOut-HardX}
%*********************************************************
%{\color{blue} Coordinator: Arianna Gleason}

There are many important imaging tools which can be leveraged to document, explore and understand tracks from nuclear recoils recorded in minerals. A key aspect is the spatial resolution of the technique. To date, many pioneering optical microscopy methods can achieve few $\mu$m resolution in 2D. Electron microscopy methods, e.g., SEM and TEM, can enable spatial resolution of hundreds of nm in 2D as well. However, enabling 3D image reconstructions of a mineral sample to search for tracks induced by Dark Matter or neutrinos with these methods is challenging. X-ray based imaging methods may enable rapid 2D and 3D volume reconstructions. Various 3rd and 4th generation lightsources around the world have the required X-ray brilliance and coherence needed to perform this task. These synchrotron and X-ray free electron laser (XFEL) lightsources permit the use of coherent diffractive X-ray imaging (CDXI) techniques.

Hard X-ray radiography and phase contrast imaging (PCI) have been widely used in the past as established techniques at modern synchrotron radiation sources to image optically opaque samples with high spatial resolution (see, e.g., Ref.~\cite{Davis1995}). PCI requires a coherent light source. Here, the sensitivity to the phase shift is introduced by an object which enhances the visibility of structures otherwise invisible in X-ray radiography based on absorption (see, e.g., Ref.~\cite{Mokso2007}). When combined with the femtosecond duration of an XFEL pulse, such a technique allows imaging of matter changing rapidly in both space and time. Advancements have been made to improve spatial resolution (~$\sim 100\,$nm) using new X-ray optics such as beryllium compound refractive X-ray lenses (Be CRLs)~\cite{BeCRLs}, which have been optimized for the XFEL environment and can withstand the full XFEL beam. The spatial resolution of the PCI method is mainly limited by the pixel size of the detector and the bandwidth of the incoming X-rays. The slight polychromaticity reduces the contrast for high spatial frequencies. For a self-amplified spontaneous emission (SASE) bandwidth of $\Delta \lambda / \lambda = 2 \times 10^{-3}$, we expect a reduction in phase-contrast transfer by 50\,\% or more for a spatial frequency above $\mu_{\rm max} = 5440\,{\rm mm}^{-1}$, corresponding to a length scale of $d_{\rm min} \approx 200\,$nm. However, with the expected beam seeding technologies soon to be available and implemented with LCLS~II, any polychromaticity will be removed such that sub-100\,nm resolutions will be routinely available.

Another form of CXDI with a setup similar to that of the aforementioned PCI, is X-ray ptychographic microscopy, which combines the advantages of raster scanning X-ray microscopy with the techniques of coherent diffraction imaging, offering wavelength-limited spatial resolution. Unlike PCI, ptychographic microscopy does not have a resolution limited by the detector pixel size. It provides reliable, real-time images of mesoscale objects with nanoscale spatial resolution. To do this, samples are scanned in a raster pattern (see, e.g., Ref.~\cite{Edo2013}) and images require successively illuminating overlapping regions of the sample with the X-ray probe and recording the diffraction patterns; see Fig.~\ref{fig:ptycho1}. Iterative algorithms can reconstruct the image (see, e.g., Refs.~\cite{GSFienup2008,MaidenRodenburg2009}). Note however that X-ray ptychographic microscopy requires raster scanning the sample with overlapping illuminated regions, which takes time and may require high precision motors. 

\begin{figure}
    \includegraphics[height=4.8cm]{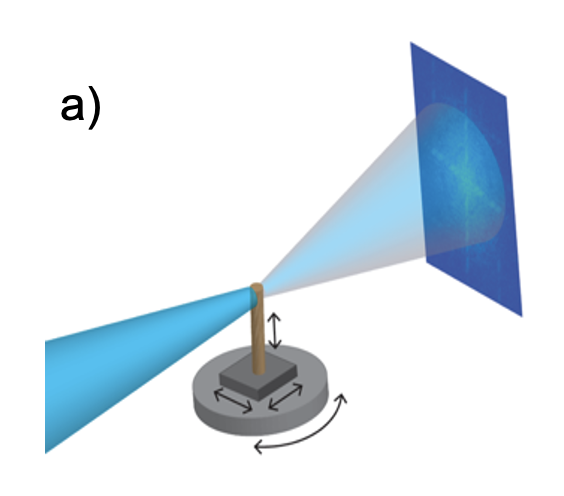}
    \includegraphics[height=4.8cm]{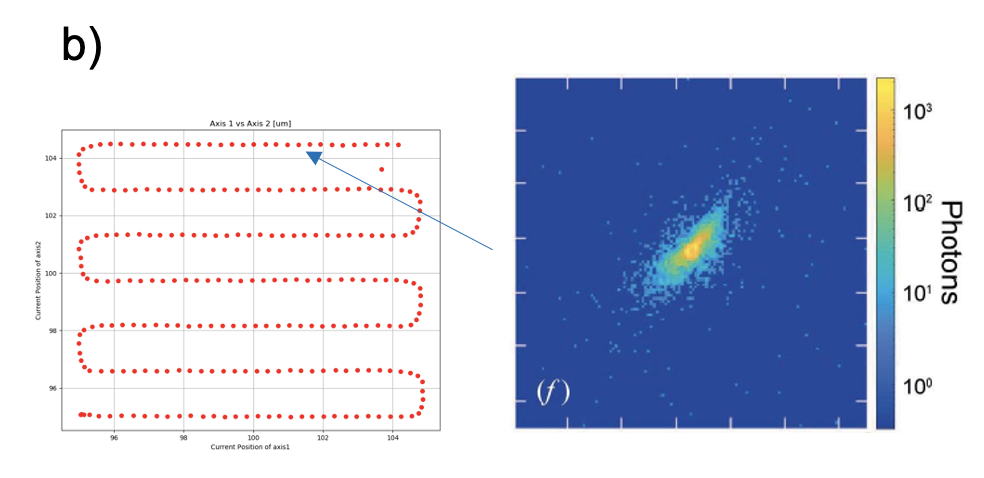}
    \caption{{\it Left}: a) Schematic of typical X-ray ptychography experiment where the X-ray beam (blue) probes a sample. The sample can be rotated and data collection repeated to establish multiple projections for building up the required tomography views to create a 3D reconstruction. {\it Right}: b) Example raster scan positions which allow collection of raw data diffraction patterns at a series of spatially overlapping points. }
    \label{fig:ptycho1}
\end{figure}

X-ray holography is another type of CXDI, however unlike PCI or ptychography, the phase information is encoded in the diffraction pattern via interference with a reference. This technique is particularly well-suited for dynamic imaging since the reconstructed image does not drift and maintains high contrast -- the motion of features between multiple images may be better than a few nm (see, e.g., Ref.~\cite{Guehrs2010}). Time-resolved, single-shot imaging is possible up to several frames during a dynamic process down to sub-ps timing. Recent efforts~\cite{Gork2018} have demonstrated so-called in-flight X-ray holography, where careful positioning of a reference is no longer required. Image fidelity is still high, at resolvable length scales down to $\sim 20\,$nm with unambiguous mapping of structures at the nanoscale and depth-of-field around $200\textit{--}400\,$nm. Additionally, the need for large sample-to-detector distances is removed (here only $0.7\,$m required distance) and detectors with larger pixels can be used (here $75\,\mu$m pixels on pnCCD). However, Ref.~\cite{Guehrs2010} used soft X-ray probes, whereas for Dark-Matter- or neutrino-induced tracks in minerals, we would likely require harder X-ray wavelengths of 7\,keV at least. Nonetheless, the methodology and technique developed in Ref.~\cite{Guehrs2010} is promising.

%*********************************************************
\subsection{AFM, SEM/FIB, and TEM/STEM} \label{sec:ReadOut-NM}
%*********************************************************
%{\color{blue} Coordinator: Kai Sun}

\subsubsection{AFM} \label{sec:ReadOut-NM-AFM}

\begin{figure}
    \centering
    \includegraphics[width=0.5\linewidth]{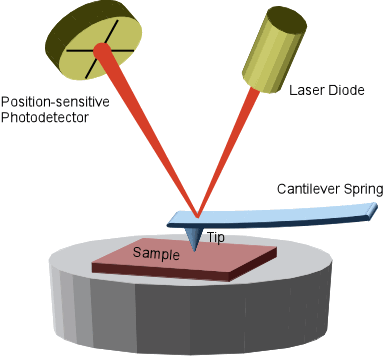}
    \caption{The optics of an AFM~\cite{HelmHP}.}
    \label{fig:nm1}
\end{figure}

Atomic Force Microscopy (AFM) is one type of the family of Scanning Probe Microscopy (SPM) techniques which includes AFM, Magnetic Force Microscopy (MFM), Electric Force Microscopy (EFM) and Scanning Tunnelling Microscopy (STM). Among those, AFM is the one which can be used for the study of non-conductive samples like minerals. The resolution of AFM can be on the order of fractions of a nanometer, more than 1000 times better than the optical diffraction limit. The information is gathered by ``feeling'' or ``touching'' the surface with a mechanical probe. Piezoelectric elements that facilitate tiny but accurate and precise movements on (electronic) command enable very precise scanning. The optics of an AFM is shown in Fig.~\ref{fig:nm1}. 

There are two main imaging modes for AFM, {\it static} (contact) modes and {\it dynamic} (non-contact or tapping) modes. Contact mode involves the tip making continuous contact with the surface of the sample as it moves across the sample. The contours of the sample are measured either by the deflection of the cantilever or by the feedback signal maintaining a constant position of the cantilever. A more flexible cantilever is generally used in contact mode which is prone to noise and therefore one must keep the interaction force low.  Alternatively, tapping mode can be used to address the problems of the probe sticking in contact mode. This is achieved by oscillating the cantilever up and down as it scans the surface of the sample. The tapping mode reduces any damage caused to the sample and tip compared to contact mode. An AFM can reach a lateral resolution of 0.1 to 10\,nm. AFM can normally be used for: 1) force measurement, 2) topographic imaging, and 3) manipulation. In the context of mineral detectors, AFM can, e.g., be used for mineral dating as mentioned in section~\ref{sec:MineralDetectors-FissioTrackAnalysis}; see also sections~\ref{sec:MineralDetectors_DM} and~\ref{sec:Studies-JAMSTEC} for a discussion of studies demonstrating AFM readout (after chemical etching) of the damage features of few-keV nuclear recoils as would be induced by neutrino or Dark Matter interactions.

\subsubsection{SEM/FIB} \label{sec:ReadOut-NM-SEM}

\begin{figure}
    \centering
    \includegraphics[width=0.6\linewidth]{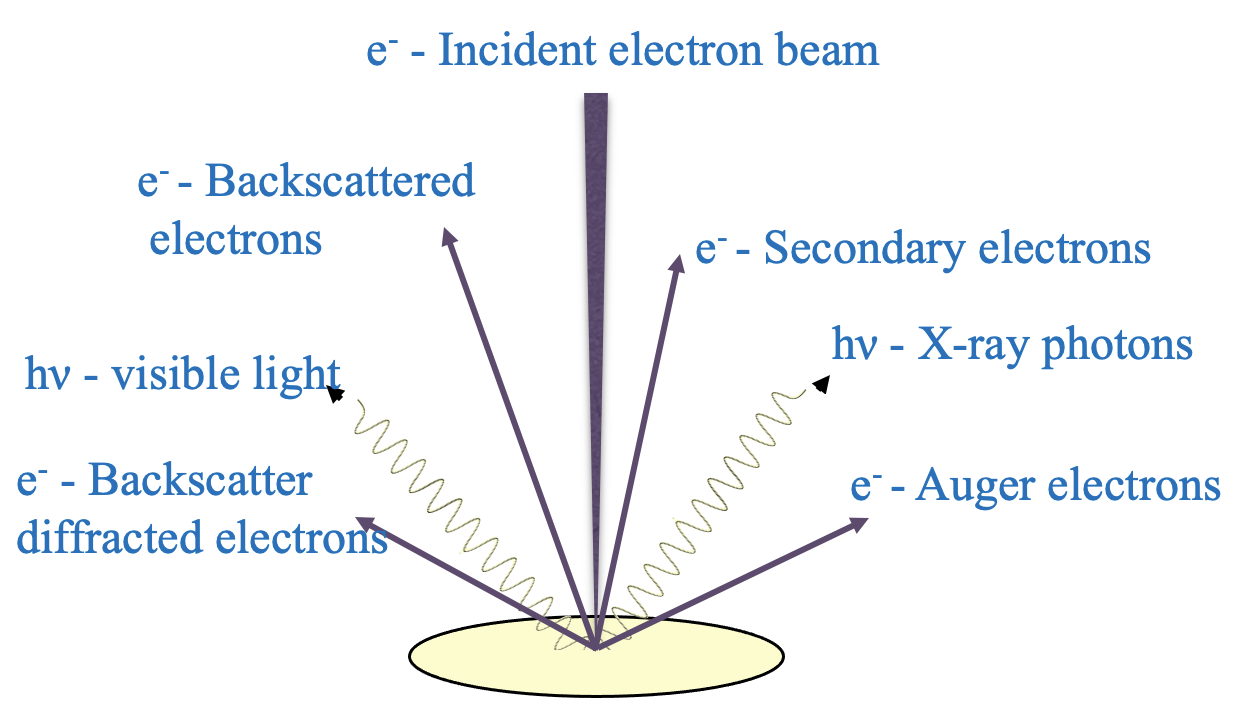}
    \caption{Schematic drawing showing the interaction of an electron beam with a solid.}
    \label{fig:nm2}
\end{figure}

\begin{figure}
    \centering
    \includegraphics[width=1\linewidth]{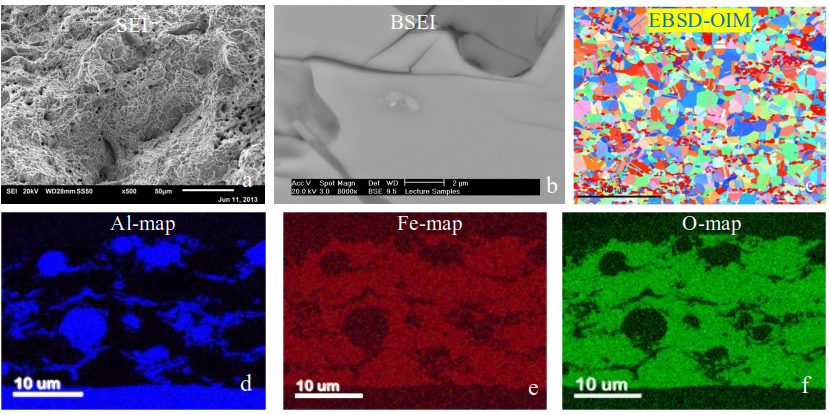}
    \caption{Applications of a SEM in materials characterization. A) Secondary Electron Image (SEI) showing the surface morphology; b) BackScattered Electron Imaging (BSEI) showing the composition information; c) Orientation Image Microscopy (OIM) map based on Electron BackScatter Diffraction (EBSD) showing orientation maps of the grains; d)-f) Al-, Fe- and O-maps, respectively, showing the element distributions.}
    \label{fig:nm3}
\end{figure}

Scanning Electron Microscopy (SEM) uses a focused energetic electron beam scanning over a surface of a solid. The electron beam will interact with the solid generating several signals as shown in Fig.~\ref{fig:nm2}. In a modern SEM instrument, several detectors are installed that can collect those different signals from which images (at a resolution better than 10\,nm), diffraction, and spectra from the scanned region will be obtained. As those signals can reflect morphology, composition or crystallinity of the solid, the microstructure and chemistry of the solid are then characterized as shown in Fig.~\ref{fig:nm3}.

\begin{figure}
    \centering
    \includegraphics[trim=0cm 5cm 0cm 0cm, clip, width=0.4\linewidth]{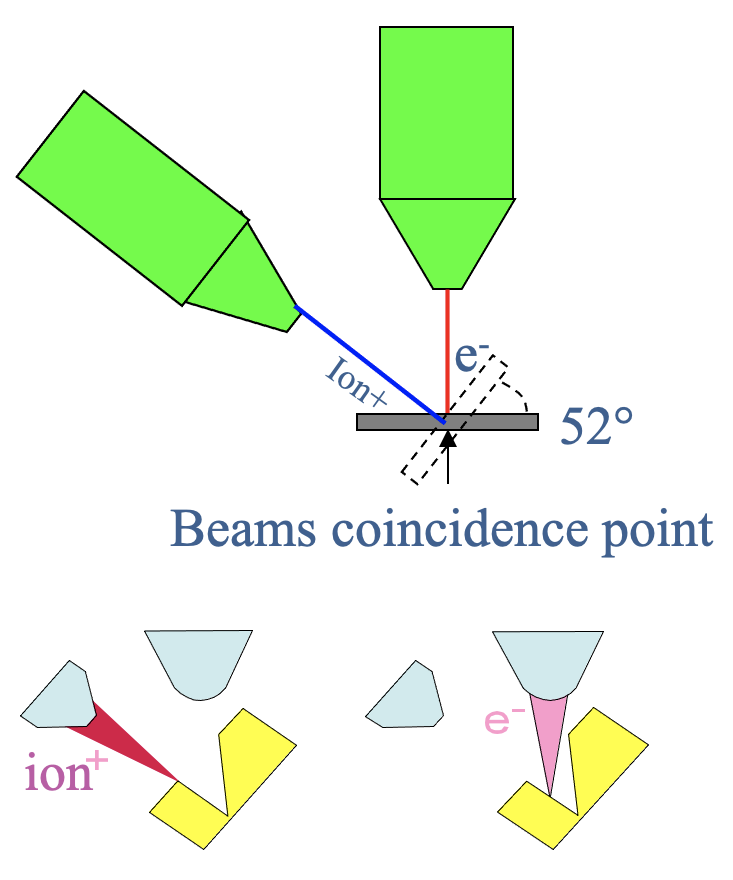}
    \hfill
    \includegraphics[trim=0cm -2cm 0cm 11cm, clip, width=0.4\linewidth]{Figures/nm4}
    \caption{Schematical drawing showing the geometrical setting of a dual-beam FIB.}
    \label{fig:nm4}
\end{figure}

\begin{figure}
    \centering
    \includegraphics[width=1\linewidth]{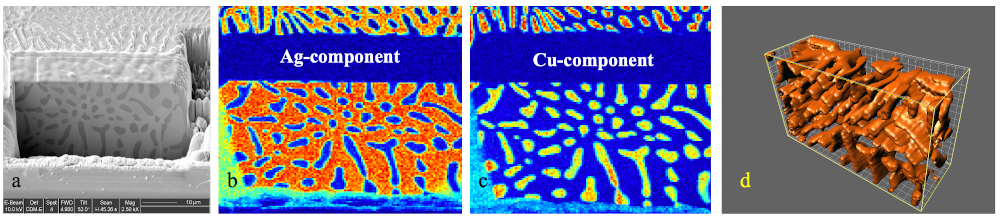}
    \caption{3D tomography of a Ag-Cu alloy by FIB series sectioning: a) SEM-SEI; b) and c) SEM-EDS element maps showing Ag and Cu components and d) 3D reconstructed tomogram showing Ag component in 3D (courtesy Thermo-Fisher).}
    \label{fig:nm5}
\end{figure}

Nowadays, it is very common to have a SEM combined with a Focused Ion Beam (FIB) in a single instrument, which is then called ``dual-beam FIB''. Fig.~\ref{fig:nm4} shows the geometrical setups of a dual-beam FIB. Normally, an electron beam will come from the top and a focused ion beam angled (here by $52\,^\circ$) from it. A commercialized dual-beam FIB normally uses Ga as its ion beam. Recently, an inert gas (like Xe or Ar) is used as its ion source and the instrument is thus called plasma FIB. One advantage is that a plasma-FIB can offer much larger beam current. Dual-beam FIBs have been widely used for cutting and patterning. By combination of SEM-based techniques with FIB ion beam cutting (or sectioning), 3D datasets can be obtained which can visualize the 3D microstructures and chemistry of a material as shown in Fig.~\ref{fig:nm5}. This can expose regions that are hidden under the surface directly, revealing all structural details in a wide range of material types.

However, there are limitations on the sample dimensions when performing FIB series sectional 3D tomography. For a Ga-source FIB, it is efficient to make cuts at no more than 100\,$\mu$m length scale, while for a Xe plasma source FIB, the linear sample dimensions can reach 100\,$\mu$m. More recently, a laser beam has been introduced to a dual-beam FIB which is then called {\it LaserFIB}~\cite{Tordoff:2020}. A LaserFIB can machine hard or soft, conducting or insulating material at unprecedented speed. Final polishing to reveal even more details can be done using the regular FIB in the same instrument.

\begin{figure}
    \centering
    \includegraphics[width=0.75\linewidth]{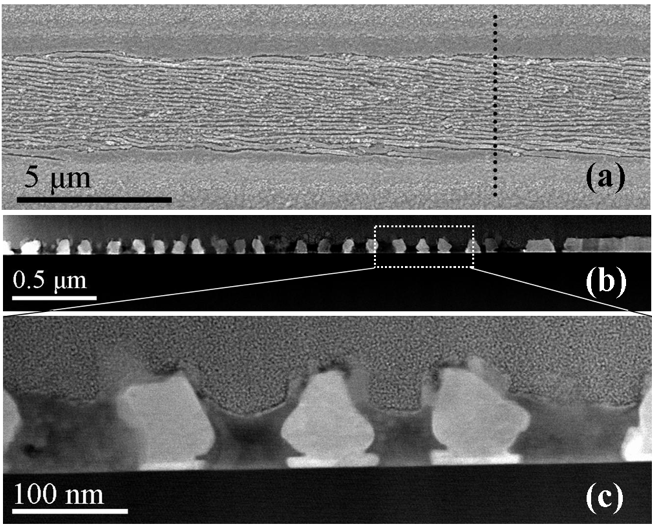}
    \caption{(a) SEM SEI showing the surface of a SbTe/BiTe thin film grown on sapphire after one-time laser scan. The vertical dotted line indicating the FIB cutting direction; (b) low magnification Scanning Transmission Electron Microscopy-High Angle Annular Dark-Field (STEM-HAADF) image with a region outlined by a dotted frame for magnified imaging shown in (c). Figure taken from Ref.~\cite{Li:2014}.}
    \label{fig:nm6}
\end{figure}

Figure~\ref{fig:nm6} shows examples of SEM and TEM images of specimen prepared by dual-beam FIB. This is a FIB-must sample for cross-sectional TEM preparation as it consists of a very soft, thin film layer (SbTe) grown on a very hard substrate (sapphire with its hardness just next to diamond); mechanical cutting and grinding methods will not work for such a sample.

\begin{figure}
    \centering
    \includegraphics[width=1\linewidth]{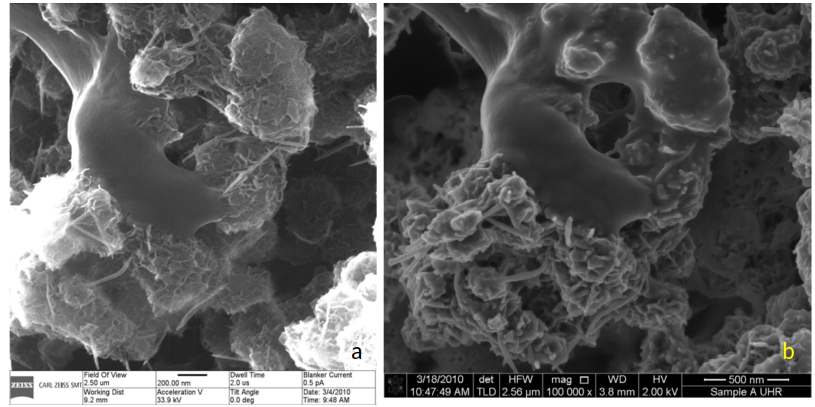}
    \caption{Comparison of (a) HIM image and (b) SEM-SEI images; figure taken from Ref.~\cite{Arey:2010}.}
    \label{fig:nm7}
\end{figure}

When a He ion source is used instead of an electron beam, the instrument is called {\it He-Ion Microscope} (HIM)~\cite{Gnauck:HIM}.  Like a SEM, a HIM is for surface imaging and chemical analysis. Compared with SEM, HIM can provide much higher resolution images ($\sim$0.3\,nm resolution at energies of $25\textit{--}30\,$keV) with better material contrast and improved depth of focus, see Fig.~\ref{fig:nm7}. Analysis of material composition can be performed using Rutherford backscattering spectrometry~\cite{Arey:2010}.

\subsubsection{TEM/STEM} \label{sec:ReadOut-NM-TEM}

Transmission Electron Microscopy (TEM) has been key in the characterization of the microstructure and microchemistry of almost all types of materials in materials science, geology, and biology due to its high spatial resolution (currently down to sub-{\AA}ngstr{\"o}m) and versatile techniques available. Nowadays, most TEMs can be operated in two different modes, Conventional TEM (CTEM) mode and Scanning TEM (STEM) mode. In CTEM mode, an electron beam generally formed as a broad or parallel beam is used for imaging. The beam does not move (unless being moved manually); CTEM is also called broad beam mode or fixed beam mode. In STEM mode, an electron beam is focused to be a sharp probe as small as sub-{\AA}ngstr{\"o}m. A scanning coil system is attached which can drive the probe beam over a defined region; STEM is also called probe mode or scanning beam mode. Functionally, CTEM and STEM are almost equal. They can both be operated for imaging, diffraction, spectroscopy and even holography if a field emission gun is used with a biprism attachment. 

\begin{figure}
    \centering
    \includegraphics[width=1\linewidth]{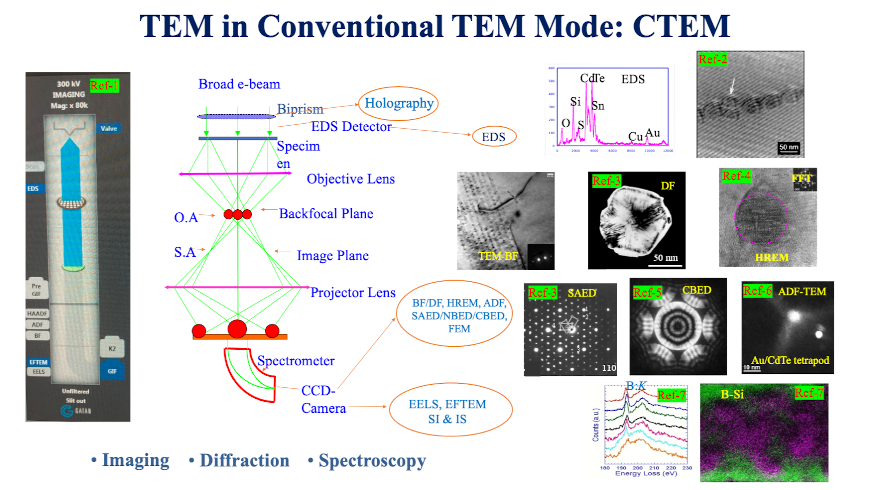}
    \caption{Schematical drawing showing optical settings of a TEM operated in CTEM mode as well as available techniques and their applications in materials research. (Figure credits as marked in the insets: Ref-1:~\cite{Gatan}, Ref-2:~\cite{Dunin-Borkowski:1998}, Ref-3:~\cite{Sun:2002}, Ref-4:~\cite{Sun:1999}, Ref-5:~\cite{Tanaka}, Ref-6~\cite{Bals:2004}, Ref-7:~\cite{Sun:2004})}
    \label{fig:nm8}
\end{figure}

\begin{figure}
    \centering
    \includegraphics[width=1\linewidth]{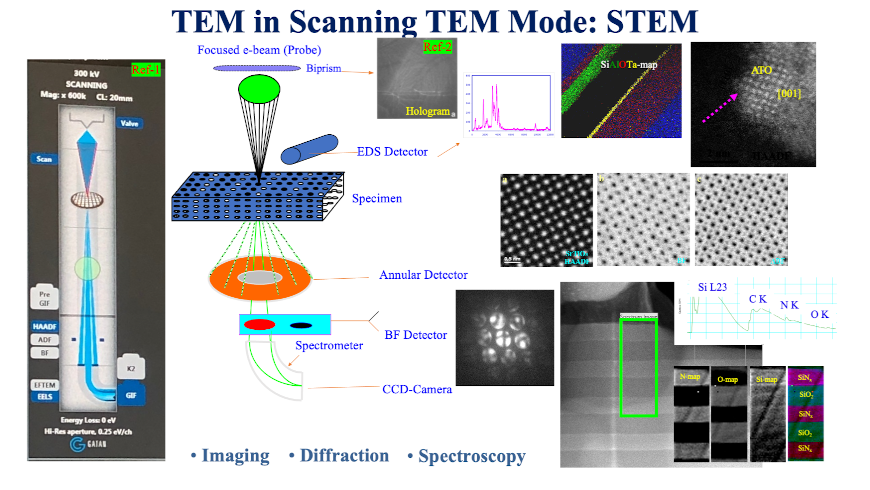}
    \caption{Schematical drawing showing optical settings of a TEM operated in STEM mode as well as available techniques and their applications in materials research. (Figure credits as marked in the insets: Ref-1:~\cite{Gatan}, Ref-2:~\cite{Mankos:1995})}
    \label{fig:nm9}
\end{figure}

Figure~\ref{fig:nm8} shows normal optical settings of a TEM running in CTEM mode and summarizes the techniques available in CTEM mode and some example applications of those techniques. In CTEM mode, one can perform Bright-Field (BF), Dark-Field (DF), High Resolution Electron Microscopy (HREM), Annular Dark-Field (ADF) imaging, elemental mapping using Electron Energy Loss Spectroscopy (EELS), and obtain a hologram by holography. One can also perform Selected Area Electron Diffraction (SAED), Nano-Beam Electron Diffraction (NBED), Convergent Beam Electron Diffraction (CBED), and X-ray Energy Dispersive Spectroscopy (EDS). Figure~\ref{fig:nm9} shows optical setting of a TEM running in STEM mode. Like CTEM, STEM can be used for imaging, diffraction, spectroscopy, and holography. Comparing with CTEM, one can perform Annular-Bright Field (ABF) imaging and element mapping using X-ray signals. STEM has some advantages over CTEM, in particular, STEM is better for small-area analysis and it is not so thickness-sensitive in imaging.

\begin{figure}
    \centering
    \includegraphics[width=1\linewidth]{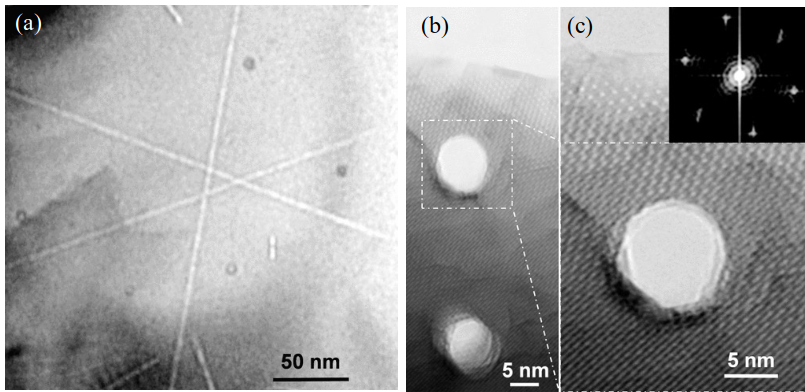}
    \caption{(a) Randomly oriented neutron-induced-fission tracks embedded in fluorapatite as observed by TEM. There are no fundamental differences in the nature of the tracks created by the spontaneous fission of $^{238}$U or the neutron-induced fission of $^{235}$U; (b) Plan view HRTEM image of two tracks induced by 2.2\,GeV Au ions showing a highly porous core. The tracks are deliberately produced along the c-axis of a fluorapatite single crystal. The Airy pattern shown in the FFT image [as an inset in (c)] is caused by the electron diffraction from the highly porous track, acting as an aperture. Figure taken from Ref.~\cite{Li:2010}.}
    \label{fig:nm10}
\end{figure}

\begin{figure}
    \centering
    \includegraphics[width=0.9\linewidth]{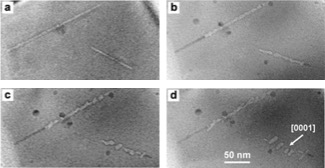}
    \caption{Crystallographic effect. In-situ TEM images showing the preferential motion of fission track segments along [0001] of fluorapatite and the slower fragmentation of tracks along this direction during thermal treatment. (a) Before annealing, and (b) after 1, (c) 17, and (d) 60\,min heated at $700\,^\circ$C. Figure taken from Ref.~\cite{Li:2010}.}
    \label{fig:nm11}
\end{figure}

TEM has been a tool widely used in fission track studies as mentioned in section~\ref{sec:MineralDetectors-FissioTrackAnalysis} (for example, Fig.~\ref{fig:FT_TEM} shows BF and HREM images of fission tracks in apatite). By in-situ heating in a TEM, fission tracks in fluorapatite were confirmed to be porous tubes from direct observation which showed that thermal annealing could induce track fragmentation in fluorapatite as shown in Figs.~\ref{fig:nm10} and~\ref{fig:nm11}.

\begin{figure}
    \centering
    \includegraphics[width=1\linewidth]{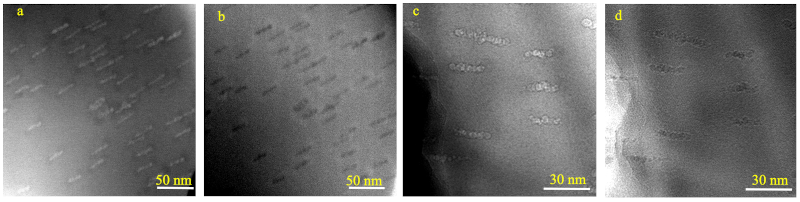}
    \caption{5\,MeV Au ion-bombardment-induced tracks in F-apatite. a) STEM-BF and b) STEM-HAADF images; c) EFTEM-BF and d) EFTEM thickness images.}
    \label{fig:nm12}
\end{figure}

In an unpublished study, STEM BF and STEM HAADF images as well as energy-filtered BF and thickness images performed in CTEM-EFTEM mode were used for imaging 5\,MeV Au ion irradiation induced fission tracks in a F-apatite as shown in Fig.~\ref{fig:nm12}. Those images confirm that the fission tracks are amorphous tubes, not empty holes.

\begin{figure}
    \centering
    \includegraphics[width=1\linewidth]{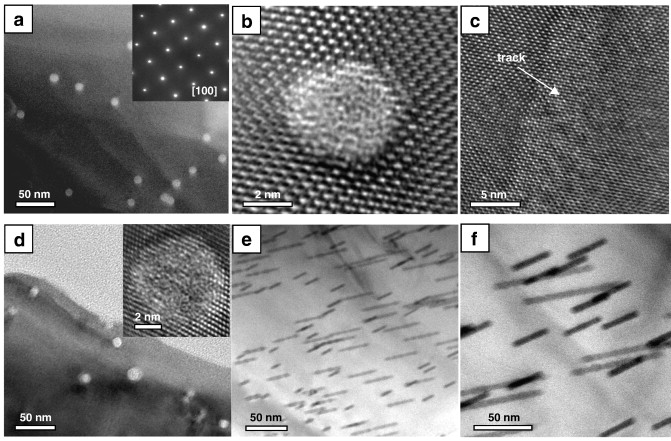}
    \caption{TEM images of zircon after exposure to 10\,GeV Pb ions for a fluence of $5 \times 10^{10}\,$ions/cm$^2$: (a–c) reference sample irradiated at room temperature and ambient pressure, (d–f) sample pressurized (7.5\,kbar) and heated ($250\,^\circ$C). Bright-field micrographs (a, d–f) of cross-sections and projections of tracks, and high-resolution images (b, inset of d) give evidence of the amorphous damage structure of the tracks. Note that amorphous domains that are surrounded by crystalline domains will still display lattice fringes; however, the change in contrast in (c) is the evidence for the presence of amorphous material. The electron diffraction pattern [inset of (a)] confirms that the irradiation was performed parallel to the c-axis. Figure taken from Ref.~\cite{Lang:2008}.}
    \label{fig:nm13}
\end{figure}

In another study~\cite{Lang:2008}, fission-track formation was simulated under crustal conditions by exposing natural zircon, at a pressure of 7.5\,kbar and a temperature of $250\,^\circ$C, to a beam of relativistic heavy ions. The latent tracks were investigated using HREM, and the diameters of several hundred tracks were measured as shown in Fig.~\ref{fig:nm13}. It was found that, based on the number of measurements, this represents a statistically significant difference between the tracks at ambient vs. high-pressure/temperature conditions. The slightly larger size of the tracks at elevated pressure can be understood in terms of the increased efficiency of the damage process in a strained crystal lattice. This slight variation in track diameter ($\sim$0.2\,nm) at high pressure probably will not affect the dimensions of etched tracks.

\begin{figure}
    \centering
    \includegraphics[width=\linewidth]{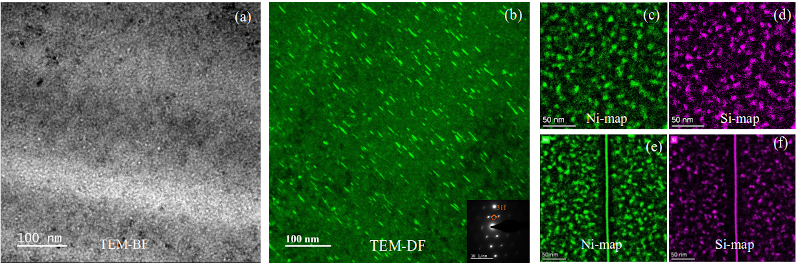}
    \caption{Neutron irradiation induced damaged microstructures in a 316 steel: (a) voids, (b) dislocation loops, and (c-f) element maps showing the chemistry of precipitates and (e)-(f) radiation induced grain boundary segregations~\cite{Sun:unpublished}.}
    \label{fig:nm14}
\end{figure}

In nuclear materials research, TEM has also been a major tool in characterizing microstructures and microchemistry of materials under neutron or high energy ion irradiation. Figure~\ref{fig:nm14} shows voids, dislocation loops, precipitates and Radiation Induced grain boundary Segregations (RIS) by TEM imaging. Those are 4 structures features that define the materials mechanical properties and thus the performance. 

\begin{figure}
    \centering
    \includegraphics[width=\linewidth]{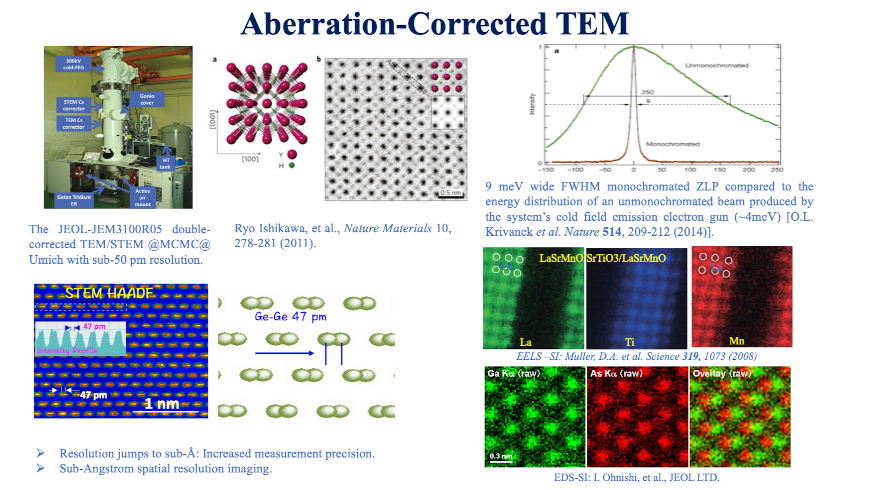}
    \caption{Abberation-corrected STEM.}
    \label{fig:nm15}
\end{figure}

In recent years, great achievements have been made in the development of TEM. In particular, aberration corrections allows to achieve sub-{\AA}ngstr{\"o}m spatial resolution. Figure~\ref{fig:nm15} outlines the improvement in spatial resolution of aberration-corrected TEMs. An aberration-corrected TEM/STEM can now easily reach a spatial resolution better than 1\,{\AA}ngstr{\"o}m (better than 0.5\,{\AA}ngstr{\"o}m in most 300\,keV STEMs). Direct imaging of single H atoms, the smallest among all atoms, has been demonstrated with such STEMs, and atomic chemical imaging has been achieved by combined advanced EDS detectors and new energy filters with STEM imaging.

The advances in TEM resolution achieved via aberration correction might open an interesting avenue for mitigating radiogenic backgrounds in mineral detector searches for neutrinos and Dark Matter. As discussed in section~\ref{sec:MineralDetectors-ART}, $\alpha$-decays of radioactive trace elements such as $^{238}$U, $^{235}$U, or $^{232}$Th leave $\alpha$-recoil tracks in minerals. The stable end products of these naturally occurring radioactive decay series are $^{206}$Pb (for $^{238}$U), $^{207}$Pb (for $^{235}$U), and $^{208}$Pb (for $^{232}$Th). This means there should be a Pb atom associated with $\alpha$-recoil tracks (possibly also other elemental atoms). Aberration-corrected atomic resolution STEM-HAADF imaging plus spectroscopy techniques like EELS or EDS can easily detect such elements if they exists in any tracks. It would be interesting to demonstrate the feasibility of detecting the Pb ions associated with radiogenic background tracks -- such detection would allow one to differentiate damage features from the decays of $^{238}$U, $^{235}$U, or $^{232}$Th from nuclear recoils induced by the interaction of Dark Matter and neutrinos independent of the physical size (e.g., track length) of the associated damage features, potentially constituting a powerful means for background rejection for rare events searches with mineral detectors. 

%*********************************************************
\section{Status of feasibility studies} \label{sec:Studies}
%*********************************************************

A number of experimental studies towards probing the feasibility of mineral detection of neutrinos and Dark Matter are either already underway or are currently being planned at research institutes in Europe, Asia, and America. This section discusses the studies of the respective research groups. Currently, the main focus of most groups' efforts is to establish and develop readout techniques for the damage features induced by nuclear recoils in minerals. Towards this end, laboratory-grown or natural mineral samples are irradiated with keV--MeV ions or with MeV neutrons - while the former is experimentally easier since achieving high damage track densities and precisely controlling the ion energy is comparatively straightforward, neutron irradiation has the advantage that it produces nuclear recoils throughout the entire sample volume, more closely mimicking the signal produced by neutrinos or Dark Matter interacting with the atomic nuclei in the sample. After irradiation, the samples are then imaged by the readout technique of choice. Once a given readout technology for nuclear-recoil-induced damage in a given mineral is established, the next step is to establish a relation between the nuclear recoil energy and the observed damage, in particular, to establish the nuclear recoil energy threshold of a given mineral+readout combination and its energy resolution. Once this relation is established, the final step is to understand how a given readout technique can be scaled up to larger sample volumes, including how the data analysis can be automatized, and how a given readout technique can be combined with other techniques that have better/worse resolution and can handle larger/smaller sample volumes. Of course, these different steps need not be realized in the order presented here and different aspects can be developed in parallel. 

%*********************************************************
\subsection{SLAC} \label{sec:Studies-SLAC}
%*********************************************************
%{\color{blue} Coordinator: Chris Kenney and Sander Breur}

Team: {\it C.~Kenney, P.A.~Breur, A.~Gleason, J.~Segal}
\vspace{0.3cm}\\
The team at SLAC National Accelerator Laboratory (SLAC) has only recently become interested in using ancient mineral samples for Dark Matter and neutrino science as a natural complement to cosmological and real-time experiments in which the laboratory participates. We intend to explore three distinct techniques: Atomic Force Microscopy (AFM), Electron beam Tomography (ET), and ptychography or Coherent X-ray Imaging (CXI), which each offer distinct advantages and disadvantages. 

\subsubsection{Electron Beam Tomography}
SLAC and Stanford University operate the Stanford SLAC Cryo-EM Center (S2C2)~\cite{cryo-em} on the SLAC site, which is an open user facility and contains five electron microscopes with energies of 300\,keV along with an expert staff to assist in sample preparation and data collection. Samples can be rotated through angles from -45 to +45 degrees, and three-dimensional  ET images of the object can be reconstructed. Although ET is usually employed to measure biological samples, the main requirement is that the sample be sufficiently thin for transmission of the electron beam. For mica with a density of about 3\,g/cm$^3$ this limits the sample thickness to about 100\,nm, which is enough to contain tracks expected from WIMP-like Dark Matter with masses under 10\,GeV/c$^2$. The ET method is non-destructive and leverages a strong established hardware and software ecosystem. ET is capable of sub-nanometer resolution - for example the 0.67\,nm achieved in~\cite{Sun:2022}. The challenges with ET are sample preparation and mass throughput. There are additional electron microscopes on the main Stanford campus.
 
 \subsubsection{Coherent X-ray Imaging}
Three-dimensional coherent X-ray imaging using SLAC's Linear Coherent Light Source (LCLS) is a second focus of the SLAC team. The LCLS~II~\cite{slac03} upgrade provides 10,000 more pulses and hence X-rays per second than LCLS~I. CXI ptychography is a relatively recent imaging technique that is progressing quickly and is enabled by the next generation of light sources, see section~\ref{sec:ReadOut-HardX}. It offers the possibility of imaging macroscopic volumes of minerals with $\mathcal{O}(10)\,$nm spatial resolution. Like ET, CXI is non-destructive and could thus be used as a first scan to identify regions of interest which can then be imaged using other methods. Using synthetic control samples of mica and silicon, a first measurement was attempted with the LCLS-I beam. A team of SLAC scientists, partnered with an external user group, tested surrogate Dark Matter mineral samples in mica and silicon single-crystal samples. This test was carried out at the X-ray Pump Probe (XPP) end-station at LCLS. Using 8\,keV X-rays, passed through a Si-monochromator, X-ray pulses at 120\,Hz where raster scanned across a $10\,\mu{\rm m} \times 20\,\mu$m area in a transmission geometry. Each sample was 20\,$\mu$m thick. Despite the 1.7\,mJ of X-ray energy in each pulse, we did not see adequate scatter, above an air scatter signal, from the single-crystal samples. Future work to improve the phase contrast and increase diffraction {\it Q} range to higher spatial resolution is required.  

\subsubsection{Atomic Force Microscopy}
A third method the SLAC team is investigating is using Atomic Force Microscopy (AFM) to scan a set of ancient mineral samples as has been reported by other research groups, see section~\ref{sec:Studies-JAMSTEC}. One significant advantage of AFM relative to ET or CXI is that the sample does not need to be thinned, which greatly facilitates handling and cleaning of the surface. Sensing nanometer-scale topography is well within the capability of AFM. Our plan is to gain experience and demonstrate the ability to identify the locations where Dark-Matter- or neutrino-induced damage tracks intersect the sample's surface. We will explore various chemical etches which are selective between crystalline and amorphous states. The general vision is to prepare sample surfaces, perform a shallow selective chemical etch and image with AFM to identify a set of potential damage sites. We will then perform more detailed imaging of the candidate sites, perhaps using multiple readout modalities. The substrates would then undergo a sputter etch or, for materials such as mica, cleaving, to remove approximately 1\,nm of material from the surface in a uniform manner. Then, the samples would be {\it selectively} etched to a depth of about 0.5\,nm and the candidate sites re-imaged using AFM and perhaps other techniques. This process would be iterated to a depth which would contain all the damage track lengths of interest. This would be on the order of 100\,nm in total depth with a total of about 100 (sputter etch)-image cycles. It is expected that this would provide a track length and recoil energy spectrum with approximately 1\,nm resolution as the depth would be measured in 1\,nm increments and the lateral extent via AFM, which can provide better than 1\,nm in-plane resolution. The thickness removed per step and the number of steps can be optimized for various science cases. For instance, a search for higher mass WIMP-like Dark Matter might employ 30 steps of 15\,nm each. We note that achieving and maintaining an atomically smooth and flat sample surface is a significant challenge.
 
\subsubsection{Data Challenge and prospects}
The central challenge in any of these methods is the initial imaging of a large area and or volume at high resolution as this requires significant time and resources to realize the full science reach of the paleo-concept. Critically, this is expected to be an extremely sparse dataset, which means the vast majority of the resources are expended in the initial pass. For the scenario of imaging 10\,mg of material with 1\,nm resolution, this translates into scanning about 200\,cm$^{2}$ or $\sim 10^{16}\,$nm$^{2}$. Theoretically, about 10,000~tracks are expected over this area due to the combination of Dark Matter, neutrinos, and radioactive decay backgrounds or, in other words, the ``track hit areal fraction'' would be of the order of $10^{-12}$. Hence, the rapid processing of the image data to find and trigger on track candidates is essential. SLAC has deep expertise in using FPGAs and other processor types to manipulate and trigger large information flows for high energy physics, astronomical, and light-source based experiments. Preliminary design of such as subsystem will be one task of our group in the near term. A false positive fraction of 90\% is acceptable for the initial trigger. Follow-up region-of-interest scans would need to be aligned to the initial global scan at the micron level, which is commonplace for tools used in photolithography, precision packaging, etc. and can be done optically. 

To fully exploit the science potential of mineral detectors would require operating on the order of 1,000 AFMs in parallel for periods on the order of a year. A single AFM typically produces data at a rate of about 10\,megabytes per second and close to 1\,terabyte per day. A thousand AFMs would generate a data flow of approximately 1\,petabyte per day. Again, the sparse nature of this data must be emphasized as only around 500\,bytes per day is expected to be associated with science candidate tracks. Trigger algorithms operating in FPGAs locally at the edge should be able to handle this; similar technology is being pursued for LCLS and other projects.

As each local volume of material is independent on the micron level in terms of its damage tracks, a mineral sample can be subdivided into many subsamples each of which can be imaged separately. This allows an almost arbitrary degree of parallelism and makes the problem tractable. To explore the WIMP-like Dark Matter parameter space down to cross sections of $10^{-45}\,{\rm cm}^2$ and masses below 10\,GeV/c$^2$ necessitates scanning the equivalent of order 1,000\,AFM-years. That is, 1,000 AFMs imaging for one year and with their data stream being processed by a similar number of FGPAs. This scale it typical of a medium size high-energy physics experiment. Ideally, the community could form an international collaboration to pursue this.     
 
\subsubsection{A First Step Forward}

\begin{table}
    \centering
 \begin{tabular}{p{0.8in}|p{0.8in} p{0.8in} p{0.8in} p{0.6in} p{0.9in} } 
Material & Thickness [$\mu$m] & Dose [ions/cm$^2$] & Energy [keV] & Angle [degrees] & Track Length [nm] \\ \hline 
Silicon & 525 & $10^{10}$ & 50 & 7 & 30 \\ 
Silicon & 525 & $10^{10}$ & 50 & 30 & 30 \\ 
Silicon & 525 & $10^{10}$ & 200 & 7 & 120 \\ 
Silicon & 525 & $10^{11}$ & 1,000 & 7 & 500 \\ 
Silicon & 10 & $10^{10}$ & 50 & 7 & 30 \\ 
Silicon & 10 & $10^{10}$ & 50 & 30 & 30 \\ 
Silicon & 10 & $10^{10}$ & 200 & 7 & 120 \\ 
Silicon & 10 & $10^{11}$ & 1,000 & 7 & 500 \\ 
Muscovite & 12 & $10^{10}$ & 50 & 7 & $\sim 30$ \\ 
Muscovite & 12 & $10^{10}$ & 50 & 30 & $\sim 30$ \\ 
Muscovite & 12 & $10^{10}$ & 200 & 7 & $\sim 120$ \\ 
Muscovite & 12 & $10^{11}$ & 1,000 & 7 & $\sim 500$ \\ 
\end{tabular}
    \caption{Parameters of control samples fabricated at SLAC.}
    \label{tab:slac01}
\end{table}

An incremental approach is foreseen, starting with a single commercial AFM to demonstrate the feasibility of the paleo method. If that is successful, then engineering to create a design that scales to 1,000s of customized AFMs that is financially possible would be needed. At that point, one would build and operate several generations of AFM farms with progressively larger numbers of machines. 

As a first step of the experimental effort at SLAC, we have fabricated a set of single-crystal silicon and muscovite samples with damage tracks to explore the viability of using AFMs, electron tomography, and coherent X-ray imaging. The parameters of the control sample set are given in Table~\ref{tab:slac01}. Note that track lengths for muscovite are approximate.

The thin samples are for transmissive imaging using X-rays and electrons, while the thick ones will be examined via AFM and other surface-interrogative methods. Key goals of the SLAC team over the next year are (i) to examine surface and bulk material quality as it pertains to background artifacts, (ii) to demonstrate the ability to image and identify artificially-formed damage tracks and (iii) to compare the predicted versus observed background tracks in natural mineral samples. If the above tasks yield promising results, then an initial modest-sensitivity Dark Matter and neutrino study will follow. 
 
%*********************************************************
\subsection{JAMSTEC} \label{sec:Studies-JAMSTEC}
%*********************************************************
%{\color{blue} Coordinator: Shigenobu Hirose}

Team: {\it S.~Hirose, N.~Abe, N.~Hasebe, Y.~Hoshino, N.~Ishikawa, T.~Kamiyama, Y.~Kawamura, K.~Murase, K.~Oguni, H.~Sakane, K.~Suzuki}
\vspace{0.3cm}\\
The feasibility studies of the team at Japan Agency for Marine-Earth Science and Technology (JAMSTEC) are aimed at observations and readouts of nuclear recoil tracks in minerals based on currently available devices as summarized in Table~\ref{table:jamstec_summary}. 

\begin{table}[h]
\centering
\begin{tabular}{lll}
Irradiation experiments & Observations / readouts\\
\hline
2 MeV He @SHI-ATEX & SEM (Fig.~\ref{fig:JAMSTEC_SHI-ATEX_Gypsum})\\
$\mathcal{O}(1)$MeV neutrons @HUNS & SEM (Fig.~\ref{fig:JAMSTEC_HUNS_Muscovite_SEM}), TEM(Fig. \ref{fig:JAMSTEC_HUNS_Muscovite_TEM}), X-ray CT (Fig.~\ref{fig:JAMSTEC_HUNS_CT})\\
200 MeV Xe @JAEA & SEM(Fig. \ref{fig:JAMSTEC_JAEA_Biotite_SEM}), TEM ({Fig.~\ref{fig:JAMSTEC_JAEA_Biotite_TEM}}) \\
$\mathcal{O}(1)$keV/u ions @Kanagawa U. & Etch+AFM (Fig.~\ref{fig:JAMSTEC_Kanagawa_Muscovite_Etch+AFM})\\
200 MeV Au @JAEA & Etch+Optical (Fig.~\ref{fig:JAMSTEC_JAEA_Muscovite_Etch+Opt})
\end{tabular}
\caption{Summary of JAMSTEC's feasibility studies}
\label{table:jamstec_summary}
\end{table}

\subsubsection{Direct observations of samples with the damage tracks}
References~\cite{Drukier:2018pdy,Baum:2018tfw} proposed to use helium ion microscopy (HIM) to read out latent damage tracks directly (i.e. without etching). Since HIM was not available at JAMSTEC, we instead used electron microscopy (EM). EM basically operates the same way as HIM, having a capability of resolving $\mathcal{O}(1)\,$nm as well as subsurface imaging with backscattering or transmission.

\begin{figure}[h]
 \centering
 \includegraphics[width=1\textwidth]{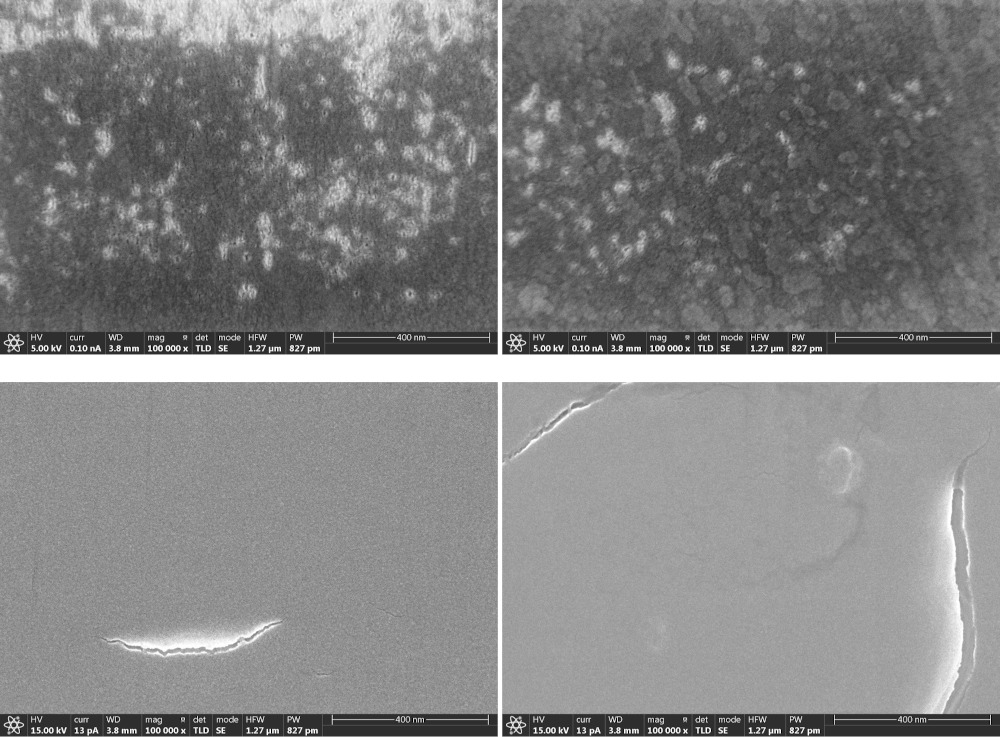}
 \caption{SEM SE images of gypsum samples irradiated by 2\,MeV He ions with a dose of $10^{11}\,$cm$^{-2}$ at SHI-ATEX. {\it Left}: Irradiated region. {\it Right}: Non-irradiated (reference) region. The voltage and the current differ, respectively, between the upper (5\,kV, 0.1\,nA) and the lower (15\,kV, 13\,pA).}
 \label{fig:JAMSTEC_SHI-ATEX_Gypsum}
\end{figure}

First, 2\,MeV He ion irradiation experiments were performed at SHI-ATEX, intended to model $\alpha$-particles from the $\alpha$-decay. The samples were observed by scanning electron microscopy (SEM) with secondary electrons (SE) after irradiation on their mechanically polished or cleaved surfaces. As shown in Fig.~\ref{fig:JAMSTEC_SHI-ATEX_Gypsum}, there was no notable difference between the irradiated region and the non-irradiated region. The magnification was limited up to $10^4$ because clear images could not be obtained beyond it due to charging (in spite of 3\,nm thick O-coating). This magnification was not enough to recognize nm-scale surface structures.

\begin{figure}[h]
 \centering
 \includegraphics[width=1\textwidth]{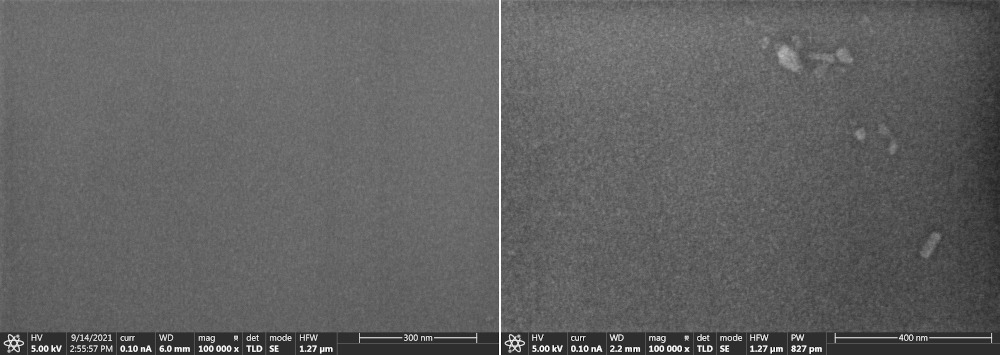}
 \caption{SEM SE images of muscovite samples irradiated by $\mathcal{O}(1)\,$MeV neutrons at HUNS. {\it Left}: Irradiated sample. {\it Right}: Non-irradiated (reference) sample.}
 \label{fig:JAMSTEC_HUNS_Muscovite_SEM}
\end{figure}

\begin{figure}[h]
 \centering
 \includegraphics[width=1\textwidth]{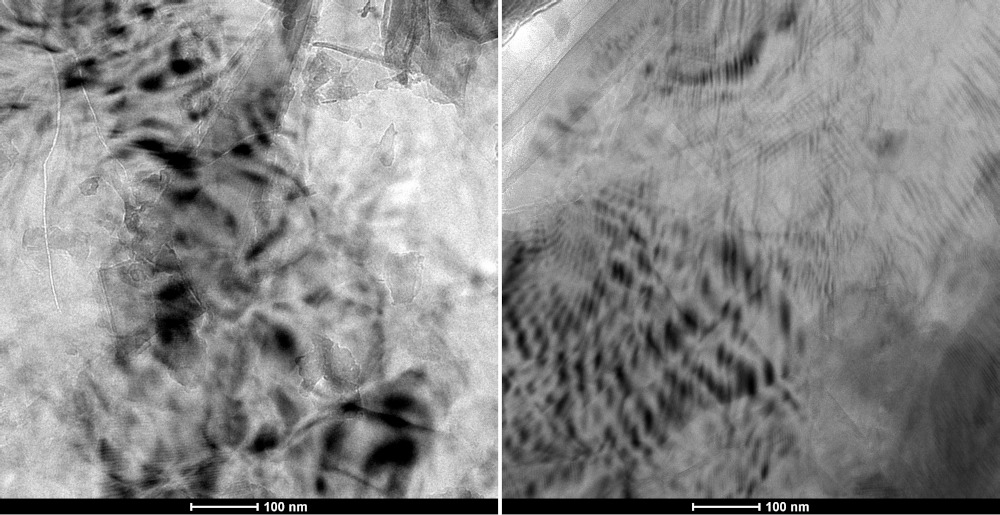}
 \caption{TEM images of muscovite samples irradiated by $\mathcal{O}(1)\,$MeV neutrons at HUNS. {\it Left}: Irradiated sample. {\it Right}: Non-irradiated (reference) sample.}
 \label{fig:JAMSTEC_HUNS_Muscovite_TEM}
\end{figure}

Next, neutron irradiation experiments were performed at Hokkaido University Neutron Source (HUNS), where the neutron energy distribution had a peak around $\mathcal{O}(1)\,$MeV. The nuclear recoils by neutrons occurs inside samples without any surface effect, in contrast to irradiation of ions. Figures~\ref{fig:JAMSTEC_HUNS_Muscovite_SEM} and~\ref{fig:JAMSTEC_HUNS_Muscovite_TEM} show, respectively, SEM and transmission electron microscopy (TEM) images of cleaved surfaces of muscovite samples. No notable difference between the irradiated sample (the dose was estimated as at least $10^{15}$\,n\,cm$^{-2}$ ) and the non-irradiated sample was seen in both SEM and TEM images. Here, the magnification in SEM was limited to $10^4$. In the TEM images, many complicated structures are observed since the TEM samples were prepared by grinding with pestle and mortar. Furthermore, X-ray computed tomography (CT) imaging was attempted, see the results in Fig.~\ref{fig:JAMSTEC_HUNS_CT}. The voxel size of these commercially available X-ray CT is now down to sub-microns, which however was not enough to resolve the nuclear recoil tracks. 

\begin{figure}[h]
 \centering
 \includegraphics[width=1\textwidth]{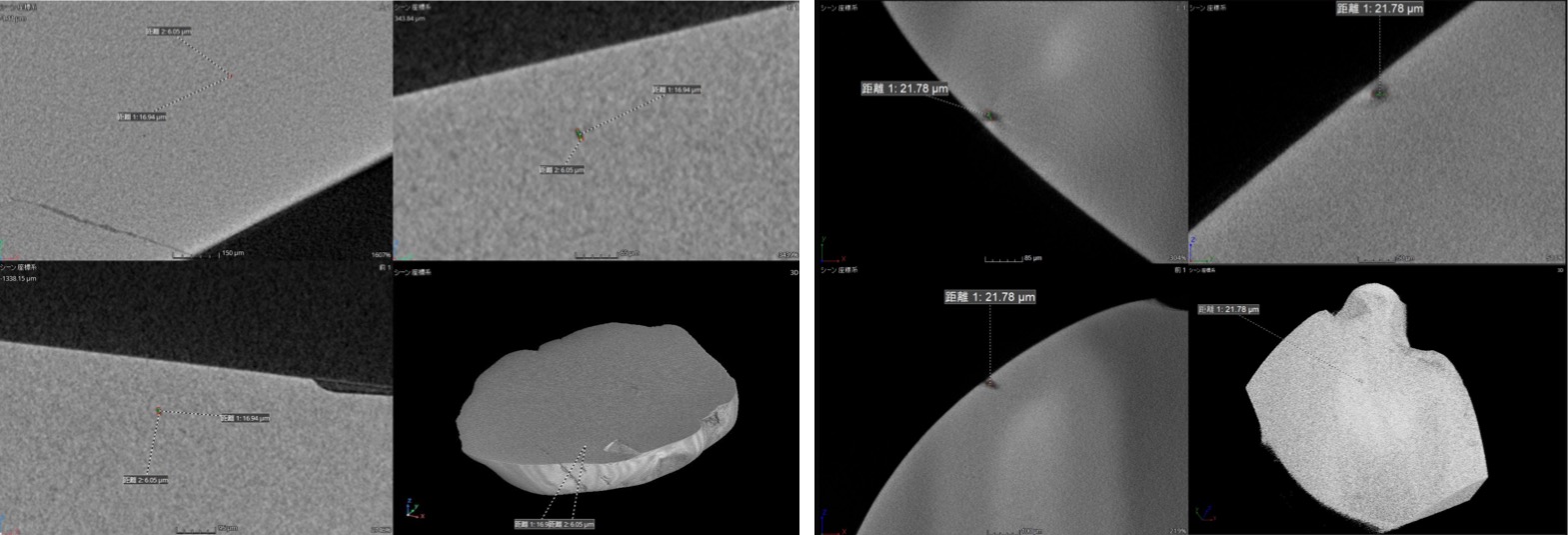}
 \caption{X-ray CT images samples irradiated by $\mathcal{O}(1)\,$MeV neutrons of $10^{14}$\, n\,cm$^{-2}$ at HUNS. {\it Left}: Olivine (taken by Phoenix Nanotom~M with voxel of 2\,$\mu$m; courtesy of JMC Corporation). {\it Right}: Garnet (taken by Comet Yxlon FF20~CT with voxel of 0.6\,$\mu$m; courtesy of Comet yxlon).}
 \label{fig:JAMSTEC_HUNS_CT}
\end{figure}

\begin{figure}[h]
 \centering
 \includegraphics[width=1\textwidth]{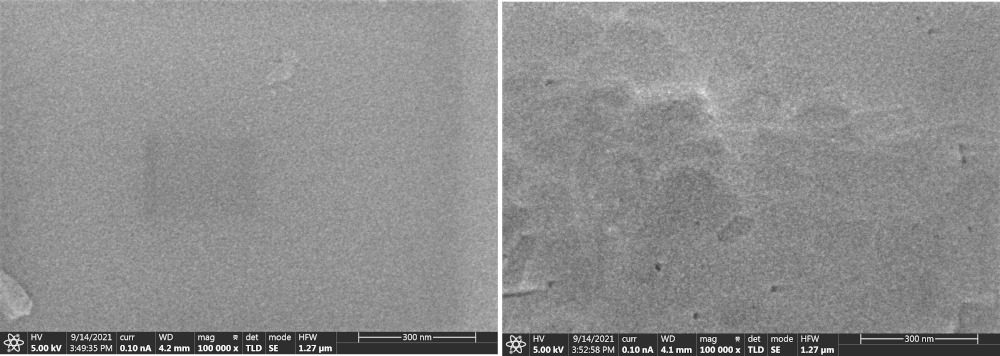}
 \caption{SEM SE images of biotite samples irradiated by 200\,MeV Xe ions with a dose of $10^{11}$\,cm$^{-2}$ at JAEA. {\it Left}: Irradiated region. {\it Right}: Non-irradiated (reference) region.}
 \label{fig:JAMSTEC_JAEA_Biotite_SEM}
\end{figure}

\begin{figure}[h]
 \centering
 \includegraphics[width=1\textwidth]{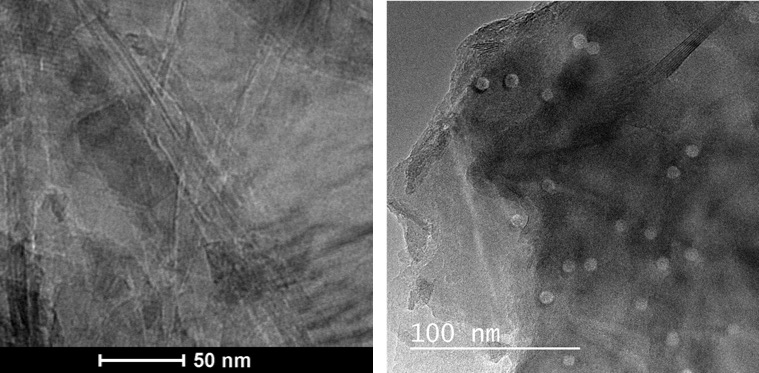}
 \caption{TEM images of biotite samples irradiated by 200\,MeV Xe ions with a dose of $10^{11}\,$cm$^{-2}$ at JAEA. {\it Left}: Ground after irradiation. {\it Right}: Irradiated after grinding.}
 \label{fig:JAMSTEC_JAEA_Biotite_TEM}
\end{figure}

It is well known that tracks created by swift heavy ions (SHIs) can be observed by TEM (see, e.g., Ref.~\cite{Ishikawa:2020}). We performed irradiation experiments with SHIs (here 200\,MeV Xe ions), expecting clear observable signatures of irradiation on the surface. The samples were irradiated at room temperature in a tandem accelerator at JAEA-Tokai (Japan Atomic Energy Agency, Tokai Research and Development Center, Tokai, Japan). At first, we could not see any signature of irradiation on the surface by SEM (Fig.~\ref{fig:JAMSTEC_JAEA_Biotite_SEM}) or TEM (Fig.~\ref{fig:JAMSTEC_JAEA_Biotite_TEM}, left). As for the former, probably the limited magnification of $10^4$ was not sufficient to see any surface structures. As for the latter, we suspect that when we prepared samples, we failed to pick up the very thin surface layer where tracks were formed by the SHIs\footnote{The ion range was calculated as $\sim 16\,\mu$m by SRIM.}. Although irradiation damages were not detectable in the experiments mentioned above, the tracks were clearly observed by TEM when the samples were first ground and then irradiated (Fig.~\ref{fig:JAMSTEC_JAEA_Biotite_TEM}, right).

\subsubsection{Observations of etched samples with AFM}
Readouts of keV/u nuclear recoil tracks in muscovite has been established by the pioneering work of Snowden-Ifft~{\it et al.}~\cite{Snowden-Ifft:1995zgn}. They etched cleaved surfaces of muscovite samples with Hydrofluoric Acid (HF) and measured depths of etch pits by Atomic Force Microscopy (AFM).

To reproduce the results of Ref.~\cite{Snowden-Ifft:1995zgn}, we began to conduct comprehensive keV/u ion irradiation experiments at Kanagawa University with various ion species and energies. The samples were annealed at $475\,^\circ$C for 5 hours to erase pre-existing $\alpha$-recoil tracks and then cleaved before irradiation. The dose was controlled to be $2\times 10^7\,$ cm$^{-2}$. After irradiation, the samples were etched with 48\%~HF at room temperature\footnote{We did not explicitly control the etchant temperature at this time. We are now controlling the temperature using a heat bath.} for 1\,hour and were observed with AFM.

\begin{figure}[h]
 \centering
 \includegraphics[height=0.3\textheight]{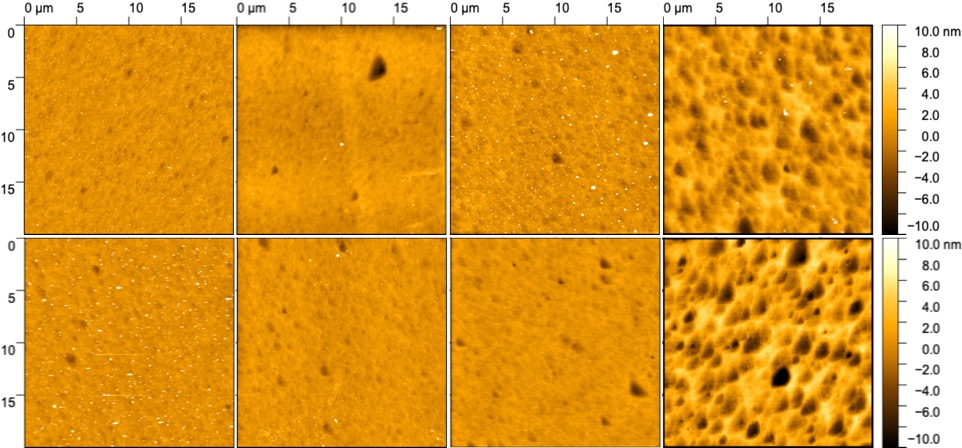}
 \caption{AFM images of etched muscovite irradiated by various species and energies of ions at Kanagawa University. {\it Upper}: from left to right, O$^{+}$ 10\,keV, Si$^{+}$ 20\,keV, Ar$^{+}$ 13\,keV, and Kr$^{+}$ 28\,keV. {\it Lower}: from left to right, O$^{+}$ 100\,keV, Si$^{+}$ 200\,keV, Ar$^{+}$ 40\,keV, and Kr$^{+}$ 84\,keV. The dose was $2\times10^7$\,cm$^{-2}$ for all cases.}
 \label{fig:JAMSTEC_Kanagawa_Muscovite_Etch+AFM}
\end{figure}

The results so far are summarized in Fig.~\ref{fig:JAMSTEC_Kanagawa_Muscovite_Etch+AFM}. The typical size and depth of etch pits are, respectively, a few $\mu$m and several nm. 
Here we have confirmed that the Snowden-Ifft's readout method of keV/u nuclear recoil tracks does work as well as that an implanted ion does not necessarily create an etch pit. For example, in the case of 84\,keV Kr ions, the number of pits is consistent with the number of incident ions ($\sim 80$ ions in the field of view) while in the case of 200\,keV Si ions, the number of pits is much less, around $10$ or so. To explain this, Ref.~\cite{Snowden-Ifft:1995rip} proposed an etch track model, in which the probability $P$ that an etching defect is created in any 10\,{\AA} layer of muscovite along the ion's path depends on the stopping powers as
\begin{align}
    P = 1 - \exp\left(-(k_\text{n}S_\text{n}+k_\text{e}S_\text{e})/\cos{\theta}\right),
\end{align}
where $S_\text{n}$ and $S_\text{e}$ are the nuclear and the electronic stopping powers, respectively, and $\theta$ is the zenith angle of the ion. $k_\text{n}$ and $k_\text{e}$ are numerical parameters, which were determined as $k_\text{n} \sim 2 k_\text{e}$ from their experiments, indicating that the $S_\text{n}$ was twice as effective as $S_\text{e}$ at creating etching defects.
Actually, the nuclear stopping power of an 84\,keV Kr ion is about six times as large as that of a 200\,keV Si ion in the etch zone\footnote{From other experiments, we estimated that the etchant can attack the surface layer of about $25\,$nm in our case.} as shown in Fig.~\ref{fig:JAMSTEC_Kanagawa_Muscovite_Etch+AFM_TRIM}.

\begin{figure}[h]
 \centering
 \includegraphics[width=1\linewidth]{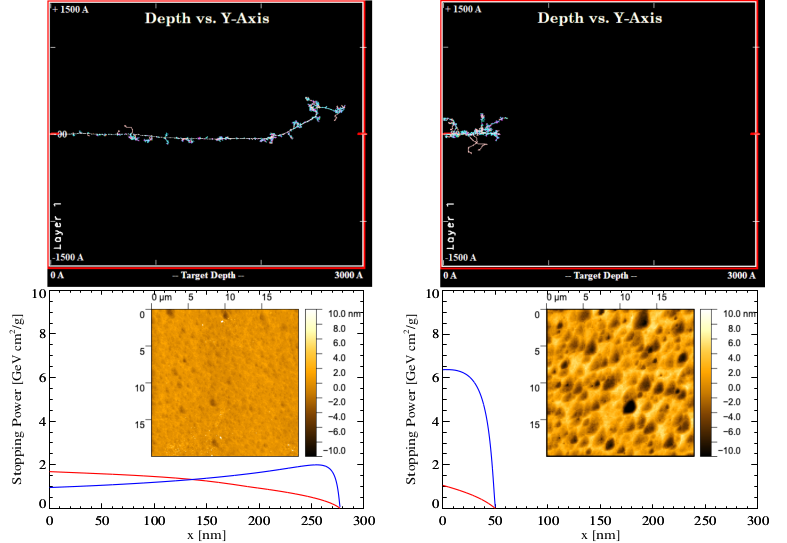}
 \caption{{\it Left}: TRIM simulation of Si $200$\,keV implanted into muscovite (upper) and corresponding Bragg curves (red: electronic, blue: nuclear) based on SRIM tables (lower). {\it Right}: Same for Kr $84$\,keV.}
 \label{fig:JAMSTEC_Kanagawa_Muscovite_Etch+AFM_TRIM}
\end{figure}

The etching is of course useful to visualize the latent tracks created by SHIs as established in the fission track dating. Figure~\ref{fig:JAMSTEC_JAEA_Muscovite_Etch+Opt} shows clear tracks in optical images of muscovite samples irradiated by 200 MeV Au ions at JAEA, where the samples were etched by the condition of fission track dating (48\% HF at 32${}^\circ$C for 5 minutes). 

\begin{figure}[h]
 \centering
 \includegraphics[width=1\textwidth]{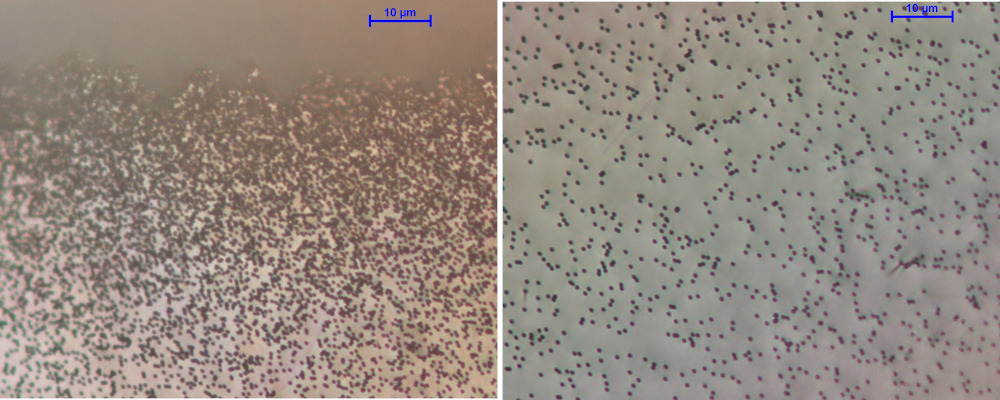}
 \caption{Observations with optical microscopy of cleaved and etched muscovite samples after irradiation by 200\,MeV Au ions with a dose of $10^{11}\,$cm$^{-2}$ at JAEA. {\it Left}: High fluence region (the edge of highly irradiated region). {\it Right}: Low fluence region (a bit away from the highly irradiated region). Note that the individual tracks are not visible in the highly irradiated region (the upper part of the left image), where they overlap each other.}
 \label{fig:JAMSTEC_JAEA_Muscovite_Etch+Opt}
\end{figure}

\subsubsection{Summary and perspective}
The direct observations/readouts of latent damage tracks in minerals with SEM or TEM were not successful. For SEM, charging prevented high-magnification imaging of samples, which may be improved by a charge-neutralizing flood gun in HIM. For TEM, the preparation procedure of thin samples ($\sim 100\,$nm thick) was not appropriate, but there should be room for improvement here. On the other hand, we confirmed that the Snowden-Ifft readout method using etching and AFM did work. We are going to extend their work by examining more micas, where the background of neutron recoils becomes critical, which is now being evaluated using \texttt{paleoSpec}.

\FloatBarrier

%*********************************************************
\subsection{Toho and Nagoya University} \label{sec:Studies-Toho}
%*********************************************************
%{\color{blue} Coordinator: Tatsuhiro Naka}

Team: {\it T.~Naka, Y.~Ido, T.~Shiraishi, T.~Kato, S.~Kazama, Y.~Ito, Y.~Kozaka, Y.~Asahara, Y.~Igami, Y. ~Kouketsu, K.~Michibayashi}
\vspace{0.3cm}\\
The group at Toho University's is one of the main groups for the directionally-sensitive Dark Matter search (the NEWSdm experiment~\cite{NEWS:2016fyf}) with super-fine grained nuclear emulsion called the nano imaging tracker (NIT) and has extensive expertise in nuclear emulsion technologies. NIT is capable of detecting the nano-metric scale tracks from Dark-Matter-induced nuclear recoils. These tracks are read out by a scanning system based on optical microscopes. This scanning system should also be applicable for a number of physics cases of paleo-detectors, e.g., the various Dark Matter searches discussed above as well as searches for neutrinos, monopoles, or Q-balls. Scanning large areas of paleo-detectors for defects is crucial to achieve the envisaged sensitivity.

\begin{figure}
 \centering
 \includegraphics[width=1.0\textwidth]{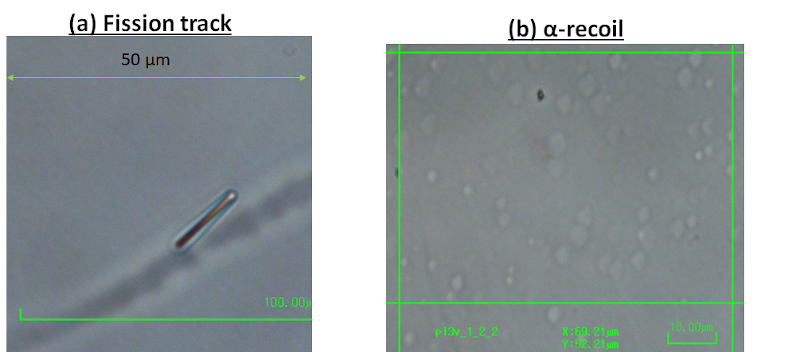}
 \caption{Optical microscope image of (a) fission track (left) and (b) $\alpha$-recoil damage (right) in muscovite mica.}
 \label{fig:Toho_mica_track}
\end{figure}

In Fig.~\ref{fig:Toho_mica_track}, optical microscopy images of a fission track (the black track with a size of $\sim 10\,\mu$m) and $\alpha$-recoil tracks (white randomly distributed spots) in muscovite mica are shown. The optical image and contrast should depend on the intensity of energy deposition by the incident particles and the mechanism, such as ionization or atomic collision. For the monopole~\cite{Price:1986ky} and Q-ball (here, we assume charged Q-balls)~\cite{Hong:2016ict,Arafune:2000yv}, observed tracks are expected to have high contrast similar to fission tracks because of high ionization losses. For nuclear recoil tracks produced by WIMPs or neutrinos scattering of nuclei in the crystal, one instead expects images similar to those from $\alpha$-recoils. For the optical microscopy readout, understanding how those candidates would be observed is of essential importance.

\subsubsection{Optical Microscope Scanning system and the Speed}

\begin{figure}
 \centering
 \includegraphics[width=0.7\textwidth]{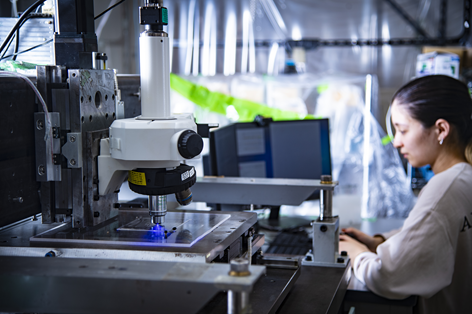}
 \caption{High resolution automatic optical microscope scanning system (PTS) for the nano-tracking with the nuclear emulsion.}
 \label{fig:Toho_PTS}
\end{figure}

A scanning system called the PTS (Post Track Selector)~\cite{Katsuragawa:2017ziw,Shiraishi:2021jce} is already demonstrated, an example picture of the system is shown in Fig.~\ref{fig:Toho_PTS}. The current PTS system has a readout speed of 10\,cm$^3$/month/system, and 3 systems are currently being operated. Those systems are continuously upgraded and an additional system is now being constructed. If the current system would be applied to mineral detectors, a scanning speed of $10^{7}\,$cm$^2$/month/system in area or $10\,$g/month/system in mass is expected. This speed is roughly $10^{6}$ times higher than what is currently possible with AFM and any other high-resolution microscope system. Therefore, optimizing a scanning system based on the optical microscope for paleo-detectors is of essential importance. And, as additional idea, combined analysis between the optical microscope and higher resolution microscope such as AFM will be important. For example, optical microscopes can be used as first event triggers thanks to their high scanning speed allowing one to process a large sample size, and one can then obtain more detailed information on the sub-volumes of the sample containing candidate events by imaging them, e.g., with an AFM. Such a concept has already been demonstrated for the analysis of emulsion detectors using optical microscopes and hard X-ray microscope~\cite{Op_XMS:2015}.

\subsubsection{Study for high contrast imaging}

\begin{figure}
 \centering
 \includegraphics[width=1\textwidth]{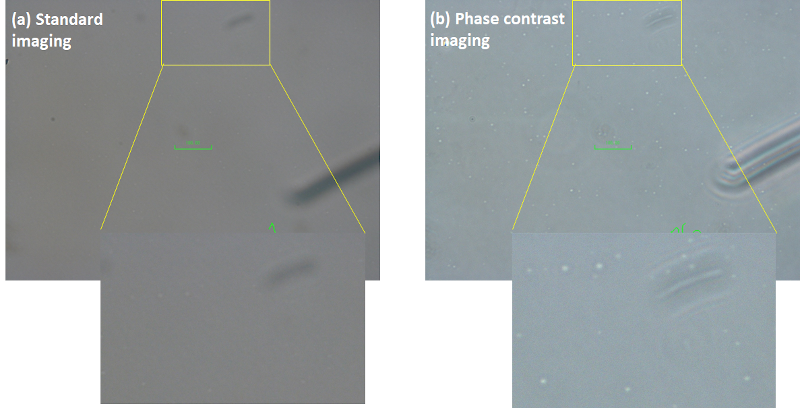}
 \caption{Comparison images of standard optical imaging (a; left) and phase contrast imaging (b; right) for the 40\,min etching with 48\% HF. White spots are pits due to $\alpha$-recoils. The scale bar in both images is $100\,\mu$m.}
 \label{fig:Toho_phase_contrast}
\end{figure}

Studying high contrast imaging with optical microscopes is important to utilize mineral detectors to search for Dark Matter. AFM and any other high resolution microscope can observe etch pits with nm scale resolution, however, it is difficult to achieve high scanning speeds. With optical microscopes, it is possible to observe larger pits produced by long-time etching, but then spatial resolution and information about the track length and orientation is lost. To enhance the contrast in optical microscopy, phase contrast imaging is a promising technique. Phase contrast imaging is the technique to obtain the phase-shift information as the contrast obtained by using a Zernike phase plate to shift the wavelength by $\lambda/4$; the contrast depends on the reflective index of objects. This is especially advantageous to image the small differences in reflective index of a medium, such features typically lead only to small contrast in standard microscopy. As a demonstration, we observed $\alpha$-recoil tracks in muscavite mica after 40\,min of HF etching, see Fig.~\ref{fig:Toho_phase_contrast}. The contrast of $\alpha$-recoil tracks with standard microscopic imaging was very low even after etching with HF (48$\%$) for 40\,min. However, the contrast in phase contrast imaging was drastically improved.Further studies of using phase contrast imaging are now on-going.    

\subsubsection{Calibration}
For the mineral detector searches for Dark Matter and other exotic particles, calibration of etch-pit formation and the pits' properties such as size, depth, and contrast depending on the etching condition is important to make more realistic and quantitative prediction. For example, low-energy nuclear recoils induced by Dark Matter can be calibrated by ion-implantation experiments, see Fig.~\ref{fig:JAMSTEC_Kanagawa_Muscovite_Etch+AFM}. Those events will also be investigated by the optical microscope system at Toho University, and we will evaluate the detection efficiency. Finally, as a next step, neutron irradiation studies of samples are planned to produce nuclear recoils distributed in the volume of a sample rather than the surface defects produced by ion implantation studies.

\begin{figure}
 \centering
 \includegraphics[width=0.5\textwidth]{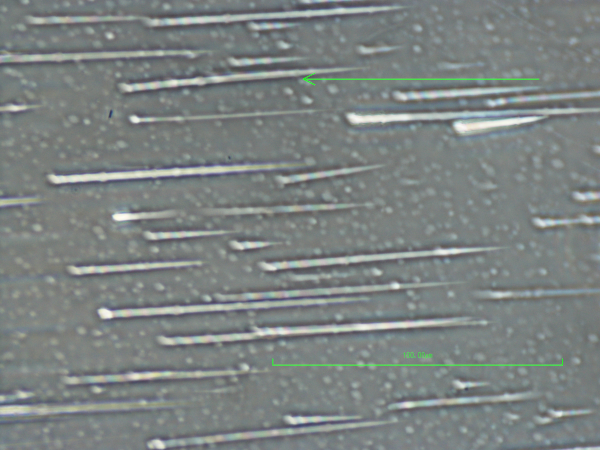}
 \caption{Phase contrast optical microscope image of muscovite mica irradiated with a 500\,MeV/u Fe ion beam, etched for 80\,min with 48\% HF. White lines are due to Fe ion beam and distributed white spots are due to $\alpha$-recoil defects. The scale bar is 100\,$\mu$m.}
 \label{fig:Toho_Fe_mica}
\end{figure}

For the monopole and the Q-ball search, a heavy ion beam is utilized for calibration studies. Recently, we checked the track formation by such heavy ions. We used a $500\,$MeV/u Fe ion beam at HIMAC, National Institute for Quantum Science and Technology (QST), Japan. In this test, muscovite mica was irradiated parallel to the mica surface with the Fe ion beam with a dose of $\sim 10^{8}\,{\rm cm}^{2}$, and the sample was etched with HF (48$\%$). The example image is shown in Fig.~\ref{fig:Toho_Fe_mica}. In this test, a clear track from the Bragg peak with a length of several tens of microns was observed. The energy loss value is around $30\textit{--}90\,$GeV/cm in the tracks' upstream region and $\mathcal{O}(1)\,$GeV/cm in the stopping region (combined electron and nuclear stopping power). The energy loss in the stopping region is similar to expectation from a charged Q-ball with $Z_{Q}>10$ and speed of  $\beta=10^{-3}$. The data shows that tracks are formed efficiently in this region.  We will investigate in more detail the properties of the etch-pit depending on the energy loss, etching condition, annealing effect, detection efficiency, etc.

We are also studying olivine as a lower background mineral. We have exposed olivine with Kr ions and will study the etch-pit formation by TEM. Finally we will work on understanding the $^{238}$U concentrations in natural samples. These studying are now on-going, and will be an important step towards lower-background and higher-sensitivity searches for Dark Matter with minerals. 

%*********************************************************
\subsection{Karlsruhe Institute of Technology \& Heidelberg University} \label{sec:Studies-KIT}
%*********************************************************
%{\color{blue} Coordinator: Klaus Eitel}

Team: {\it K.~Eitel, A.~Elykov, U.~Glasmacher}
\vspace{0.3cm}\\
In 2019, a prospective joint project entitled ``Searching for Dark Matter particle signatures with salt minerals as Palaeo-Detectors''~\cite{HEiKA} was funded by the strategic partnership HEiKA between Heidelberg University and Karlsruhe Institute of Technology (KIT). The aims of this project were numerical simulations of WIMP-induced track formation in halite (NaCl) as well as the visualization of recoil tracks in halite by etching and optical microscopy as major ingredients to explore the discovery potential in a specific mineral from a well-known geological site. For that purpose, the Permian (260 Myr) salt deposit Unterbreizbach (Thuringia, Germany) was chosen where the salt is located at a low temperature at a depth between 400\,m and more than 1000\,m since more than 80\,Myr. The modeling of spectra and event rates for a set of different WIMP masses and a moderate exposure of 1\,kg\,Myr demonstrated in general a sensitivity of paleo-detectors competitive or even surpassing direct WIMP search experiments in a scenario where cosmic-induced background as well as recoils from $\alpha$-decays are negligible. As expected, a major result of our study for Unterbreizbach is the fact that the background rate of cosmic-induced nuclear tracks exceeds a potential WIMP signal for a (spin-independent) WIMP-nucleon cross section of $\sigma_p^{\rm SI} = 10^{-45}$\,cm$^2$ by up to almost 2 orders of magnitude (e.g.~at track lengths of 100\,nm and a WIMP mass of 50\,GeV/c$^2$) rendering a WIMP search with halite from Unterbreizbach not competitive with direct WIMP searches~\cite{Fuhrer}. As an interesting by-product of our studies, a comparison between nuclear track lengths for small energies of $\approx$100\,keV modeled with the \texttt{SRIM} and \texttt{GEANT4} packages shows differences of up to 50\%. % (\texttt{SRIM} with respect to \texttt{GEANT4}).

All analyzed Natural (NH) and Artificial Halite (AH) crystals showed a large amount of etch pits caused by crystal defects. Artificially created crystal defects were etched using two different etching solutions. 
Keeping NH, AH, and sylvinite (KCl) crystals for a defined time at a certain temperature did anneal the crystal defects. If, however, the temperature is increased, new crystal defects are formed. Annealed halite and sylvinite was used to detect cosmic rays and spallation products, effectively. Raman and UV-C fluorescence spectroscopy could be used to visualize artificial and natural crystal defects in volume. With increasing ion fluence, the Raman bands change and point to the formation of Na- and Cl-cluster. As a result, the specific type of crystal defect cannot be distinguished by the applied analytical techniques. Our studies showed that halite from the salt deposit Unterbreizbach cannot be used as a paleo-detector competitive for Dark Matter search. However, we could identify several steps which have to be taken to fully exploit the Dark Matter discovery potential of paleo-detectors.

At present, we aim to formulate and conduct several key feasibility and calibration studies for the use of minerals as paleo-detectors. The Karlsruhe Nano Micro Facility at KIT offers a wide range of state-of-the-art nm-scale manipulation and imaging technologies, including, among others, AFM, FIB-SEM, TEM and HIM. Access to these facilities provides us with the opportunity to compare several high-resolution track imaging techniques. To that end we have begun a selection process of minerals, whose samples will then be irradiated with low- and high-energy particles, giving rise to nuclear recoil tracks with a wide range of lengths. Subsequently, we will use several microscopy techniques to read out the resulting damage tracks. The observed tracks will then be classified and compared to expectations from simulations, with the ultimate goal of correlating their number and morphology with the deposited energy. In addition, we will study the effects of temperature variation on the survival of these nm-sized tracks in the selected mineral samples. Therefore, in the near future, we will tackle several of the key hurdles facing paleo-detectors in neutrino physics and Dark Matter searches.

%*********************************************************
\subsection{Queen's University} \label{sec:Studies-Queens}
%*********************************************************
%{\color{blue} Coordinator: Joe Bramante}

Team: {\it L.~Balogh, Y.~Boukhtouchen, J.~Bramante, A.~Fung, M.~Leybourne, T.~Lucas, S.~Mkhonto, A.~Vincent}
\vspace{0.3cm}\\
Queen's University's Paleodetection Group is currently exploring a number of experimental and theoretical aspects of astroparticle detection in minerals. On the experimental side, a number of suitable olivine and galena samples have been obtained, and are currently undergoing calibration using the Queen's University Reactor Materials Testing Laboratory, which houses a tunable 1-8 MeV proton beamline capable of 45 $\mu$A current, purpose built for calibrating Dark Matter detectors. Irradiated minerals will be readout using High-Resolution Transmission Electron Microscopy (HRTEM). 

Theorists working in the Queen's group are studying the energy deposition from composite Dark Matter and other highly-energetic dark sector particles, which are being calibrated for readout in olivine and galena, using \texttt{SRIM} along with analytic methods. Theorists are also reexamining single nuclear recoil signatures of Dark Matter in minerals. In contrast to prior approaches, it is being examined whether a full Monte Carlo simulation approach to the predicted track length per event will allow for a fuller characterisation of the expected spectrum, including the high-track length tail, as well as a direct translation from the experimental calibration with a proton beam to the expected signal from Dark Matter or neutrino interactions with nuclei in the material. 

%*********************************************************
\subsection{PALEOCCENE} \label{sec:Studies-VT}
%*********************************************************
%{\color{blue} Coordinator: Patrick Huber}

Team: {\it K.~Alfonso, G.R.~Araujo, L.~Baudis, N.~Bowden, B.K.~Cogswell, A.~Erickson, M.~Galloway, A.A.~Hecht, R.H.~Mudiyanselage, P.~Huber, I.~Jovanovic, G.~Khodaparast, B.A.~Magill, T.~O'Donnell, N.W.G.~Smith, X.~Zhang}
\vspace{0.3cm}\\
The PALEOCCENE concept~\cite{Cogswell:2021qlq} aims at the passive detection of low energy nuclear recoil events with color center. The cause for the nuclear recoil could be Dark Matter, neutrinos or neutrons. Color centers allow for optical readout via fluorescence microscopy and specifically the use of selective plane illumination microscopy (SPIM, see section~\ref{sec:ReadOut-Optical}) should allow the detection of single color centers in a macroscopic cubic-centimeter volume. PALEOCCENE is now a collaboration of about 20 scientists at 7 institutions in the U.S. and Europe with expertise ranging from  particle and nuclear physics, over solid state physics and nuclear engineering to optical microscopy. Initial studies have focused on CaF$_2$ due to a combination of previous studies~\cite{Pelka:2017} including theoretical modeling~\cite{MORRIS2020109293} and suitable excitation and emission wavelength in the visible spectrum~\cite{Mosbacher:2019igk}. CaF$_2$ crystals have been exposed to MeV-neutron fluences from $10^7-10^{18}\,\mathrm{n}\,\mathrm{cm}^{-2}$ from both radioactive AmBe sources and a nuclear reactor. Other samples were exposed to MeV gamma ray fluxes from Co-60 source of $(1-50)\times 10^{13}\,\mathrm{ph}\,\mathrm{cm}^{-2}$. In response to the neutron irradiation, spectral emission features at around 600\,nm and 640\,nm were identified in response to 488\,nm laser excitation~\cite{Alfonso:2022meh}. These features show a clear relation to the received dose, albeit, a non-linear one. Only up to fluences of about $2\times10^8\,\mathrm{n}\,\mathrm{cm}^{-2}$ is the emission strength proportional to the dose; at higher fluence the emission goes down to nearly zero at around $10^9\,\mathrm{n}\,\mathrm{cm}^{-2}$. The features at 600 and 640\,nm are quite narrow (about 2\,nm wide).  At the much higher gamma fluences from the Co-60 source, we find that the emission mainly happens at more than 700\,nm and a very broad feature is observed. At the very highest Co-60 fluences the features at 600 and 640\,nm are suppressed. For the reactor irradiations, it is unclear if the effect mainly stems from neutrons or from the gamma rays which are present as well at a very high rate in reactor. The spectral signature from the reactor samples is similar to the Co-60-induced features but of course much more pronounced. The preliminary conclusions from this work are that the dose/emission strength relationship is only linear at the very lowest doses and thus previous studies conducted at intermediate doses ($10^{10}\,\mathrm{n}\,\mathrm{cm}^{-2}$) can not be easily extrapolated. The spectral signature of neutron and gamma irradiation is very different and thus no confusion arises, i.e., there is very good gamma/neutron separation. 

Our studies using the mesoSPIM light sheet microscope (see section~\ref{sec:ReadOut-Optical}) have clearly identified color centers in response to gamma irradiation and have now also started seeing track events. Given the large amounts of data produced by the fast scans with the mesoSPIM, further studies are ongoing with focus on automatizing its data processing to enable to eventually reach sensitivity to the cosmic neutron background as intermediate milestone. In the future, super-resolution techniques could push the resolution in the range of tens of nanometers.

This optical detection scheme can enable Dark Matter detection and reactor neutrino detection as well as neutron detection. It also could serve as a first level screening of old rock samples to identify track candidates which then could be further investigated  using high-resolution techniques.

%*********************************************************
\subsection{University of Maryland} \label{sec:Studies-MD}
%*********************************************************
%{\color{blue} Coordinator: Reza Ebadi}

Team: {\it R.~Ebadi, M.~C.~Marshall, A.~Mathur, E.~H.~Tanin, S.~Rajendran, R.~L.~Walsworth}

\subsubsection{Directional detection of WIMP Dark Matter using solid-state quantum sensing} \label{sec:Studies-MD-Directional}
Time Projection Chamber (TPC) nuclear recoil detectors are currently the leading platforms for searching for particle Dark Matter with masses ranging approximately from GeV to TeV, also known as Weakly Interacting Massive Particle (WIMP) Dark Matter~\cite{Akerib:2022ort}. The improvement in sensitivity in recent years has been primarily driven by increasing the target mass, while keeping the backgrounds under control. In the near future, however, these detectors will be sensitive to the so-called {\it neutrino fog} – an irreducible background due to coherent elastic neutrino-nucleus scattering~\cite{Billard:2013qya}. The neutrino background hinders improved sensitivity of traditional methods for two main reasons: first, the nuclear recoil spectra due to neutrino scattering closely resembles that due to Dark Matter scattering; second, there are significant uncertainties associated with neutrino fluxes (the relevant neutrino fluxes being those from solar neutrinos, atmospheric neutrinos, and the diffuse supernova neutrino background)~\cite{OHare:2021utq}. Therefore, background subtraction schemes are challenging for traditional TPC detectors.

Directional detection of nuclear recoils offers a promising method of discriminating neutrino signals from putative WIMP Dark Matter signals~\cite{Vahsen:2021gnb}. The WIMP flux in the laboratory frame is expected to exhibit a dipolar angular distribution due to the solar system's motion towards the Cygnus constellation, the so-called {\it WIMP wind}. The angular distribution of solar neutrinos peaks toward the Sun's direction at any given time, which is at least 60 degrees apart from the Cygnus constellation throughout a year. A sufficient directional resolution, thus, allows one to distinguish between nuclear recoils induced by neutrinos and those induced by Dark Matter. Gas-phase TPC detectors have demonstrated low-threshold directional nuclear recoil detection capabilities, suitable for directional WIMP detection~\cite{Vahsen:2020pzb}. However, these detectors face a scaling up challenge, requiring extremely large volumes in order to operate at the neutrino fog level. (See Ref.~\cite{Vahsen:2021gnb} for more details, as well as a discussion of other proposed directional detectors.)

A proposed solid-state directional detector combines compactness (meter-scale) and sensitivity to the neutrino fog. The directional signal in crystalline solid-state detectors is a submicron-scale damage track correlated with the incident particle's directions~\cite{Rajendran:2017ynw}. Solid-state quantum sensing offers a promising pathway towards high-precision, high-resolution directional readout of particle-induced tracks. The proposal for such a detector is presented by Rajendran {\it et al.}~\cite{Rajendran:2017ynw}; an extended evaluation of requirements and pathway for experimental realization is outlined by Marshall {\it et al.}~\cite{Marshall:2020azl}; and a recent review~\cite{Ebadi:2022axg} describes the current state of the proof-of-principle experiments and the planned next steps. Current studies and developments are primarily focused on diamond as a target material.\footnote{Silicon carbide (SiC) may be an interesting alternative to diamond. Similar to diamond, SiC has Nitrogen-Vacancy (NV) centers that could potentially be used to map the strain induced by a nuclear recoil track. SiC can be instrumented to register nuclear recoil events based on charge, phonon, or photon collection~\cite{Griffin:2020lgd}. Compared to diamond, the advantage of SiC is that it may be more feasible to produce sufficient volume of high-quality crystals as would be required for a competitive Dark matter search.} Diamond is an integral component of emerging quantum technologies; it exhibits well-characterized quantum defects with a wide range of applications~\cite{Barry:2019sdg}. 

The proposed diamond-based detection scheme combines different techniques from multiple disciplines in order to achieve directional readout. First, the nuclear recoil registration is based on charge, phonon, or photon collection. Such nuclear (and electron) recoil detection in semiconductors is an established technology in existing Dark Matter detectors~\cite{LopezAsamar:2019smu,Rau:2020abt,DAMIC:2021esz}, with recent interest in adapting it for a diamond detector~\cite{Kurinsky:2019pgb,Essig:2022dfa,Canonica:2020omq,Abdelhameed:2022skh}. Such real-time detection allows time-stamping each event (and thus identifying relative directions of incident particle, the WIMP wind, and solar neutrinos) and spatially localizing events at the mm scale. The nuclear recoil also leaves a durable damage track in the lattice structure. Having identified a mm-scale segment of the detector that bears the damage track induced by Dark Matter or neutrinos, the damage track can be reconstructed in 3D to provide directional information. In the following, we briefly summarize the methods that have been identified as suitable for such track reconstructions. To avoid prohibitively long scanning times, this step is divided into two hierarchical stages after the mm-scale segment has been identified and removed from the meter-scale overall detector: micron-scale localization and three-dimensional nanoscale reconstruction.

\underline{Micron-scale localization} can be achieved with optical-diffraction-limited strain spectroscopy. Nitrogen-Vacancy (NV) centers are atomic-scale quantum defects within diamond lattice, spin states of which can be coherently manipulated and read out (see Sec.~3 of Ref.~\cite{Ebadi:2022axg} for a concise and pedagogical review). Notably, the NV ground state spin transitions are sensitive to local lattice strain. A widefield Quantum Diamond Microscope (QDM) allows parallel readout of NV centers' spin states (which encode strain information) with a spatial resolution limited by optical diffraction. The expected WIMP or neutrino-induced strain, averaged over a ${\rm \mu m^3}$ voxel size, is about $10^{-7}$ to $3\times10^{-6}$. In a recent work, Marshall {\it et al.}~\cite{Marshall:2021xiu} demonstrated high-precision and fast strain mapping with about a ${\rm \mu m}$ resolution over a mm-scale field-of-view. The reported sensitivity is $5(2)\times 10^{-8}/\sqrt{\rm Hz\cdot \mu m^{-3}}$, surpassing the requirement for localization of the Dark Matter track. According to the reported sensitivity, mapping a ${\rm mm^3}$ segment of diamond will take about 13 hours, much faster than time scales relevant for background contamination. A key component of the demonstrated strain spectroscopy is a dynamical decoupling measurement protocol, dubbed strain-CPMG, which is insensitive to magnetic inhomogeneities arising from the electronic and nuclear spin bath within the diamond lattice. This protocol allows for longer NV ensemble spin dephasing times and thereby enhances strain sensitivity. In the described study, $z$-resolution was not precisely controlled using the QDM setup. Hence, a full three-dimensional micron-scale resolution has yet to be demonstrated. Optical sectioning methods such as light-sheet microscopy and structured illumination microscopy provide a promising pathway to achieving sufficient $z$-resolution~\cite{Marshall:2020azl,Ebadi:2022axg}.

Alternatively, NV creation can be used to localize particle impact sites instead of strain spectroscopy~\cite{Marshall:2020azl}. This scheme uses diamonds with low density of pre-existing NV centers, but a high nitrogen impurity density, as detector material. As a result of WIMP or neutrino scattering, lattice vacancies are generated, which can become mobile under high temperatures until they find a nearby nitrogen and form stable NV centers. The localization can then be achieved through fluorescent detection of the induced NV centers. Given a high concentration of nitrogen impurities (about 200 parts per million), the travel range of vacancies can be significantly reduced, preserving the directional information that can be read out by nanoscale mapping methods.

\underline{Nanoscale reconstruction} of the damage track can be done either using superresolution NV strain spectroscopy or scanning X-ray diffraction microscopy. Superresolution techniques have enabled nanoscale magnetic imaging using NV-diamond systems. To name a few, stimulated emission depletion (STED)~\cite{STED_Early_Hell2009} and charge state depletion (CSD)~\cite{CSDmicroscopy2015} have been used to achieve $\lesssim$ 10 nm resolution in NV-diamonds. It is expected to be relatively straightforward to adapt such methods to strain sensing using a suitable measurement protocol, e.g., strain-CPMG. Nanoscale resolution in lateral and axial dimensions can be achieved using a combination of different methods. In addition to superresolution NV microscopy, scanning X-ray diffraction microscopy (SXDM) is also a feasible method for mapping nanoscale strain features. In a recent study, Marshall {\it et al.}~\cite{Marshall:2021kjk} have demonstrated key components of SXDM for the directional WIMP detection proposal. In this study sub-micron strain features were resolved, with strain resolution of $1.6\times 10^{-4}$ (note that the expected WIMP or neutrino-induced strain at a 30\,nm spot is about $1.8\times 10^{-4}$). The intrinsic strain features of a CVD diamond were also reconstructed in 3D via performing SXDM from two distinct diffracting planes at relative incident X-ray angles. In addition, limited high-resolution SXDM scans were performed in randomly selected areas of a CVD diamond, which did not reveal intrinsic strain features resembling a WIMP or neutrino signal at nanoscale; while this result is promising, a comprehensive study of false positive rates is required in the future to establish a conclusive assessment.
 
In light of these promising proof-of-principle experiments, further development of the techniques are motivated to achieve a full measurement pipeline that is not only sensitive to WIMP and neutrino-induced damage tracks, but also is cost and time effective in localizing and reconstructing signals within mm-scale diamond chips. This goal is not far from the current state of the art. In the near term, precise single ion implantation experiments will enable experimental characterization of directional sensitivity to single-particle-induced tracks. In the medium term, instrumentation of a prototype diamond-based low-threshold nuclear recoil detector will be needed. Diamond-based directional detectors, sensitive to the neutrino fog, will also rely on the availability of large volumes of diamond (cubic meter scale). Over the past decade, quantum-grade diamond growth technology has advanced significantly; a reasonable advance in the near future will enable low-cost production of repeatable, high-quality diamond chips for the WIMP detector. To summarize, the work toward directional WIMP Dark Matter detectors based on diamond will leverage or motivate advances in multiple disciplines, including atomic physics, particle physics, quantum science, and material science.

\subsubsection{Ultra-heavy Dark Matter detection using geological quartz} \label{sec:Studies-MD-Quartz}
Ultra-heavy composite states can generically arise in a dark sector that exhibits strong self-interactions. Such composite states will comprise part or all of the local Dark Matter content with low number density. If the ultra-heavy Dark Matter interacts classically with Standard Model particles, it can leave continuous damage tracks in rocks. Due to the large momenta carried by these states, the resulting tracks would generically be straight and very long. A suitable imaging technique can be used to search for such geometrically distinctive tracks in an ancient rock sample. The Dark Matter mass reach of this scheme is set by the age $T$ and total area $A$ of rock samples scanned~\cite{Ebadi:2021cte}:
\begin{equation}
    m_{\rm DM}\lesssim 1\text{ kg}\left(\frac{A}{1\text{ m}^2}\right)\left(\frac{T}{1\text{ Gyr}}\right)
\end{equation}
Paleodetection is a compelling technique in this context because it enables sensitivity to larger Dark Matter masses $m_{\rm DM}$ by maximizing exposure times. 

A search for long damage tracks in ancient mica crystals was conducted by Price and Salamon in the 1980s~\cite{Price:1986ky}. They did not observe such tracks in $\sim \text{Gyr}$ old mica samples with a total area of $A\sim 0.1\text{ m}^2$, thereby limits on Dark Matter states with mass $m_{\rm DM}\lesssim 0.1 \text{ kg}$ can be found~\cite{Acevedo:2021tbl}. The readout scheme employed by Price and Salamon includes enlarging defects with acid etching to make them visible under optical microscopy. Since this process enlarges not only signals but also backgrounds, sufficiently clean mica samples must be used, limiting feasibility of this method to scan larger sample volumes. Nevertheless, imaging technologies have advanced rapidly in the last few decades, offering new opportunities for probing larger Dark Matter masses.

Quartz, an abundant mineral with particularly well-studied backgrounds, serves as a good paleo-detector. A passing ultra-heavy Dark Matter state that imparts $\gtrsim \text{eV}$ energy to nuclei can melt quartz along its path. The melted region would not re-crystallize as it cools, leaving a cylindrical damage track in the form of amorphous silica whose radius depends on the amount of energy it deposits per unit length $dE/dx$. Such tracks would be visible through the modality of scanning electron microscopy combined with cathodoluminescence spectroscopy (SEM-CL). This technique has the advantage of a fast scanning rate with micron-scale resolution, thus probing $dE/dx\gtrsim \text{ MeV/\AA}$. The long, cylindrical form of the expected damage tracks can be distinguished from other sources of disruption to the quartz lattice (such as radioactive decays, extended growth defects, or cosmic rays), which do not persist beyond a few millimeters. The unique geometry of the expected signal can be leveraged in designing readout schemes to efficiently search for ultra-heavy Dark Matter; e.g., Ebadi {\it et al.}~\cite{Ebadi:2021cte} proposed correlating multiple layer scans to find tracks that go beyond any expected background. 

A type of quartz that is known to have low CL background is hydrothermal vein quartz. This is supported by a proof-of-principle SEM-CL scan of a vein quartz sample that is reported in Ref.~\cite{Ebadi:2021cte}. In addition, this experiment demonstrated a key aspect of the proposed readout scheme: the ability to identify radioactivity damage in quartz, which locally resembles tracks expected to be induced by ultra-heavy Dark Matter. Because high-quality vein quartz can be found abundantly, e.g. in Jack Hills, Australia, the mass reach of this method will primarily be determined by the available scanning time for SEM-CL instruments. Using the scan rate of the mentioned proof-of-principle experiment, $\sim 100\,\text{min/cm}^{2}$, as a benchmark, four SEM-CL devices over two years can scan a total area of $1\,\text{m}^{2}$, corresponding to a Dark Matter reach of $m_{\text{DM}}\lesssim 1\,\text{kg}$. Extrapolating, twenty devices over four years can scan $10\,\text{m}^{2}$, and a hundred devices over eight years can scan $100\,\text{m}^{2}$, corresponding to a mass reach of $\sim 10\,\text{kg}$ and $100\,\text{kg}$, respectively. In conclusion, the initial studies suggest a promising approach for probing an uncharted parameter space of ultra-heavy Dark Matter using SEM-CL scanning of geologically old vein quartz samples.

Future work includes calibrating both the signal (by studying the CL emission from artificial damage tracks created by a laser in synthetic quartz) and the noise (by studying the CL emission from a range of noise sources). An improved grasp of the signal-to-noise ratio of SEM-CL imaging will potentially reduce the detection threshold below what is required for melting by determining the sensitivity of the technique to long tracks of lattice distortions. We thus expect that a wide range of ultra-heavy Dark Matter parameter space, in both mass and coupling to the Standard Model, will be accessible to this methodology.

%*********************************************************
\section{Next Steps} \label{sec:LaundryList}
%*********************************************************
%{\color{blue} Coordinator: Sebastian Baum/Patrick Stengel}

There are plentiful challenges on the way towards unleashing the full potential of minerals as detectors for neutrinos and Dark Matter. In this section, we collect a number of goals that could be achieved in the next few years on this path. Let us stress that the order of this list is neither intended as a timeline nor as an ordering in importance or difficulty.
\begin{itemize}
    
    \item Currently, the standard tool to compute the latent damage in general materials from nuclear recoils in the $\mathcal{O}(0.1)\textit{--}\mathcal{O}(100)\,$keV nuclear recoil energy range is \texttt{SRIM}~\cite{Ziegler:1985,Ziegler:2010}. While this software is well tested, not least because of its importance to chip design and manufacturing, it also lacks certain features relevant for mineral detectors such as accounting for the crystal structure of the material. One possible avenue is to use molecular dynamics simulations to improve understanding of damage tracks, see, e.g., Ref.~\cite{PhysRevB.93.035202} for a study for diamond. Furthermore, connecting the crystal damage computed in software such as \texttt{SRIM} to the observable in any given readout technique is non-trivial. Studies of reading out laboratory-induced (e.g., by ion- or neutron-irradiation) crystal damage in different materials with different readout techniques over the range of nuclear recoil energies of interest can lay the foundation for developing better modeling tools.
    
    \item There is currently no universal understanding of the criteria for nuclear recoils to produce latent crystal damage. For the applications described in this work, the most important corollary is that the recoil energy threshold to form latent damage tracks in a given material that can be read out with a given readout technique is not well understood. This is another question that can be addressed in studies of reading out laboratory-induced crystal damage in different materials.
    
    \item For many of the applications discussed in this whitepaper, it is crucial to measure not only the presence of the latent crystal damage induced by nuclear recoils, but also to reconstruct, on an event-to-event basis, the energy of the nuclear recoil. In principle, the latent crystal damage does hold information about the nuclear recoil energy, e.g., via the length of a damage track or the number and distribution of color centers along a nucleus' path. Establishing the nuclear recoil energy resolution achievable with a given readout technique in a given mineral is another question that can be addressed in studies of reading out laboratory-induced crystal damage in different materials.
    
    \item To date, the expected background from radiogenic neutrons for paleo-detector searches for astrophysical neutrinos or Dark Matter has never been observed. Once microscopy techniques for the readout of nuclear recoils with few-keV recoil energy threshold are established, it should be relatively straightforward to test the predictions for radiogenic-neutron-induced recoils. This can be done by measuring the latent damage features in natural mineral samples with fission-track ages of $\gtrsim 10\,$Myr obtained from relatively uranium-rich deposits, e.g., where $^{238}$U concentrations are of a few tens of ppm.
    
    \item Similarly, cosmogenic neutron backgrounds for paleo-detector searches for astrophysical neutrinos or Dark Matter have never been observed. Similarly to the radiogenic-neutron-induced background, it should be relatively straightforward to test predictions for this background  by measuring the latent damage features in natural mineral samples with fission-track ages of $\gtrsim 10\,$Myr obtained from sites which have little shielding from cosmic rays. Furthermore, one could establish the overburden (history) and the geological understanding thereof by measuring this cosmogenic background in samples obtained from different depths and from regions with different geological histories.
    
    \item For searches for astrophysical neutrinos and Dark Matter with paleo-detectors, thermal annealing of the latent damage features might be important depending on the age and thermal history of samples and the particular mineral. Most of what is known about thermal annealing to date is based on fission tracks; extending these studies to the damage features from $\mathcal{O}(0.1)\textit{--}\mathcal{O}(100)\,$keV nuclear recoils is an important task that must be tackled in order to understand which minerals from which geological deposits are best-suited as paleo-detectors. Understanding the thermal annealing of color centers better might also be important for identifying the best materials to use for, e.g., monitoring reactor neutrinos and other nuclear safeguarding applications.
    
    \item Mineral detector searches using keV-scale nuclear recoils, e.g.\ for solar, supernova, or reactor neutrinos as well as for WIMP-like Dark Matter, are susceptible to radiogenic backgrounds. In particular, paleo-detector applications require highly radiopure samples. In order to identify the best geological sites from which to source paleo-detector samples and which minerals from such sites are best suited, radioassays of minerals from candidate sites must be carried out. Measuring the concentration of uranium, thorium, and other radioactive trace elements down to the required levels of sub-ppb (parts per billion) is challenging, but possible using, for example, Laser Ablation Inductively Coupled Plasma Mass Spectrometry (LA-ICP-MS) or high-purity germanium detectors (perhaps after neutron activation of the samples). 

    \item As mentioned in various places throughout this whitepaper, reading out latent damage features with the required spatial resolution in sufficiently large samples to lead to the envisaged sensitivity will produce enormous amounts of raw data. In order to process and analyze these data, automated data analysis techniques will be necessary. Machine learning will likely play an important role in this task of finding and characterizing patterns in image(-like) data. The development of such techniques can begin immediately as the first data from feasibility studies, e.g., those discussed in section~\ref{sec:Studies}, is already available and more such data will be produced as these studies progress.
    
\end{itemize}

%*********************************************************
\section{Summary} \label{sec:Summary}
%*********************************************************
%{\color{blue} Coordinator: Sebastian Baum/Patrick Stengel}

This whitepaper discusses a wide range of potential applications of minerals as nuclear recoil detectors. For example, minerals could be used to search for Dark Matter: one could use natural minerals from well-understood geological sites as ``paleo-detectors'' for Dark Matter as discussed in sections~\ref{sec:Physics-WIMPs},~\ref{sec:Physics-CompositeDM},~\ref{sec:Studies-SLAC}--\ref{sec:Studies-KIT}, and~\ref{sec:Studies-MD-Quartz}. Minerals that formed hundreds of millions of years ago on Earth could have recorded traces from Dark Matter interactions over geological times, enabling enormous exposures. Beyond their raw discovery potential, such searches would, if successful, also provide unique information about the properties of Dark Matter particles and their distribution in our Galaxies for a wide range of Dark Matter candidates ranging from conventional Weakly Interacting Massive Particles (WIMPs) as light as a few hundred MeV to composite Dark Matter candidates with macroscopic masses as large as 100\,kg. Another interesting possibility is to use an array of laboratory-manufactured crystal detectors as a directional direct detection experiment for GeV-scale Dark Matter candidates, see section~\ref{sec:Studies-MD-Directional}. By instrumenting each crystal in the array similar to conventional Dark Matter detectors~\cite{Schumann:2019eaa} one could register candidates for a nuclear recoil caused by Dark Matter in that crystal, and by subsequently measuring the length and direction of the damage track caused by the recoiling nucleus one can gain directional sensitivity. 

Searches for neutrinos and cosmic rays are also potentially interesting applications for mineral detectors. Standard Model interactions of neutrinos and cosmic ray particles are a guaranteed source of recoils across a range of energies similar to the nuclear recoils Dark Matter might induce. As described in sections~\ref{sec:Physics-AstroNu} and~\ref{sec:Physics-CosmicRays}, mineral ``paleo-detectors'' of neutrinos and cosmic rays could probe a variety of astrophysical phenomena. Neutrinos originating from our Sun and galactic core collapse supernovae can leave a record of interactions with nuclei in mineral detectors, perhaps allowing for a detailed study of solar evolution and revealing the star formation history in the Milky Way. Atmospheric neutrinos and cosmic ray air showers are produced by primary cosmic ray interactions in the atmosphere. Mineral detection of these cosmic ray byproducts could offer a unique probe of the cosmic ray intensity as the solar system traverses the Galaxy over 100\,Myr timescales. Since the fluxes of atmospheric neutrinos and cosmic rays penetrating the surface of the Earth also depend on the density and composition of the atmosphere, as well as the geomagnetic field, mineral detectors can also be sensitive to the evolution of the Earth and its climate. Although it is a particularly challenging application, the mineral detection of geoneutrinos described in section~\ref{sec:Physics-GeoNu} could provide a window into the thermal history of the Earth’s crust and the mantle.

Minerals might also have a unique role to play in the detection of man-made neutrinos. As discussed in sections~\ref{sec:Physics-ReactorNu} and~\ref{sec:Studies-VT}, laboratory-produced crystals could be used as detectors for reactor neutrinos - similar to solar and supernova neutrinos, reactor neutrinos give rise to nuclear recoils via coherent elastic neutrino-nucleus scattering. Since minerals are inherently passive detectors, they could offer a promising pathway to cost-effective, tamper-proof monitoring of nuclear reactors. Although not discussed in this whitepaper, minerals could also be used to detect neutrons via the nuclear recoils neutrons induce in a crystal. As neutron detectors, minerals could have further application as passive detectors for nuclear safeguarding, disarmament, and other security applications~\cite{vonRaesfeld:2021gxl,Cogswell:2021qlq}.

As discussed in this whitepaper, minerals have long been used as nuclear recoil detectors, in particular in the geosciences (see section~\ref{sec:MineralDetectors}). In the 1990s~\cite{Snowden-Ifft:1995rip,Snowden-Ifft:1995zgn}, the first efforts were made to use minerals as Dark Matter ``paleo-detectors.'' A major limitation that hampered such searches were the available readout methods. In the last decades, existing imaging techniques have been undergoing transformative developments and new techniques have been developed. For example, optical superresolution microscopy is now a routine technique at many university laboratories and coherent X-ray sources such as free-electron lasers have revolutionized the imaging possibilities at the few-nm scale. Microscopy techniques with sub-nm resolution such as electron microscopy, newly-developed helium-ion beam microscopes, and scanning probe microscopy techniques combined with fast sample preparation methods such as focused-ion beams or ultrafast lasers can image samples with ever increasing spatial resolution and sample throughput. At the same time, our capabilities of analyzing huge amounts of data, as will be required for most applications of mineral detectors discussed here, have been revolutionized in the 21st century thanks to the ever larger computing power at our disposal and the advances in machine learning techniques. The expertise in the physics and astronomy communities to deal with the data streams from modern experiments and observatories, e.g., at the Large Hadron Collider, the Vera~C.~Rubin Observatory, or the Square Kilometer Array, give us confidence that the data analysis challenge for mineral detectors will be met.

The successful development of imaging and data analysis techniques for nuclear recoil damage in minerals would also allow for a revolution in the geoscience applications that started the field of minerals as solid state nuclear track detectors half a century ago. As touched upon in section~\ref{sec:Physics-Geoscience}, such techniques could vastly expand the capabilities to date geological minerals via fission track and $\alpha$-recoil track dating. Moreover, by studying the distribution of partially annealed damage tracks, one can infer the time-temperature history of samples in their geological environments. 

As we have seen in section~\ref{sec:Studies}, a number of research groups in Europe, Asia, and North America are actively pursuing experimental studies to bring mineral detectors to fruition. It will be exciting to witness the progress of these programs in the coming years and to see what can, and perhaps what cannot, be done. While doubtlessly many challenges will be encountered along the way, now is the right time to bring modern microscopy and data analysis techniques to minerals to see what they can teach us about the Universe we live in, near and afar.

%*********************************************************
\acknowledgments
%*********************************************************
We thank the Institute for Fundamental Physics of the Universe (IFPU) Trieste for hosting the MDDM$\nu$ workshop in October 2022. Many of the discussions leading to this whitepaper took place at the MDDM$\nu$ workshop.

The team at JAMSTEC thanks the technical staff of Marine Works Japan Ltd. at JAMSTEC, Katsuyuki Uematsu, for helping us to observe samples with SEM and TEM. The JAMSTEC group is also grateful to the technical staff of the tandem accelerator at JAEA-Tokai for supplying high-quality ion beams.

The work of S.~Baum is supported by NSF grant No. PHYS-2014215, DoE HEP QuantISED award No. 100495, and the Gordon and Betty Moore Foundation Grant No. GBMF7946. 
P.~Stengel is funded by the Instituto Nazionale di Fisica Nucleare (INFN) through the project of the InDark INFN Special Initiative:  ``Neutrinos and other light relics in view of future cosmological observations" (n. 23590/2021).
The work of G.R.~Araujo is supported by the Candoc Grant No. K-72312-09-01 from the University of Zurich. 
The work of J.~Bramante is supported by NSERC. Research at Perimeter Institute is supported in part by the Government of Canada through the Department of Innovation, Science and Economic Development Canada and by the Province of Ontario through the Ministry of Colleges and Universities. 
K.~Freese is Jeff \& Gail Kodosky Endowed Chair in Physics at the University of Texas at Austin and is grateful for support. K.~Freese is supported by the U.S. Department of Energy, Office of Science, Office of High Energy Physics program under Award Number DE-SC-0022021 and by the Swedish Research Council (Contract No. 638-2013-8993). 
S.~Horiuchi acknowledges support by DOE grant DE-SC0020262, NSF grants No.~AST1908960, No.~PHY1914409 and No.~PHY2209420,
and via JSPS KAKENHI JP22K03630. 
The work of P.~Huber is supported by U.S. DOE Office of Science DE-SC0020262 and National Nuclear Security Administration Office of Defense Nuclear Nonproliferation R\&D through the Consortium for Monitoring, Technology and Verification DE-NA0003920. 
B.J.~Kavanagh thanks the Spanish Agencia Estatal de Investigaci\'on (AEI, MICIU) for the support to the Unidad de Excelencia Mar\'ia de Maeztu Instituto de F\'isica de Cantabria, ref. MDM-2017-0765 and acknowledges funding from the Ram{\'o}n y Cajal Grant RYC2021-034757-I, financed by MCIN/AEI/10.13039/501100011033 and by the European Union ``NextGenerationEU''/PRTR. 
The research of U.~Glasmacher is supported by BMBF~05K22VH1 and the HEIKA-Project (ZUK 49/{\"U}6.3.057~HEIKA).
%\textbf{}
The work of A.~Gleason is supported by DOE Early Career Award, Fusion Energy Sciences, 2019. 
The work of K.~Murase is supported by the NSF Grant No.~AST-1908689, No.~AST-2108466 and No.~AST-2108467, and KAKENHI No.~20H01901 and No.~20H05852. 
S.~Rajendran is supported in part by the U.S.~National Science Foundation (NSF) under Grant No.~PHY-1818899, by the DOE under a QuantISED grant for MAGIS, and the Simons Investigator Award No.~827042. The work at Johns Hopkins University was supported by the U.S.~Department of Energy (DOE), Office of Science, National Quantum Information Science Research Centers, Superconducting Quantum Materials and Systems Center (SQMS) under contract No.~DE-AC02-07CH11359. 
The work of N.~Vladimirov is supported by URPP Adaptive Brain Circuits in Development and Learning (AdaBD) Program. 
The research of Toho and Nagoya University is supported by JSPS KAKENHI Grant Number 20K20944. 
The work at University of Maryland, College Park was supported by the Argonne National Laboratory under Award No. 2F60042; the Army Research Laboratory MAQP program under Contract No. W911NF-19–2-0181; the DOE fusion program under Award No. DE-SC0021654; and the University of Maryland Quantum Technology Center.

%*********************************************************
\bibliographystyle{JHEP.bst}
\bibliography{theBib}
%*********************************************************

\end{document}